\documentclass[11pt,a4paper]{article}
\pdfoutput=1

\usepackage{jheppub}
\usepackage[utf8]{inputenc}
\usepackage{multirow,graphicx,tabularx,mathrsfs,booktabs}
\usepackage{slashed}
\usepackage{url}
\usepackage{hyperref}
\usepackage[normalem]{ulem}
\usepackage{xspace}

\graphicspath{{figures/}}

\def\tablename{Table}
\def\figurename{Figure}

\newcommand{\mtot}{m_{T,\text{tot}}}
\newcommand{\vp}{\phi}

\newcommand{\vpjt}{\mbox{${\vp^\dag i\,\raisebox{2mm}{\boldmath ${}^\leftrightarrow$}\hspace{-4mm} D_\mu^{\,a}\,\vp}$}}

\newcommand{\toolfont}[1]{\textsc{#1}}
\newcommand{\ord}{\ensuremath{\mathcal{O}}}

\newcommand{\ope}[1]{\ensuremath{\mathcal{O}_{#1}}}
\newcommand{\opet}[1]{\ensuremath{\tilde{\mathcal{O}}_{#1}}}

\newcommand{\lumi}{\ensuremath{\mathcal{L}}}
\newcommand{\lag}{\ensuremath{\mathscr{L}}}

\newcommand{\met}{\ensuremath{\slashed{E}_T}}

\newcommand{\gev}{{\ensuremath\rm GeV}}

\def\ie{{i.\,e.}\ }

\newcommand{\madgraph}{\textsc{MadGraph}\xspace}
\newcommand{\madminer}{\textsc{MadMiner}\xspace}

\newcommand{\sally}{\textsc{Sally}\xspace}
\newcommand{\harry}{\textsc{Harry}\xspace}

\DeclareMathOperator{\pois}{Pois}
\DeclareMathOperator{\cov}{cov}
\newcommand{\diff}{\mathrm{d}}

\begin{document}

\title{Benchmarking simplified template cross sections in $WH$ production}

\author[a]{Johann Brehmer,}
\affiliation[a]{Center for Cosmology and Particle Physics, Center for Data Science, New York University, USA}

\author[b]{Sally Dawson,}
\affiliation[b]{Department of Physics, Brookhaven National Laboratory, Upton, NY, 11973, USA }

\author[b,c]{Samuel Homiller,}
\affiliation[c]{C. N. Yang Institute for Theoretical Physics, Stony Brook University, NY, 11790, USA}

\author[d,e]{Felix Kling,}
\affiliation[d]{Department of Physics and Astronomy, University of California, Irvine, USA}
\affiliation[e]{SLAC National Accelerator Laboratory, 2575 Sand Hill Road, Menlo Park, CA 94025, USA}

\author[f]{and Tilman Plehn}
\affiliation[f]{Institut f\"ur Theoretische Physik, Universit\"at Heidelberg, Germany}

\preprint{YITP-SB-19-22}

\abstract{
  Simplified template cross sections define a framework for the measurement and
  dissemination of kinematic information in Higgs measurements. We benchmark the
  currently proposed setup in an analysis of dimension-6 effective field theory
  operators for $WH$ production. Calculating the Fisher information allows us to
  quantify the sensitivity of this framework to new physics and study its
  dependence on phase space. New machine-learning techniques let us compare the
  simplified template cross section framework to the full, high-dimensional
  kinematic information. We show that the way in which we truncate the effective
  theory has a sizable impact on the definition of the optimal simplified
  template cross sections.
}

\maketitle

\clearpage
\section{Introduction}
\label{sec:intro}

Since the discovery of the Higgs boson at the LHC in 2012, the experimental and
theoretical communities have focused intense scrutiny on the question of whether
the observed particle has exactly the properties predicted by the Standard Model
(SM)~\cite{Dawson:2018dcd}. The interpretation of the vast number of Higgs
measurements requires a theoretical framework, which should be as flexible as
possible.  From a theoretical perspective and given that there are no non-SM
particles observed at the electroweak scale, potential deviations from the SM
are best analyzed in an effective field theory (EFT)
approach~\cite{Brivio:2017vri}. Such a framework allows us to systematically
include both rate information and kinematic distributions in the
analysis~\cite{Butter:2016cvz, Brivio:2016fzo}. In this study, we assume that
the Higgs boson is in an $SU(2)_L$ doublet and work in the so-called Standard
Model Effective Field Theory (SMEFT)~ In this case deviations from the SM are
parameterized in terms of a series of $SU(3)_c\times SU(2)_L\times U(1)_Y$
invariant operators and possible new physics effects are contained in the Wilson
coefficients of these operators. The expansion involves a new high energy scale
$\Lambda$, and we truncate the expansion with dimension-6 operators, whose
effects are generically suppressed relative to SM predictions by factors of
$v^2/\Lambda^2$ or $E^2/\Lambda^2$.

To interpret Higgs measurements in a combined experimental and theory analysis
we need to a) develop an efficient strategy to extract the maximum information
from Higgs events, and b) publish this information in an efficient form. Many
analyses focus on total rate measurements, but the available global fits have
shown that kinematic information on new Lorentz structures provides the most
powerful limits on new physics, for instance, exploiting the boosted regime of
$WH$ production~\cite{Banerjee:2013apa, Butter:2016cvz, DiVita:2017vrr,
Ellis:2018gqa, Almeida:2018cld, Biekotter:2018rhp}. For the first aspect this
means that total or fiducial rate measurements alone are not useful. As for the
second question, the experimental collaborations could choose to publish full
likelihoods~\cite{ATL-PHYS-PUB-2019-029, Collaboration:2242860,
Cranmer:2013hia}, which would ensure the most accurate modeling of systematic
uncertainties and their correlations. However, despite some recent progress this
strategy has not been widely adapted. Lacking the full likelihoods, the
publication of measured event counts or cross sections in all bins of a
histogram including uncertainties and their correlations is
desirable~\cite{Maguire:2017ypu}. In reality, global fitting
projects~\cite{Leung:1984ni, Buchmuller:1985jz,
GonzalezGarcia:1999fq,Grojean:2018dqj,deBlas:2016nqo,Ellis:2018gqa} often rely
on extracting the necessary information from plots in publications and
backwards-engineering as much of the analyses as possible.

An ad-hoc solution to these issues is the use of  simplified template cross
sections (STXS)~\cite{Tackmann:2138079, Berger:2019wnu} which propose to measure
and publish cross sections in slices of simple phase-space parameters.  They
have been proposed for all major Higgs production channels and are continuously
developed alongside the corresponding analysis opportunities. Given the
established SMEFT approach we can benchmark any proposed method of extracting
and publishing kinematic event information. A particularly well-suited approach
for benchmarking is based on information geometry. Its central object, the
Fisher information, encodes the maximal precision with which continuous
parameters of a Lagrangian can be measured, and therefore allows us to directly
compare the power of single LHC observables or STXS to the power of multivariate
analysis strategies~\cite{Brehmer:2016nyr, Brehmer:2017lrt}. In our case, these
parameters are the Wilson coefficients of the dimension-6 SMEFT operators. Until
now, a major limiting factor in this approach was that detector effects and
invisible particles like neutrinos could not accurately be taken into account.
This problem was recently solved with a technique that combines matrix element
information and machine learning~\cite{Brehmer:2018eca, Brehmer:2018kdj,
Brehmer:2018hga}, automated in the \madminer{} tool~\cite{Brehmer:2019xox}.

In this paper we benchmark the $WH$ production process, which is well known for
its leading impact in extracting SMEFT coefficients~\cite{Biekotter:2018rhp,
Freitas:2019hbk}.  In Sec.~\ref{sec:eft}, we first review the SMEFT framework
and define the Wilson coefficients included in our study.  The basics of the
Fisher information approach and the \madminer machinery~\cite{Brehmer:2019xox}
are described in Sec.~\ref{sec:stats}. Physics results on the relationships
between effective operators and kinematic regions are presented in
Sec.~\ref{sec:phys}, starting with the simple kinematic distributions of
information. Motivated by the differences between effects arising at linear and
quadratic order in the Wilson coefficients, we propose an improved  STXS
definition for $WH$ production in Sec.~\ref{sec:phys_2d} and benchmark the reach
in EFT coefficients in Sec.~\ref{sec:phys_limits}. Many technical details as 
well as a discussion of systematic uncertainties are given in the appendices.

\section{Effective operators}
\label{sec:eft}

New physics beyond the Standard Model (SM) can be parameterized in
terms of the effective Lagrangian
\begin{align}
\lag = \lag_\text{SM}+ \sum_{d,k} \frac{ C_k^d}{\Lambda^{d-4}} \ope{k}^d
\; .
\end{align}
The dimension-$d$ operators $\ope{k}^d$ form a complete basis of $SU(3)_c\times
SU(2)_L\times U(1)_Y$ invariant operators containing only SM
fields~\cite{Leung:1984ni, Buchmuller:1985jz, GonzalezGarcia:1999fq}. This defines
the effective field theory to be the SMEFT. All non-SM physics effects are
contained in the Wilson coefficients $C_k^d$.  We further assume that the
operators conserve $C$, $P$, baryon and lepton numbers, and neglect all flavor
effects~\cite{Corbett:2012ja}.  With these assumptions there are $59$
dimension-6 operators.  We work in the context of the Warsaw
basis~\cite{Grzadkowski:2010es}, although one of the strengths of our approach
is that it is straightforward to go from one basis to another. Several groups
have performed global fits to determine the restrictions on SMEFT coefficients
from LEP and LHC data~\cite{Banerjee:2013apa, Butter:2016cvz, DiVita:2017vrr,
Ellis:2018gqa, Almeida:2018cld, Biekotter:2018rhp}, with various assumptions to
reduce the number of operators. In the future, such fits will need to become
increasingly sophisticated as the amount of data increases.  Our study is a step
in the direction of optimizing the flow of information into these
analyses.\bigskip

For the hard partonic process
\begin{align}
q \bar{q}' \to WH \to \ell \bar{\nu} \; b \bar{b}
\end{align}
we start with the experimental inputs in terms of the measured values of $M_W$,
$M_Z$, and $G_F$.  At tree level  in the SMEFT, these parameters are shifted
from their SM values, which introduces a dependence on $C_{HD}, ~C_{HWB},~
C_{\ell \ell}$, and $C_{H \ell}^{(3)}$~\cite{Brivio:2017vri,Brivio:2017bnu}. The
effective couplings of the fermions to the $Z$ and $W$ gauge bosons are also
shifted from their SM values.  None of these effects change the momentum
structures of the vertices and so they are of limited interest to anomalous
coupling fits to the $WH$ process at the LHC with its large QCD backgrounds.  In
this study, we also neglect dipole operators that do not interfere with the SM
prediction ($C_{dW}, C_{uW}, C_{Hu}, C_{dH}, C_{eW}$) and also $C_{Hud}$ which
contributes only to the CKM matrix. Finally, $C_{HWB}$,  $C_{\ell \ell}$, $C_{H \ell}^{(3)}$ 
are strongly limited by global fits to the Warsaw basis coefficients~\cite{Ellis:2018gqa,Berthier:2016tkq,Haller:2018nnx}, 
and so we neglect the contributions of these
operators.  The dependence on $C_{dH}$ vanishes for $m_b=0$ and we also set this
coefficient to zero.

All of these assumptions lead to a restricted set of
operators that are expected to have the dominant effects on the $WH$ process, namely
\begin{align}
\opet{HD} = \ope{H\square} - \frac{\ope{HD}}{4}
&= (\phi^\dagger \phi)\square (\phi^\dagger\phi)
  - \frac{1}{4} (\phi^\dagger D^\mu\phi)^* (\phi^\dagger D_\mu \phi) \notag \\
\ope{HW} &= \phi^\dagger\phi W_{\mu\nu}^a W^{\mu\nu a} \notag \\
\ope{Hq}^{(3)}&= (\vpjt)({\overline Q}_L \sigma^a \gamma^\mu Q_L)\; ,
\label{eq:ops}
\end{align}
where $\phi$ is the $SU(2)$ Higgs doublet, $D_\mu=\partial_\mu +ig_s T^AG_\mu^A
+ig {\sigma^a\over 2}W_\mu^a+ig^\prime Y B_\mu$, $Q_L^T=(u_L,d_L)$, $W_{\mu\nu}^
a= \partial_{\mu}W^a_{\nu} - \partial_{\nu}W^a_{\mu} +
g\epsilon^{abc}W^b_{\mu}W^c_{\nu}$, and
$\vpjt=i\phi^\dagger(\frac{\sigma^a}{2}D_{\mu} \phi) -i(D_{\mu}\phi)^\dagger
\frac{\sigma^a}{2}\phi$. The first line  of Eq. \ref{eq:ops} gives the
combination of two operators that affect the $WH$ amplitude in exactly the same
way, namely through a finite Higgs wave function renormalization.  We describe
them with one Wilson coefficient,
\begin{align}
\frac{\tilde{C}_{HD}}{\Lambda^2} \opet{HD} =
\frac{\tilde{C}_{HD}}{\Lambda^2} \ope{H\square} -
\frac{\tilde{C}_{HD}}{4 \Lambda^2} \ope{HD}\, ,
\end{align}
defining our SMEFT model parameters as
\begin{align}
\left\{ \, \tilde{C}_{HD}, \, C_{HW}, \, C_{Hq}^{(3)} \,
\right\} \;, 
\label{eq:wilson}
\end{align}
assuming the normalization $\Lambda = 1\,\mathrm{TeV}$ unless otherwise noted.
Obviously, the degeneracy between $\ope{H\square}$ and $\ope{HD}$ will be broken
by other Higgs and gauge observables contributing to a global analysis, so we
can neglect this effect in this analysis.  In principle, dimension-6 operators
can also appear in the $W$ and $H$ decays. With our assumptions, the decays only
receive SM contributions, but more generally we also know that the dominant
momentum-dependent corrections are completely dominated by the Higgs production
processes~\cite{Brehmer:2016nyr}.\bigskip

As mentioned above, the effects of dimension-6 operators can either scale like
$v^2/\Lambda^2$ or like $E^2/\Lambda^2$. While in general all three operators in
Eq.\eqref{eq:ops} allow for the second kind of scaling, it turns out that for
single Higgs production $\opet{HD}$ changes only the wave function of the Higgs
field and therefore the Higgs couplings to all other particles. In contrast,
$\ope{HW}$ changes the momentum structure of the $WWH$
vertex~\cite{Dedes:2017zog}. Even more interesting is the existence of a $q
{\overline{q}}' WH$ 4-point vertex proportional to $C_{Hq}^{(3)}$, which avoids
the $s$-channel suppression of the Standard Model $q {\overline{q}}' \rightarrow
W\rightarrow WH$ diagram and is momentum-enhanced at high
energy~\cite{Biekotter:2018rhp}.  To be more specific, we can compute the
helicity amplitudes for the on-shell process $u \bar{d} \to W^{+,\lambda}H$,
where $\lambda$ is the $W$-helicity~\cite{Nakamura:2017ihk}. The leading powers
of the ratio $M_W/\sqrt{s}$ contributing to the amplitude squared for $WH$
production in the Standard Model and including the three operators are:
\begin{center}
\begin{tabular}{c | cccc}
  \toprule
  $W$ polarization & SM & $\tilde{C}_{HD} $ & $C_{HW}$ & $C_{HQ}^{(3)}$ \\
  \midrule
  $\lambda=0$ & $1$  &	 $1$  & $1$ & $\dfrac{s}{\Lambda^2}$  \\[4mm]
  $\lambda=\pm$ &$ \dfrac{M_W}{\sqrt{s}}$ & $\dfrac{M_W}{\sqrt{s}}$ & $\dfrac{\sqrt{s} \, M_W}{\Lambda^2} $ &
  $\dfrac{\sqrt{s}\,M_W}{\Lambda^2}$\\
  \bottomrule
\end{tabular}
\end{center}
\label{eq:formula}
This structure obviously causes problems with a universal definition of
sensitive phase-space regions even in terms of the relatively simple observable
$m_{WH} \equiv \sqrt{s}$, since the operators contribute differently in
different phase-space regions.
 %

\section{Signal and backgrounds}
\label{sec:analysis}

We simulate the Higgs production process
\begin{align}
pp \to W H \to \ell \nu \; b\bar{b},
\end{align}
at 13~TeV with \textsc{MadGraph5\_aMC@NLO}~\cite{Alwall:2014hca} at tree level
using the PDF4LHC15 PDF set~\cite{Butterworth:2015oua} (\texttt{lhaid=90900})
and the default dynamical scale choice of \madgraph, namely, the transverse mass
of the $2 \to 2$ system formed from a $k_T$ clustering~\cite{Hirschi:2015iia}.
We include in the matrix elements all diagrams with an intermediate $W$ and $H$,
though non-resonant and $t$-channel contributions are generally unimportant. The
electroweak contributions from diagrams with a $Z$ or $\gamma^*$ are neglected,
as they are largely removed by the analysis cuts.

The higher-dimensional operators are implemented at tree level in the Warsaw
basis~\cite{Grzadkowski:2010es, Brivio:2017vri} using the \textsc{SMEFTsim}
package~\cite{Brivio:2017btx} with the $\{M_W, M_Z, G_F\}$ input scheme and
assuming $U(3)^5$ invariance in the flavor sector.  Because we want to compare
the linear dimension-6 contributions with the dimension-6 squared terms, we
generate all amplitudes to dimension 6 and do not truncate the cross
section.\bigskip

The leading QCD backgrounds to our process are
\begin{align}
pp \to W b\bar{b}, \qquad
pp \to t\bar{b}\,(\bar{t}b),
\qquad \text{and} \qquad
pp \to t\bar{t}\, .
\end{align}
We also simulate them with \textsc{MadGraph5\_aMC@NLO}~\cite{Alwall:2014hca} at
tree level.  At generator level, we apply cuts that mimic a typical experimental
analysis~\cite{Aaboud:2018zhk, Sirunyan:2018kst}.  We require all leptons to
have $p_{T,\ell} > 10\,\text{GeV}$, $\met > 25\,\text{GeV}$, and tagged $b$-jets
to have $p_{T,b} > 35\,\text{GeV}$.  The leptons and $b$-jets are required to
lie within $|\eta_{\ell, b} | < 2.5$, and be well separated, $\Delta R_{bb, \ell
b} > 0.4$. We also require a Higgs mass window $ 80$~GeV$ <m_{b\bar{b}}
<~160$~GeV, which has no effect on the signal, but allows for efficient sampling
of the backgrounds. To reduce the semi-leptonic $t\bar{t}$ background, we
require additional light jets to lie outside the central region, demanding
$p_{T,j} < 30$~GeV and $\Delta R_{bj, \ell j} > 0.4$. We ignore the fully
leptonic and fully hadronic $t\bar{t}$ backgrounds since they lead to different
final states.

Prior to analysis, the parton-level events are processed to approximate the most
important detector effects. The energies of the $b$-jets are smeared with a
Gaussian transfer function chosen to approximate the $h \to b\bar{b}$ mass
resolution from Ref.~\cite{Sirunyan:2018kst}.  We also smear the transverse
components of the missing transverse energy according to the most recent ATLAS
performance~\cite{Aaboud:2018tkc}. We assume a flat $b$-tagging probability of
$70\%$, and neglect charm and light-flavor mis-tagging probabilities. For
simplicity, we neglect any reconstruction inefficiencies for electrons and
muons. Further details on the treatment of detector effects and a comparison
with alternative methods is presented in App.~\ref{sec:detector}.

\section{Statistical analysis}
\label{sec:stats}

The task of this paper is to quantify the information in the $WH\rightarrow \ell
\nu b{\overline{b}}$ process and to identify observables and phase-space regions
that allow us to probe new physics effects parameterized in terms of the Wilson
coefficients given in Eq.\,\eqref{eq:wilson}. While our reference process is
relatively simple, we know from Sec.~\ref{sec:eft} that the three relevant
operators contribute very differently to the event kinematics. To describe their
effects we need to quantize the sensitivity of the $WH$ phase space to a set of
continuous model parameters. We start by discussing the challenges of such an
analysis in a high-dimensional observable space, show how optimal observables
can be constructed for this process, and review the Fisher information as an
appropriate and convenient object to encode our analysis sensitivity.

\subsection{Score as the optimal observables}
\label{sec:stats_score}

In a typical LHC measurement we link $n$ observed events $\{x\}$, each of which
is characterized by a phase space vector $x_i$, to a vector of model parameters
$\theta = (\tilde{C}_{HD}, C_{HW}, C_{Hq}^{(3)} )$. The central quantity that
describes their relation is the likelihood function, the probability density of
the observed data as a function of the model parameters. It is given
by~\cite{Cranmer:2006zs}
\begin{align}
  p_\text{full}(x|\theta) = \pois(n | \lumi \, \sigma(\theta)) \; \prod_i
  p (x_i|\theta) \,.
  \label{eq:extended_likelihood}
\end{align}
The first term with $\pois(n|\lambda) = \lambda^n e^{-\lambda} / n!$ describes
the probability of observing $n$ events given an integrated luminosity $\lumi$
and predicted cross sections $\sigma(\theta)$. The remaining factors describe
the kinematic information in each event $x_i$ and are equal to the fully
differential cross section normalized to one:
\begin{align}
  p(x|\theta) = \frac 1 {\sigma(\theta)} \frac {\diff^k \sigma(x|\theta)} {\diff x^k} \,.
\end{align}
The Poisson term is relatively straightforward to compute. Calculating
the event-wise kinematic likelihoods $p(x|\theta)$ explicitly, however, requires inverting the chain
of Monte-Carlo tools used to simulate LHC events and integrating over
an extremely high-dimensional space, which is impossible in
practice~\cite{Brehmer:2018kdj, Brehmer:2018eca}.

The most common solution is to restrict the data $x$ to one or a few hand-picked
kinematic variables such as invariant masses, transverse momenta, or angular
correlations, which integrates out some kinematic information and typically
reduces the power of the analysis~\cite{Brehmer:2016nyr, Brehmer:2017lrt}. The
STXS also follow this approach. We compare them to an alternative method, which
defines statistically optimal observables for all directions in parameter space.
We start by assuming that the values of the Wilson coefficients $\theta_i$ are
small, that is, we consider the parameter space region close to the SM. In this
case one can explicitly construct the most powerful observables, which turn out
to be~\cite{Brehmer:2018eca}
\begin{align}
  t_i(x) = \frac {\partial  \log p(x | \theta) } {\partial \theta_i} \Biggr |_{\theta = 0} \,.
  \label{eq:score}
\end{align}
The vector $t(x)$ is the gradient in model space, described by the Wilson
coefficients $\theta_i$.  Each component $t_i(x)$ is a function of phase space
$x$, in other words, the set of basic kinematic observables like reconstructed
energies, momenta, angles, and so on. The high-dimensional observable vector $x$
can be compressed to these functions in each event without losing any
statistical power: the $t_i(x)$ are the sufficient statistics. Furthermore,
confidence limits on the Wilson coefficients can be calculated from histograms
of the components $t_i$ in the same  manner as with histograms of one or two
familiar kinematic variables. Such an analysis of the $t_i(x)$ is guaranteed to
lead to optimal statistical limits in the neighborhood of the SM.

In particle physics, the $t(x)$ are known as Optimal Observables and have been
calculated in a parton-level approximation for many
processes~\cite{Atwood:1991ka, Davier:1992nw, Diehl:1993br}, an approach closely
related to the Matrix Element Method~\cite{Kondo:1988yd}. In the statistics
community $t(x)$ is originally known as the
\textsl{score}~\cite{fisher1935detection}; we follow this older nomenclature.

Calculating the score is not straightforward: it is defined through the
likelihood function $p(x|\theta)$, which is intractable when we use a realistic
simulation of shower and detector effects. Recently, however, a new technique
was developed that solves this problem with a combination of matrix-element
information and machine learning: the \sally algorithm introduced in
Refs.~\cite{Brehmer:2018hga, Brehmer:2018kdj, Brehmer:2018eca} allows us to
train a neural network to estimate $t(x)$ as a function of the observables
$x$.\footnote{See Ref.~\cite{Dawson:2018dcd} for an alternative formulation and
a toy example, Ref.~\cite{Brehmer:2019bvj} for a comparison with the Matrix
Element Method and the traditional Optimal Observable approach, and
Ref.~\cite{Brehmer:2019xox} for a tool that makes it easy to use this
technique.} This way the score defines optimal observables not only at the
parton level, but including the effects of invisible and undetected particles,
parton showers, and detector response.

We use the implementation of the \sally algorithm in
\madminer~0.5~\cite{Brehmer:2019xox} to extract matrix-element information from
our Monte-Carlo simulations, train neural networks to estimate the score, and
calculate the expected limits on the Wilson coefficients. In addition to the
four-momentum components, we choose to include a number of higher-level
observables to aid in the training of the network, like the $p_T$ and $\eta$ of
each particle, the $\Delta R$, $\Delta \phi$, invariant mass, transverse mass,
and transverse momentum of each particle pair, altogether 48 observables for our
$WH$ process. We simulate $4 \cdot 10^6$ signal and $16 \cdot 10^6$ background
events (after all cuts) and train fully connected neural networks with three
hidden layers of 100 units and tanh activation functions. The \sally{} loss
function~\cite{Brehmer:2018kdj} is minimized with the Adam
optimizer~\cite{2014arXiv1412.6980K} over 50 epochs with a learning rate that
decays exponentially from $10^{-3}$ to $10^{-5}$ and a batch size of 128. We
train ensembles of five networks, using the ensemble mean as prediction and the
ensemble variance as a diagnostic tool to check the consistency of the
predictions.

\subsection{Fisher information}
\label{sec:stats_fisher}

Benchmarking an analysis strategy such as the STXS requires a meaningful, but
convenient metric. We propose to use the Fisher information, which we review in
this section. Consider the expected significance with which a parameter point
$\theta$ can be excluded in a measurement. It is
essentially~\cite{Wilks:1938dza, Wald} given by the expected log likelihood
ratio
\begin{align}
-2 \mathbb{E} \left[ \log \frac{p_{\text{full}}(x|\theta)}{p_\text{full}(x|\hat{\theta})} \right] \; ,
\end{align}
where $\hat{\theta}$ is the maximum-likelihood estimator of the Wilson
coefficients.  The log likelihood ratio is a generalization of the
$\chi^2$ test statistic to non-Gaussian distributions.  If the best
fit point is close to the SM and the tested theory parameters $\theta
\sim C/\Lambda^2$ are small, we can expand the log likelihood ratio,
\begin{align}
 -2\; \mathbb{E} \left[\log \frac {p_{\text{full}}(x|\theta)} {p_{\text{full}}(x|0)} \right]
 = \underbrace{- \mathbb{E} \left[\frac{\partial^2 \log p_{\text{full}}(x|\theta)} { \partial \theta_i \;\partial \theta_j} \right]}_{\equiv I_{ij}}
 \; \theta_i \; \theta_j + \ord(\theta^3)\,,
 \label{eq:fisher_info_def}
\end{align}
where $\mathbb{E}[\cdot]$ now denotes the expectation value assuming events
distributed according to the SM.  The leading term in the Taylor expansion
defines the Fisher information $I_{ij}$. This matrix has $n \times n$ entries,
where $n$ is the number of theory parameters (plus nuisance parameters when
discussing systematic uncertainties; see Appendix~\ref{sec:systematics}). It 
quantifies the expected sensitivity of a measurement, where larger components correspond
to more precise measurements, and has several useful properties:
\begin{itemize}
\itemsep0em
\item According to the Cram\`er-Rao bound, the covariance matrix of any unbiased
estimator, or the precision of a measurement, is bounded by the inverse Fisher
information:
\begin{align}
  \cov [\theta_i, \theta_j] \geq (I^{-1})_{ij} \,.
\end{align}
So the larger an eigenvalue of the (by definition diagnolizable) Fisher
information matrix, the more precisely we can measure the direction in
parameter space given by the corresponding eigenvector. The diagonal
elements correspond to the sensitivity to a particular theory parameter when
keeping all other parameters fixed.
\item The Fisher information is independent of the parameterization of
  the observables.
\item It transforms covariantly under parameter transformations, so it
  can easily converted between model space bases.
\item  It is additive for different processes and phase-space
regions: the full sensitivity of multiple analysis is given by the sum of the
Fisher information matrices of several channels or phase-space regions.
This also allows us to study the distribution of information over phase
space.
\item It has a geometric interpretation as a metric in parameter
  space, defining distances
\begin{align}
  d(\theta_1, \theta_0)^2 = I_{ij} \; (\theta_1-\theta_0)_i \; (\theta_1-\theta_0)_j
  \label{eq:fisher_distance}
\end{align}
that quantify the expected reach of a measurement. These distances
directly correspond to expected confidence levels~\cite{Brehmer:2019xox}.
For example, the 95\% CL
contours for two free parameters corresponds to a distance $d = 2.447$.
\item Finally, the Fisher information formalism allows us to easily include
  systematic uncertainties, which are modeled with nuisance
  parameters and profiled over in the statistical
  analysis~\cite{Brehmer:2016nyr}.
\end{itemize}
For a more detailed discussion of this formalism we refer the reader to
Refs.~\cite{Brehmer:2016nyr, Brehmer:2019xox}.

If we insert Eq.~\eqref{eq:extended_likelihood} into the definition of
the Fisher information in Eq.~\eqref{eq:fisher_info_def} we find
\begin{align}
  I_{ij}
    = - \mathbb{E} \left[\frac{\partial^2 \log p_{\text{full}}(x|\theta)} { \partial \theta_i \;\partial \theta_j} \right]
    =   \frac{\lumi}{\sigma}  \frac{\partial \sigma}{\partial \theta_i} \frac{\partial \sigma}{\partial \theta_j}
  + \frac{\lumi \sigma}{N} \sum_{x \sim p(x|0)} \; t_i(x) \; t_j(x)
\label{eq:fisher_info}
\end{align}
where $N$ is the number of events in the Monte-Carlo sample. This means that
after we train a neural network to estimate the score $t(x)$ as discussed in the
previous section, we can calculate the full Fisher information, corresponding to
the maximal reach based on all observables including correlations. As
mentioned above, the Fisher Information also easily allows to include systematic
uncertainties, which are discussed in App.~\ref{sec:systematics}.

In addition, we can calculate the information in the distribution of one or two
kinematic variables with histograms~\cite{Brehmer:2016nyr, Brehmer:2017lrt}. We
will use this to calculate the sensitivity of the STXS. Finally, as a
cross-check we can calculate the parton-level information directly from event
weights, using the technique introduced in Ref.~\cite{Brehmer:2016nyr}. This
information neglects shower and detector effects and assumes that we can exactly
reconstruct the parton-level final state, including flavor information and the
full four-momenta of invisible particles. While this is not a realistic
scenario, we will use this method to verify the machine-learning results in
App.~\ref{sec:background_reweighting}.

\subsection{Beyond the leading Fisher information}
\label{sec:squared}

By definition, the Fisher information on dimension-six Wilson coefficients
measures their leading effects, \ie it is linearized in powers of $1/\Lambda^2$.
Similarly, the score vector $t(x)$ defined in Eq.~\eqref{eq:score} is only
statistically optimal in the parameter region where the Wilson coefficients are
small. Further away from the SM, the $\ord(1/\Lambda^4)$ terms become important
and eventually dominate. The same happens if the interference between
dimension-6 amplitudes and the SM at order $\ord(1/\Lambda^2)$
vanishes\cite{Panico:2017frx}. In this situation, the Fisher information
approximation is no longer accurate, and while an analysis based on the score
will still lead to correct confidence limits, they might no longer represent the
best possible limits. One way to discuss this case is to use machine-learning
techniques to estimate the full likelihood or likelihood ratio function to all
orders in $1/\Lambda^2$~\cite{Brehmer:2018hga, Brehmer:2018kdj, Brehmer:2018eca,
Stoye:2018ovl}. In a related way, using the geometric interpretation of the
Fisher information, one could calculate distances along geodesics that capture
these higher-order effects as well~\cite{Brehmer:2016nyr}.

Here we follow a different approach. In the limit where the linear term vanishes
or is very small, we can still discuss the Fisher information and define optimal
observables by assuming that the effects from one operator squared (${}\sim
\theta_i^2$), dominate over both interference terms (${}\sim \theta_i$) and
cross terms (${}\sim \theta_i \theta_j$). In this approximation we can write the
likelihood in terms of the squared Wilson coefficients $\Theta_i \equiv
\theta_i^2 $ as new parameters. We note that while the $\Theta_i$ are perfectly
well defined statistical quantities, their physics interpretation requires the
additional condition $\Theta_i \geq 0$. In a global EFT fit this leads to a
series of interesting effects~\cite{SFitter_top} which will have no impact on
our discussion in Sec.~\ref{sec:phys_1d}.

To analyze the impact of the contributions at quadratic order, ${\cal
{O}}({1/\Lambda^4})$,  we again define the score as a function of $\Theta$
according to Eq.~\eqref{eq:score} and the Fisher information as a function of
$\Theta$ according to Eq.~\eqref{eq:fisher_info}. In practice, this means we
replace the derivative with respect to $\theta_i$ by a derivative with respect
to $\Theta_i$, which is proportional to the second derivative with respect to
$\theta_i$. We will label confidence limits based on the score in terms of the
$\Theta$ as \harry{}\footnote{We leave it to the reader to imagine a carefully
chosen acronym.}.

\section{New physics in kinematics}
\label{sec:phys}

The physics question we are going to tackle in this study is how the effects of
the three operators in Eq.\,\eqref{eq:ops} are distributed over phase space and
how we can define LHC observables to best constrain them. In particular, we need
to determine if the established simplified template cross sections serve this
purpose, or if they can be improved.

The Fisher information is an ideal tool to analyze questions on where in phase
space we can search for new physics signals and how we can extract this
information from kinematic analyses~\cite{Brehmer:2016nyr, Brehmer:2017lrt}. The
first question we discuss is the impact of detector effects, which can hide
information that in principle exists at the parton level but is not accessible
in a realistic measurement~\cite{Brehmer:2018eca, Brehmer:2018kdj}. As described
in Sec.~\ref{sec:stats} we can use machine learning to calculate the actually
observable information at the detector level. We will illustrate this aspect in
Sec.~\ref{sec:phys_det}. Next, in Sec.~\ref{sec:phys_1d} we will discuss how the
sensitivity to different dimension-6 operators is distributed over phase space,
both linearized in the Wilson coefficients and including the Wilson coefficients
squared. In Sec.~\ref{sec:phys_2d} we will study the information in
two-dimensional and multivariate distributions.  We analyze how much of the
available information is captured by simplified template cross sections and
discuss their proposed form. Finally, we calculate expected exclusion limits for
this channel based on the use of STXS as well as a multivariate
analysis~\cite{Tackmann:2138079, Berger:2019wnu}.

\subsection{Information after detector effects}
\label{sec:phys_det}

\begin{figure}[t]
\includegraphics[width=.32\linewidth]{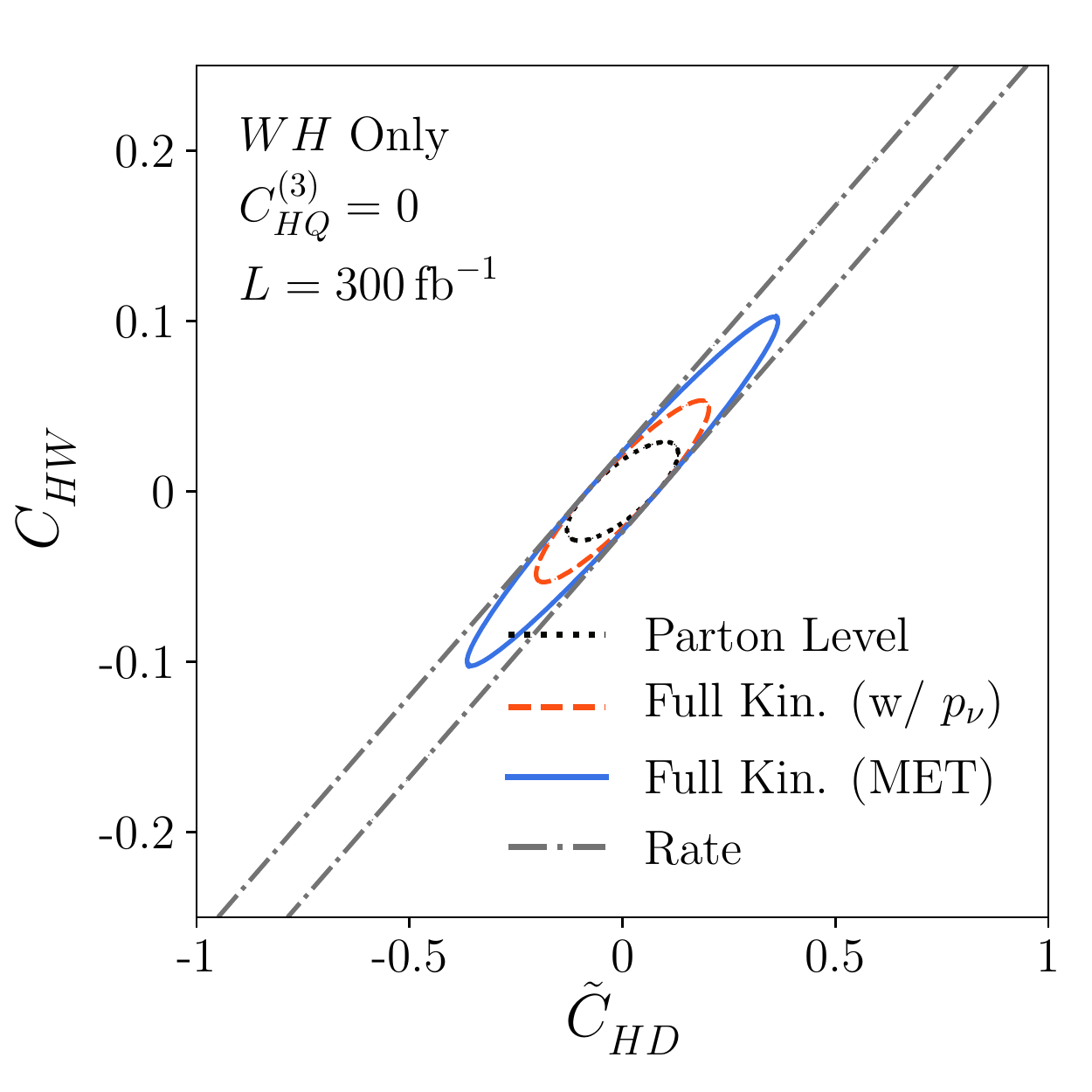}
\includegraphics[width=.32\linewidth]{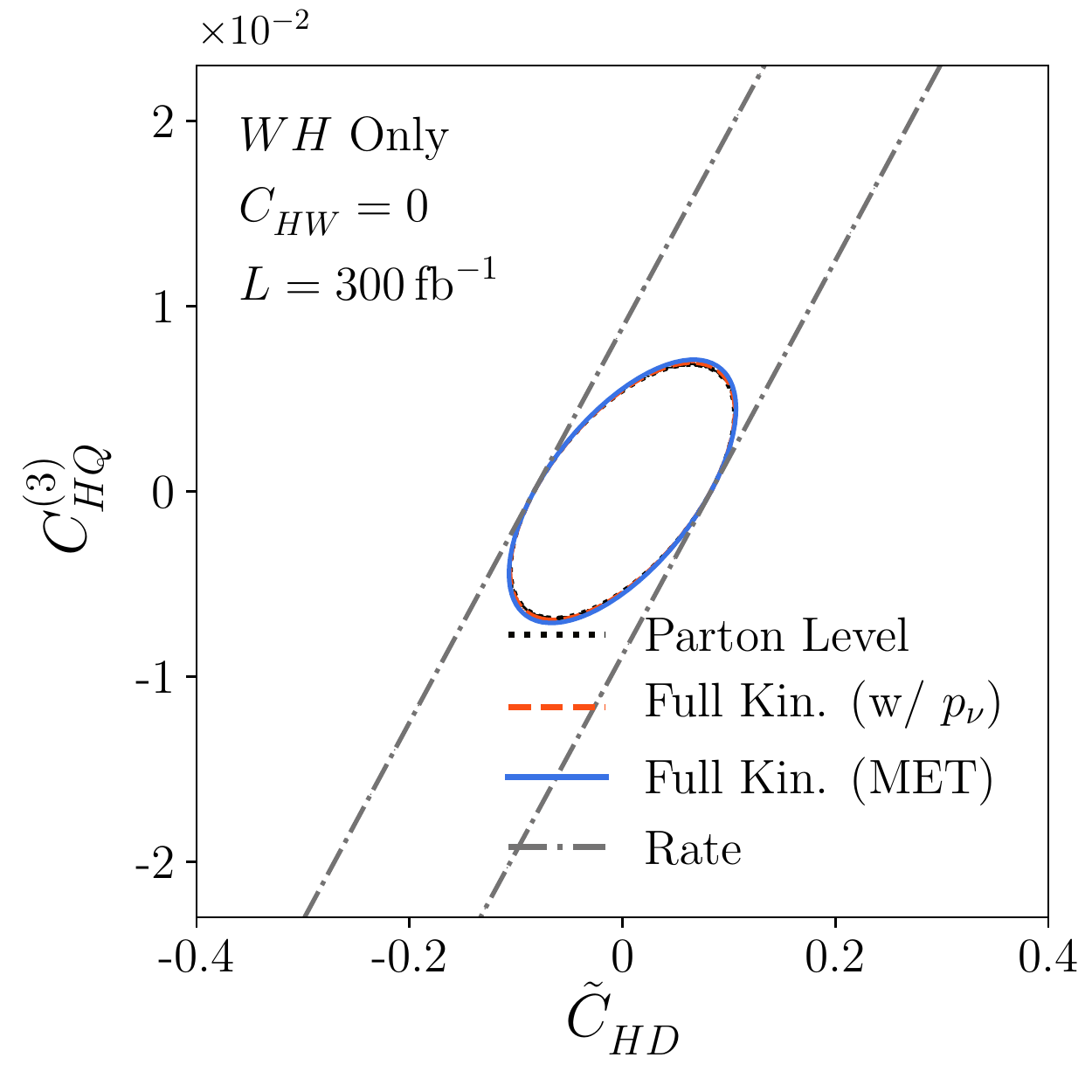}
\includegraphics[width=.32\linewidth]{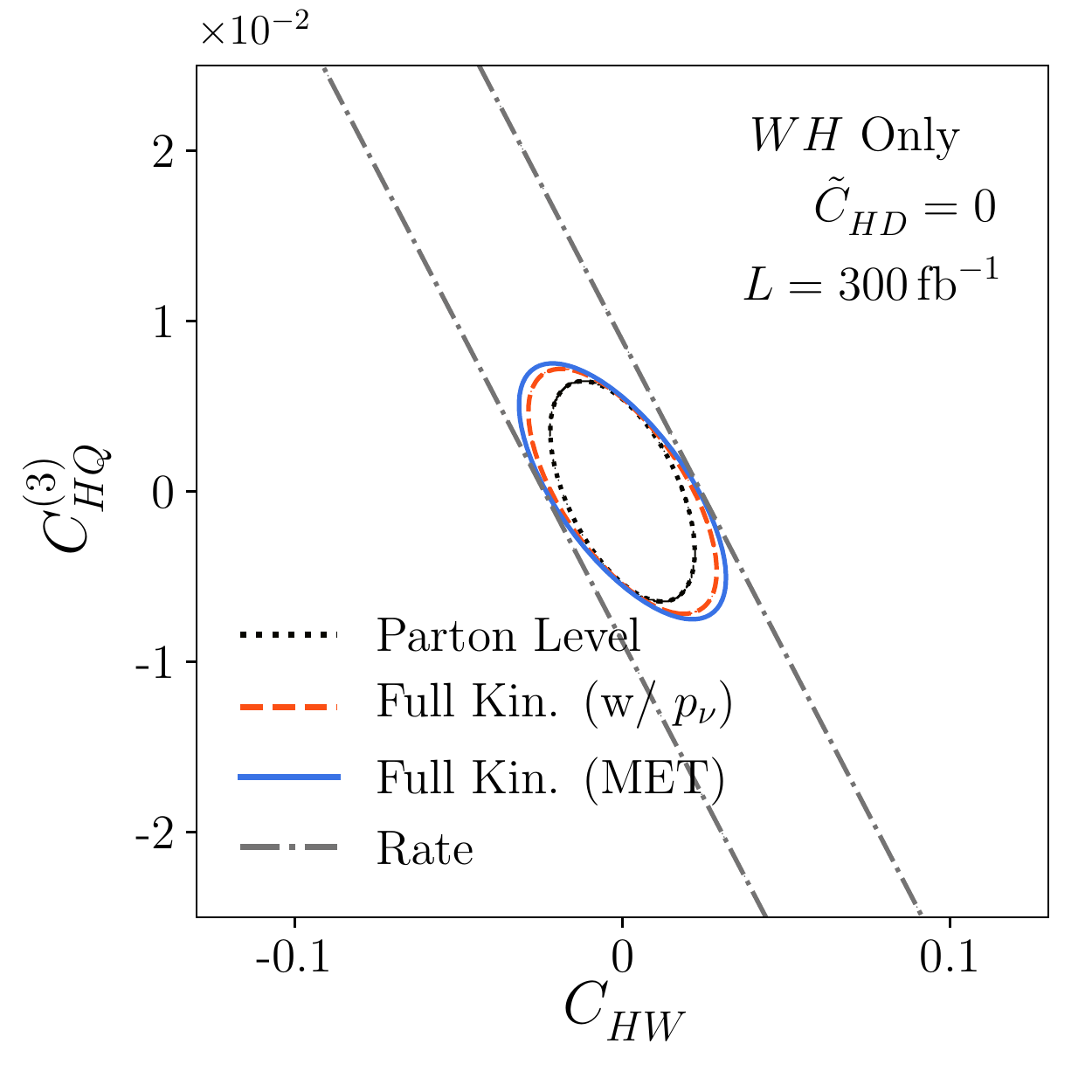} \\ [-1mm]
\includegraphics[width=.32\linewidth]{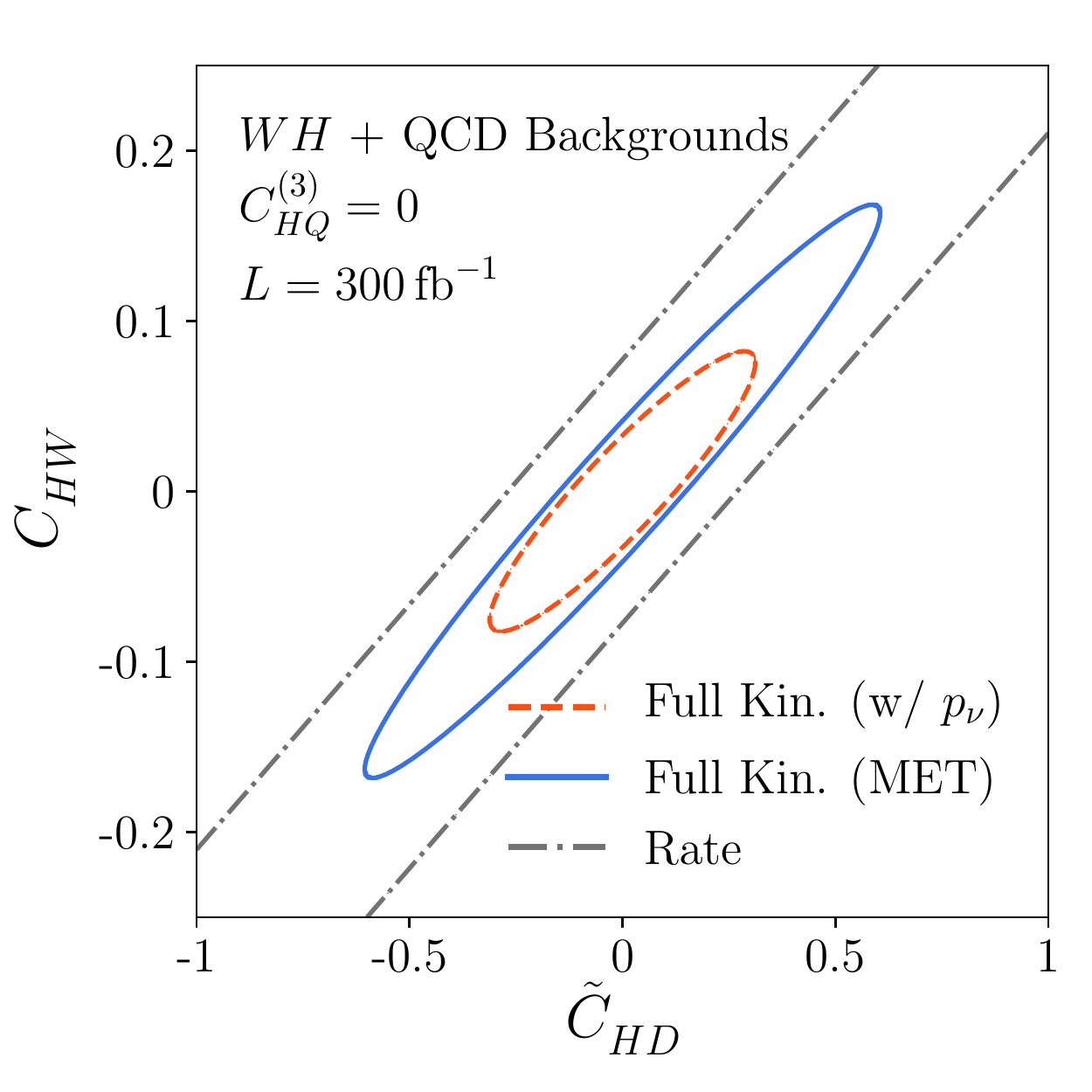}
\includegraphics[width=.32\linewidth]{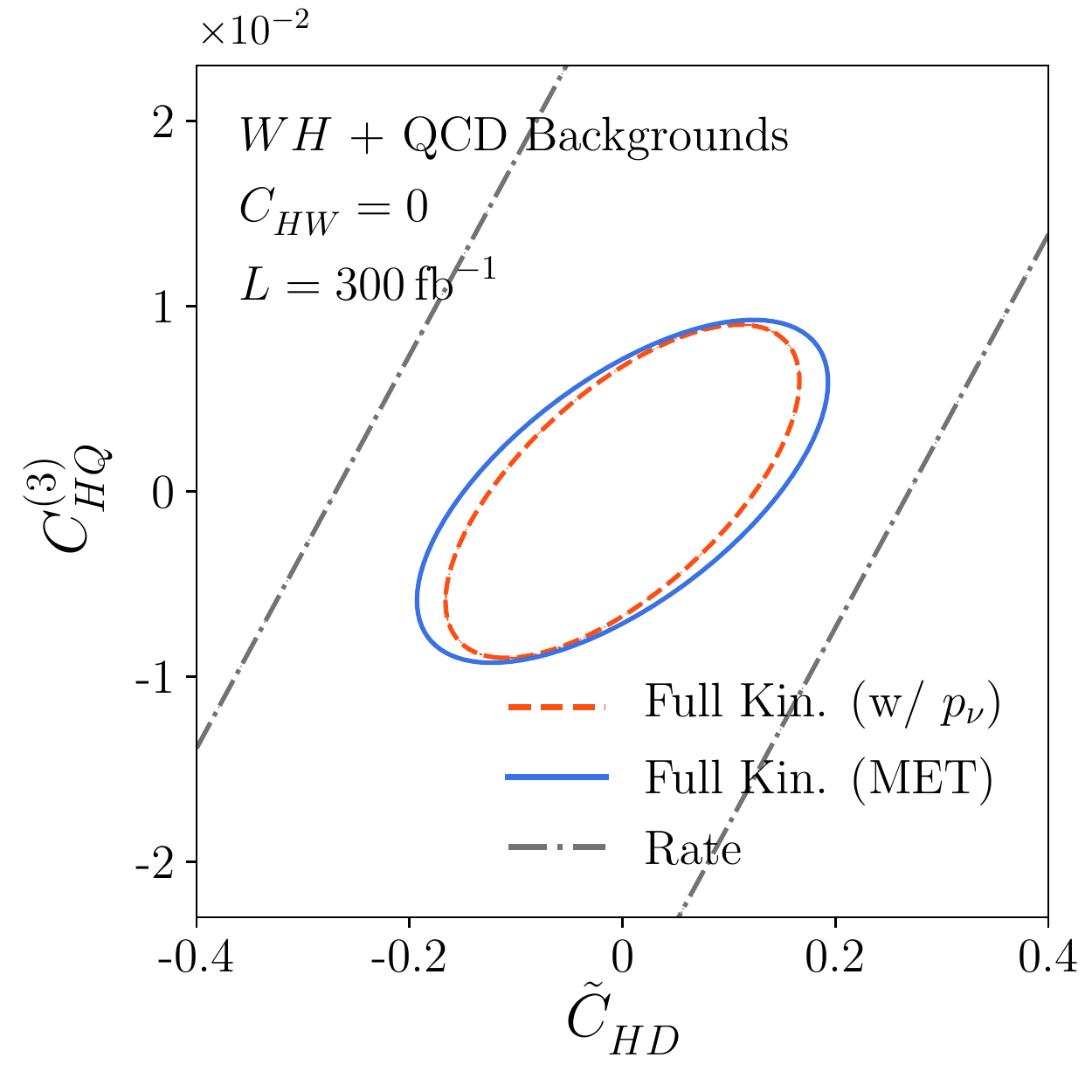}
\includegraphics[width=.32\linewidth]{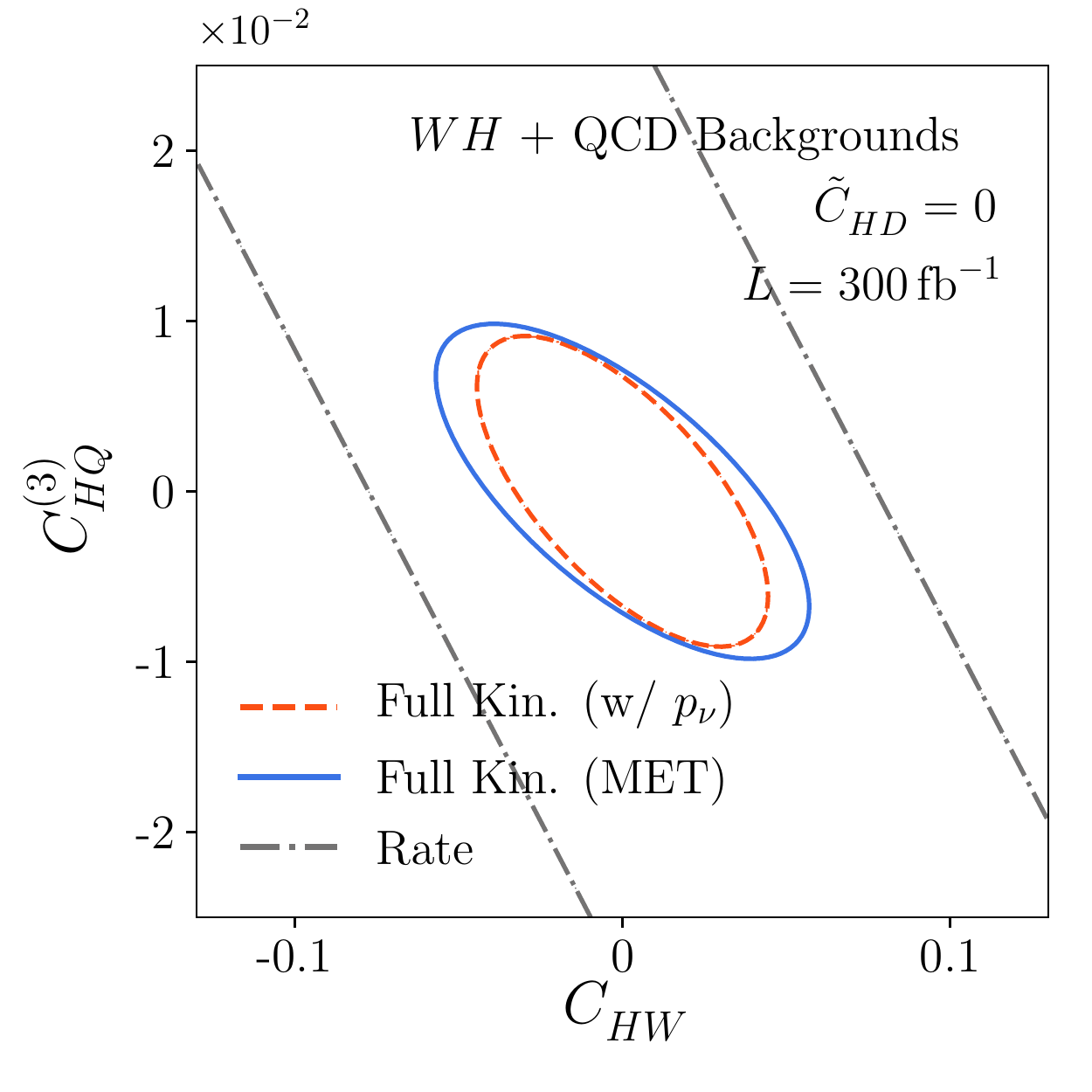}\\ [-8mm]
\caption{95\% CL limits based on the (linearized) Fisher information
  on pairs of Wilson coefficients with $\Lambda = 1$~TeV, for $WH$
  production only (top) and including the QCD background (bottom). We
  include full parton-level information without detector effects
  (dotted black), detector effects, but retaining the full neutrino
  4-momentum (dashed orange), and  finally retaining only the transverse neutrino
  momentum (solid blue). We also show the information using only  the
  signal rate in grey.
}\label{fig:validation}
\end{figure}

Fisher information calculations for fully reconstructable signals and their
irreducible backgrounds, also including leading detector effects, have been
studied in the past~\cite{Brehmer:2016nyr, Brehmer:2017lrt}. Using \madminer we
can also study signatures which cannot be fully reconstructed. Specifically in
the $WH$ production process, we study the information based on the smeared
missing transverse energy rather than the full neutrino momentum. In this
section we analyze the loss of information due to these detector effects,
comparing three sets of observables: first, keeping the full information on all
initial and final state particles at parton level, second, using all final state
particles after detector smearing, and third, keeping only the observable
missing transverse momentum in place of the neutrino momentum, including all
detector-level smearing.

To gain some intuition about the impact of detector effects on the information
content of the hard process we analyze two of the three operators given in
Eq.~\eqref{eq:wilson} at a time, setting the third Wilson coefficient to zero.
We also limit ourselves to linearized differential cross sections in each of the
Wilson coefficients at this point, the effects of the squared terms will be
discussed in detail later.

In Fig.~\ref{fig:validation} we show how the full information about the
dimension-6 signal at the parton level is reduced when we move from parton-level
truth to realistic observables step by step. The dotted black line corresponds
to retaining the full parton-level information without detector effects that is
used for the evaluation of the matrix elements. This includes unobservable
degrees of freedom such as the initial-state parton flavors, particle helicities
and the un-smeared momenta. Removing this information and introducing a
realistic smearing of the particle momenta washes out possible structures, for
example asymmetries in the rapidity distributions, as can be seen when going
from the black to the orange line. In this step we still assume that we can
measure the full neutrino four-momentum. Finally, we remove the unobserved
longitudinal neutrino momentum and neutrino energy, which changes the orange
line into the blue line.  While the longitudinal neutrino momentum can in
principle be reconstructed if we assume an on-shell $W$ boson, this
reconstruction is spoiled by the detector smearing.

In the top panels  of Fig. \ref{fig:validation}, we neglect the QCD backgrounds
and consider only $WH$ production with linear dimension-six contributions. The
loss of information has the strongest impact on the correlated measurements of
$\tilde{C}_{HD}$-$C_{HW}$, because the phase-space effects of both operators are
very SM-like. In this situation we rely on the best possible understanding of
the full final state. In the $\tilde{C}_{HD}$-$C_{Hq}^{(3)}$ and
$C_{HW}$-$C_{Hq}^{(3)}$ planes, we know that the phase-space effects are more
dramatic, so the additional information from the third neutrino direction is
less relevant.  We also show the information from the total rate measurement,
which is not only much poorer, but it also shows a perfect flat direction for
each pair of operators.

In the lower panel of Fig.~\ref{fig:validation} we show the same change in
available information at detector level, now in the presence of the QCD
backgrounds. In this situation, a complete understanding of the final state
neutrinos is desirable for understanding small changes in the full phase
space distribution due to BSM effects, and hence the missing information
affects all three operators.

\subsection{Distribution of information}
\label{sec:phys_1d}

\begin{figure}[t]
\includegraphics[width=.49\linewidth]{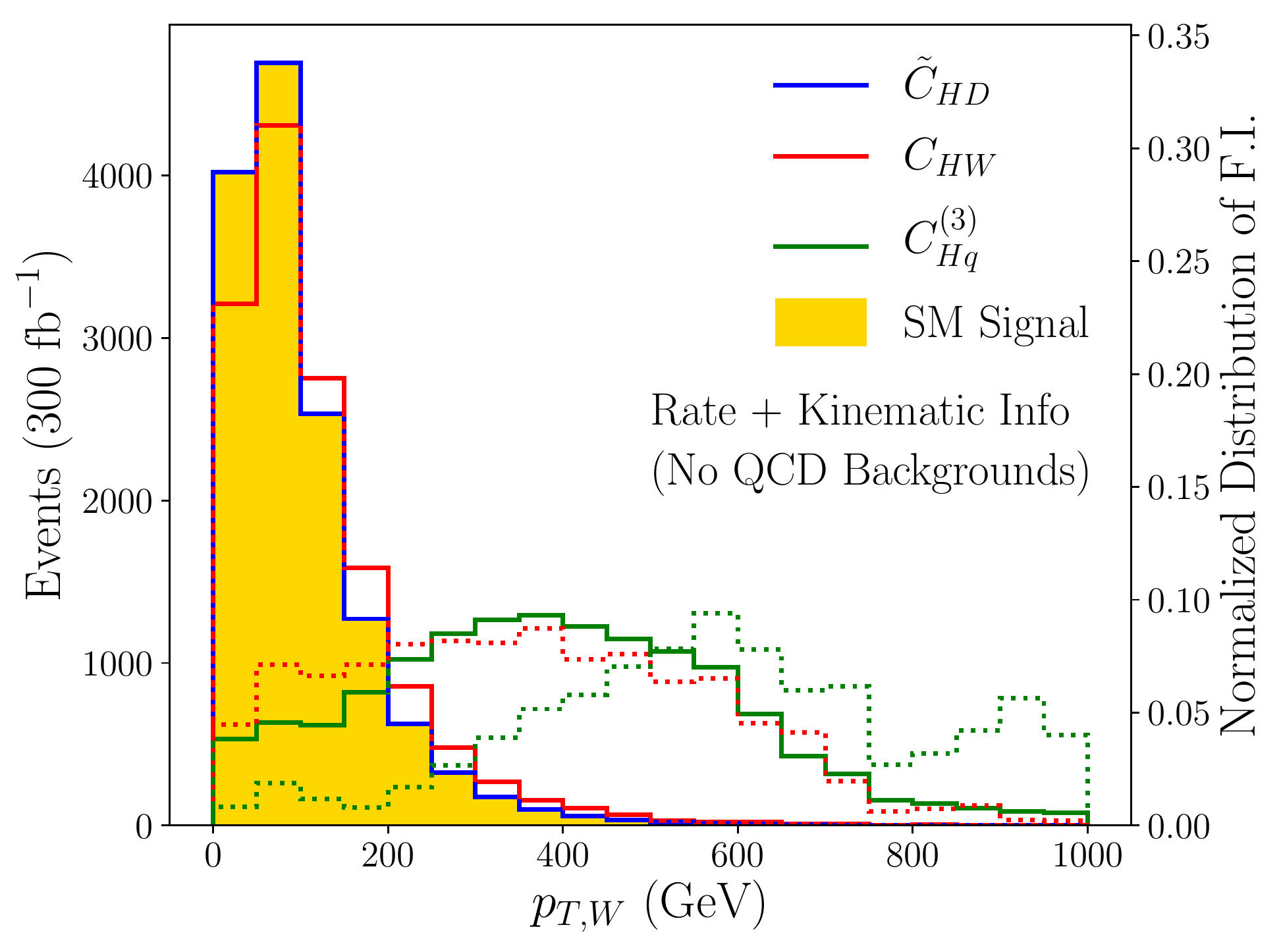}
\includegraphics[width=.49\textwidth]{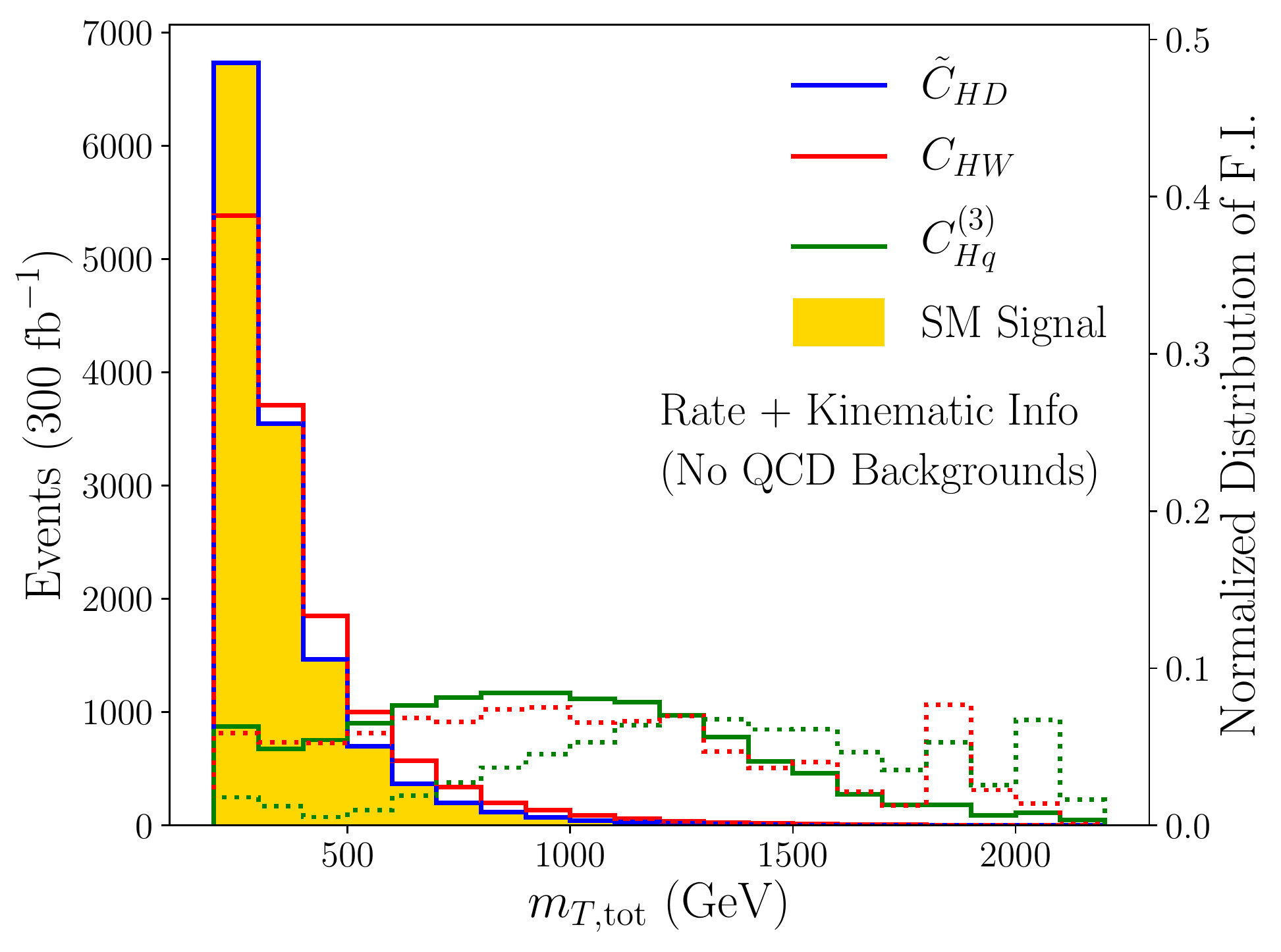}
\caption{Left axis: SM signal distribution. Right axis: normalized
  distributions of the diagonal elements of the Fisher information,
  representing the information on the three Wilson coefficients in bins
  of $p_{T,W}$ (left panel) and $m_{T,\mathrm{tot}}$ (right panel).
  The solid lines are linearized in the Wilson coefficients, while the
  dotted lines show the information on the dimension-6 squared terms.
}
\label{fig:sig_only}
\end{figure}

From Sec.~\ref{sec:eft} we know that the three dimension-6 operators
we are considering have distinctly different effects on the $WH$
production process. If we want to exploit the momentum enhancement of
$C_{Hq}^{(3)}$, it is clear that the leading kinematic variable for the
partonic $2\to 2$ signal is the partonic energy or
equivalently~\cite{Biekotter:2016ecg}
\begin{align}
p_{T,W} \approx p_{T,H}
\quad \text{or} \quad
m_{WH} \; .
\end{align}
This simple description as a $2 \to 2$ process is broken by two
aspects, the $2 \to 3$ structure of the $Wb\bar{b}$ continuum
background and angular correlations reflecting the $W$ polarization.

\begin{figure}[t]
\includegraphics[width=.49\linewidth]{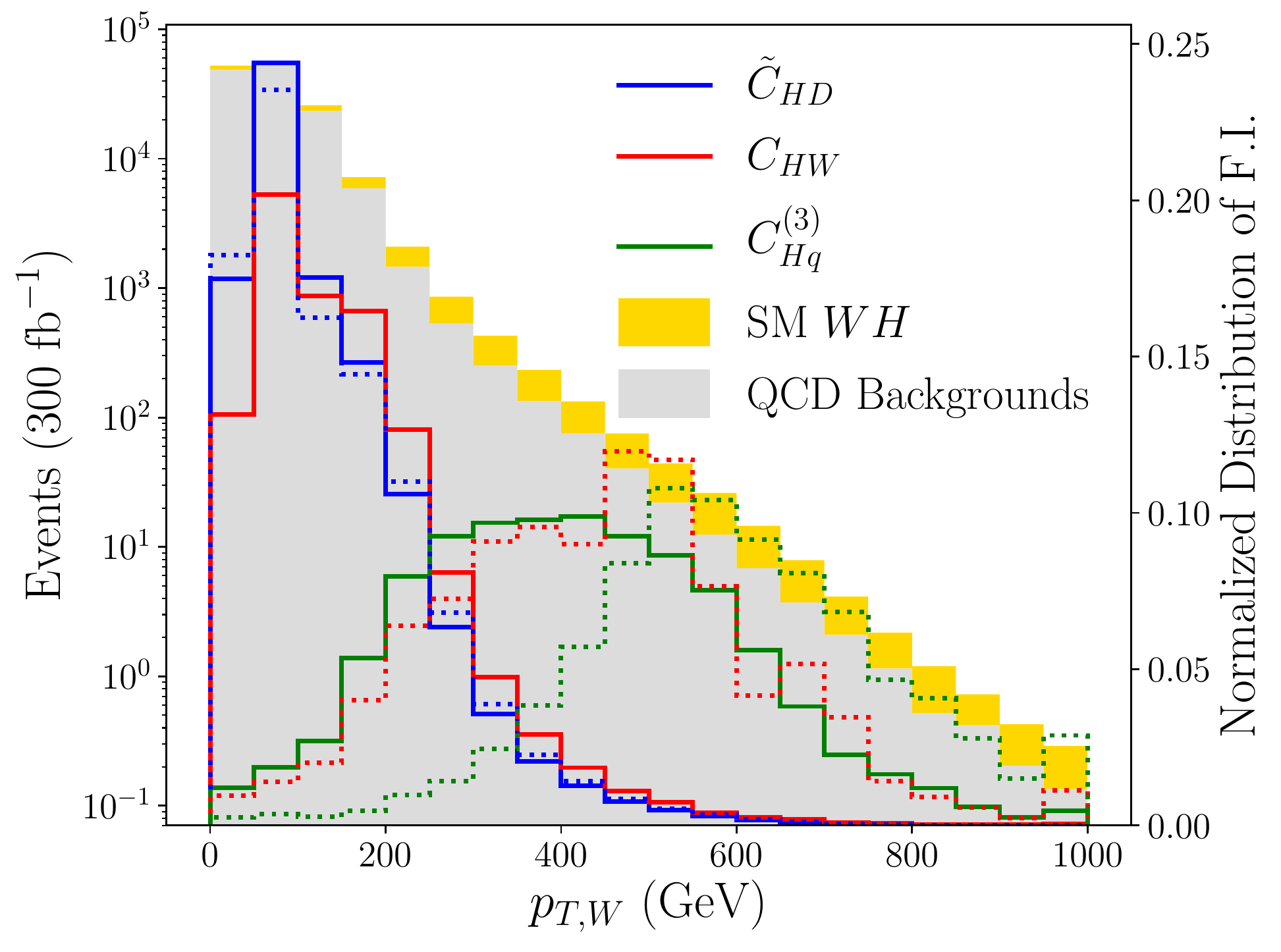}
\includegraphics[width=.49\linewidth]{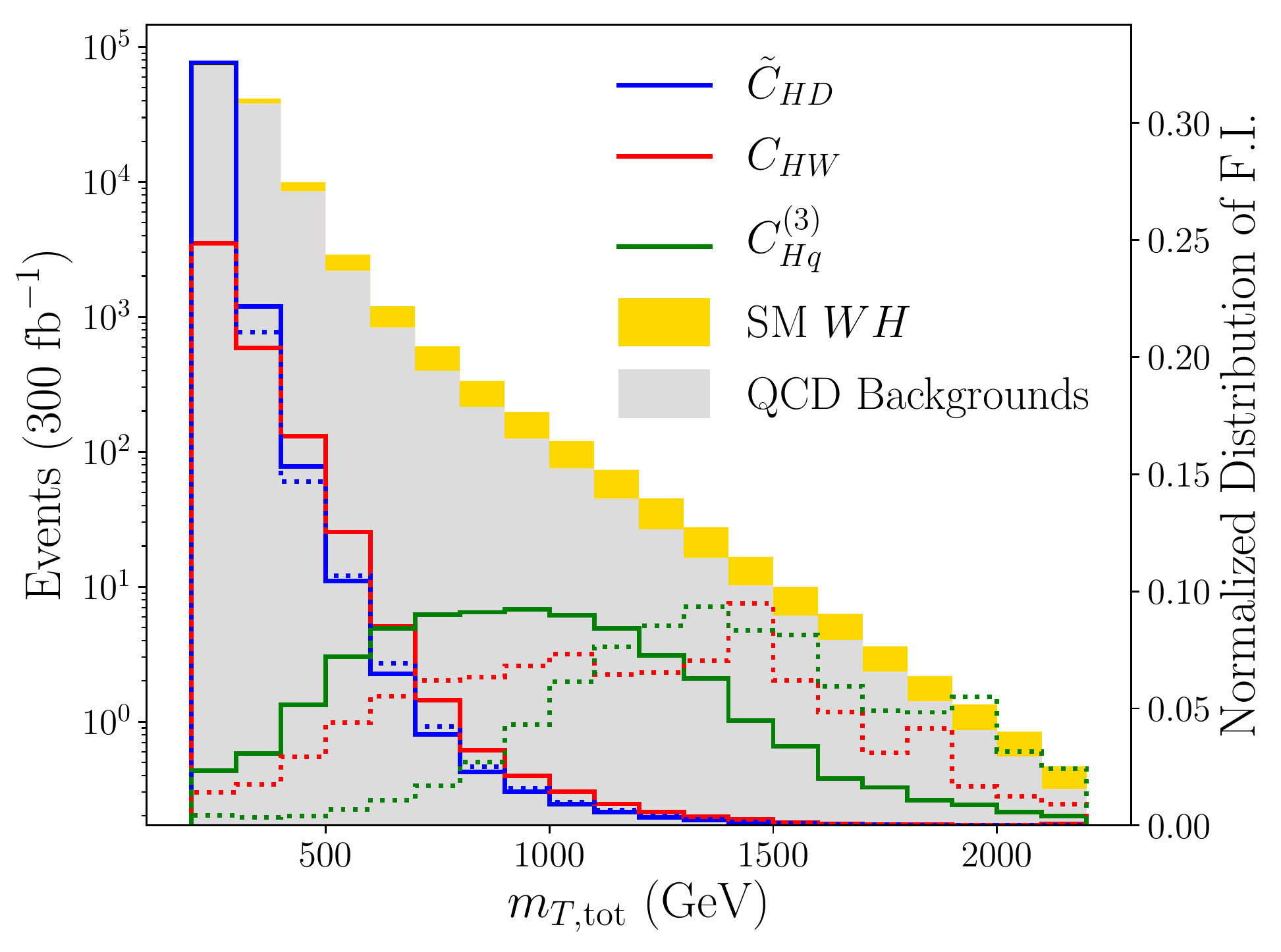}
\caption{Left axis: stacked histograms of the SM signal on top of the
  QCD backgrounds for $p_{T,W}$ (left) and $m_{T,\text{tot}}$ (right).
  Right axis: normalized distributions of the diagonal elements of the
  Fisher information, representing the information on the three Wilson
  coefficients. The solid lines are linearized in the Wilson
  coefficients, while the dotted lines show the information on the dimension-6 squared
  terms.}
\label{fig:distributions1}
\end{figure}

In the left panel of Fig.~\ref{fig:sig_only}, we show the $p_{T,W}$
distribution for the $WH$ signal, together with the information distribution. We
start with the normalized distribution of the diagonal element of the Fisher
information introduced in Sec.~\ref{sec:stats_fisher}, by definition linearized
in the Wilson coefficient, as solid lines. These curves correspond to the diagonal
elements of the Fisher Information matrix computed using only events that lie within
the bin boundaries for a chosen kinematic variable. These diagonal elements are 
inversely proportional to the estimated limit on the theory parameter, and the full 
information (and optimal limit) can be obtained by adding up the information in each
bin. The distributions thus demonstrate where in phase space the constraining power
on each Wilson coefficient arises, and where the experimental analyses should be focused.

We see that the information on all three operators reflects their respective Lorentz 
structures. The information on $\opet{HD}$ mimics the distribution of the signal, 
and the information in $\ope{HW}$ is energy-suppressed because it couples to the 
transverse $W$-modes.
In contrast, $\ope{Hq}^{(3)}$ is visible at large momentum transfer because of
the 4-point vertex.

In the same figure we also show the information distribution as dotted lines for
the dimension-6 squared terms only for each of the three operators. In this case
the interference with the Standard Model does not act as a projector onto the
$W$-polarization structure of the Standard Model. In these plots we treat the
squared Wilson coefficients as unconstrained parameters; physically, they are
restricted to be non-negative, but this does not change our conclusions. What we
see clearly is that the sensitive phase-space region for $\ope{HW}$ now peaks
around 400~GeV, well outside the dominant phase-space region of the SM signal.
Also the peak in the information on $\ope{Hq}^{(3)}$ moves from around 400~GeV
to 600~GeV, dominated by the squared 4-point vertex.

The right panel of Fig.~\ref{fig:sig_only} shows the distribution and
information distribution over the total transverse mass of the final state,
\begin{align}
\mtot
= \sqrt{\left( E_T^{bb\ell} + \met \right)^2
- \left| \mathbf{p}_T^{bb\ell} + \slashed{\mathbf{p}}_T \right|^2 } \; ,
\label{eq:mttot}
\end{align}
where $E_T^{bb\ell} = \sqrt{|\mathbf{p}_T^{bb\ell}|^2 + m^2_{bb\ell}}$
and $\mathbf{p}_T^{bb\ell}$ is the vector sum of the transverse
momenta of the lepton and the $b$-jet, with similar conclusions.\bigskip

\begin{figure}[t]
\includegraphics[width=.49\linewidth]{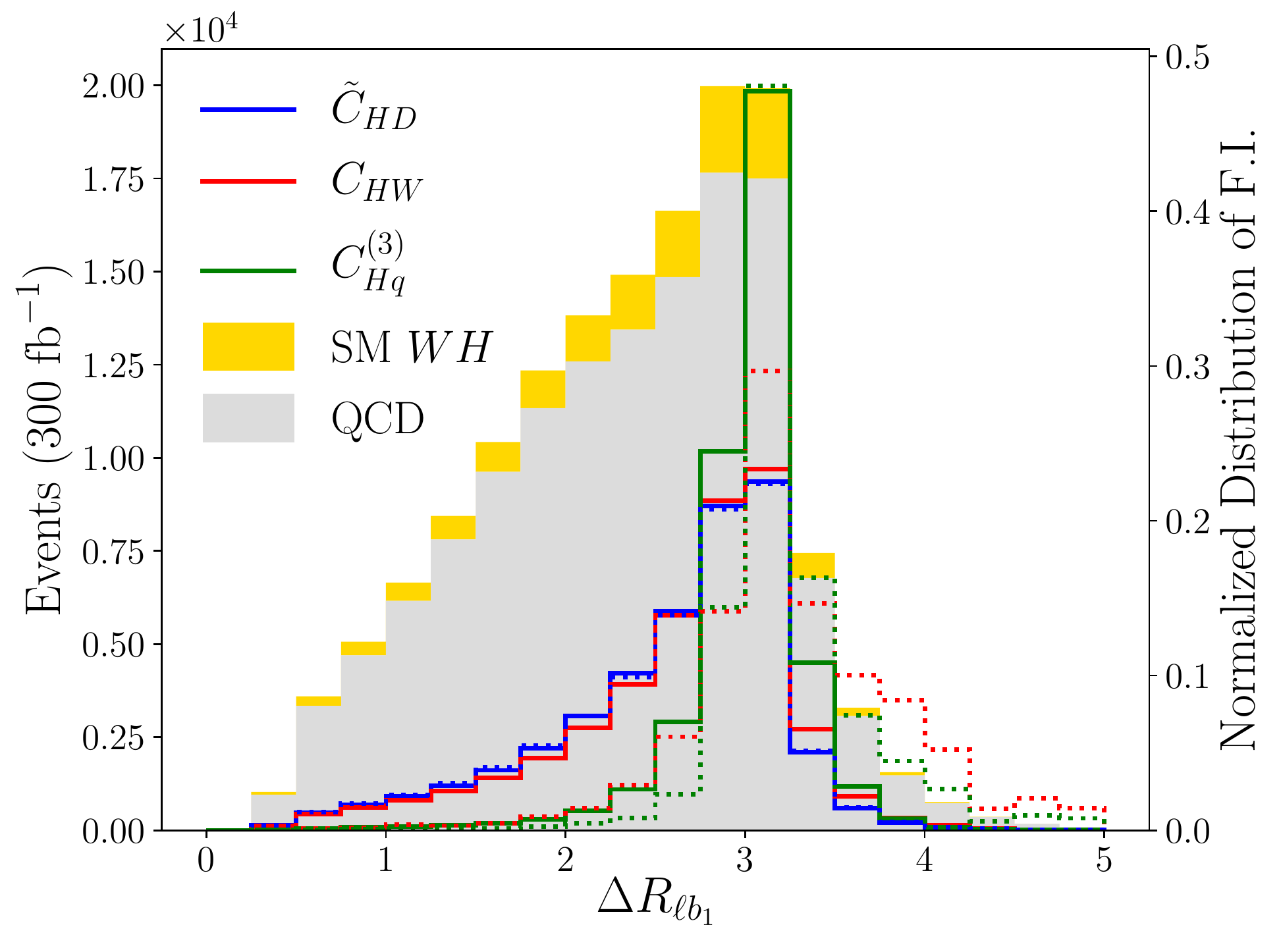}
\includegraphics[width=.49\linewidth]{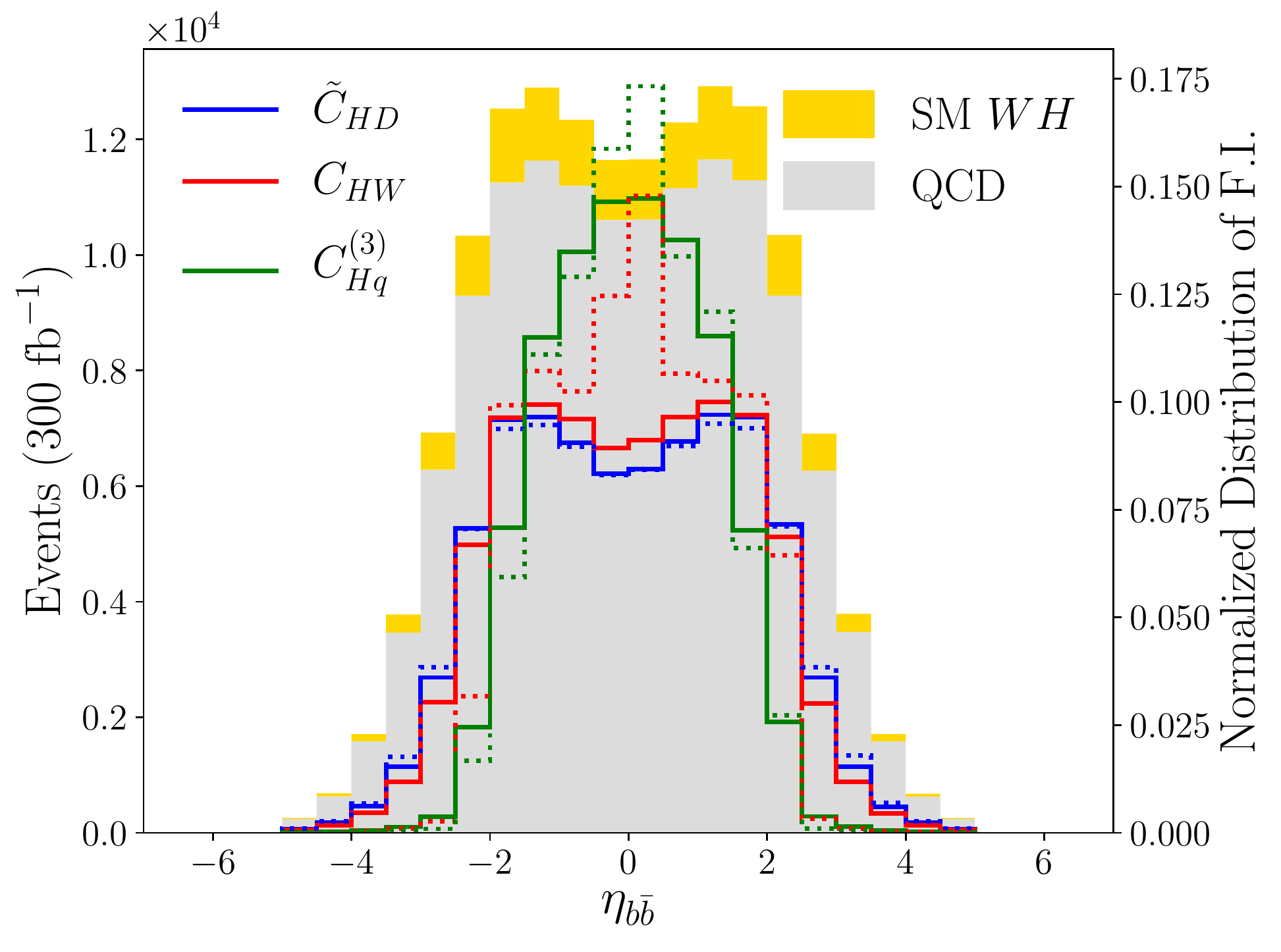}
\caption{Left axis: stacked histograms of the SM signal on top of the
  backgrounds for different kinematic observables. Right axis:
  normalized distributions of the diagonal elements of the Fisher
  information, representing the information on the three Wilson
  coefficients.}
\label{fig:distributions2}
\end{figure}

In Fig.~\ref{fig:distributions1}, we demonstrate  how including the continuum
QCD background changes the picture. First, we see that regions of low $p_T$ or
low $\mtot$ are swamped by the QCD backgrounds, so they do not lend themselves
to extracting momentum-enhanced contributions. For $\opet{HD}$ there is no
momentum enhancement, and the information in the two distributions behaves
exactly as before. Similarly, the information distribution on $\ope{HW}$ in the
linear case is primarily in the low bins, where the event rate is larger, but is
largest for $p_{T,W} \gtrsim 300$ or $\mtot \gtrsim 700$~GeV if we consider the
dimension-6 squared term. In contrast, $\ope{Hq}^{(3)}$ gains only very little
information from the low-$p_{T,W}$ or low-$\mtot$ regime. We note that for
operators only affecting the total rate, the distribution of information over
$p_{T,W}$ is completely consistent with the distribution of the statistical
significance of the $VH$ production over the QCD background, as studied in
Ref.~\cite{Plehn:2013paa}.

Here we pause to comment on the validity of the EFT expansion.
While for each operator, only the combination $C_i/\Lambda^2$ is observable,
matching with a perturbative theory at higher scales requires $C_i \lesssim 1$,
and thus more precise measurements are intrinsically probing potentially higher
scales. In order for the constraints to be self-consistent, it is important that we do 
not try to set constraints on the EFT parameters using data that is outside the range 
of validity of the EFT expansion. This could be done rigorously by implementing
a cut demanding that some kinematic variable tracing the momentum transfer not
exceed some multiple of $\Lambda$, but fortunately, in Fig.~\ref{fig:distributions1}, 
we see that all of the Information arises from events with 
$p_{T,W} \lesssim 1\,\mathrm{TeV}$, so such a cut is unnecessary for our analysis
here.

Finally, in Fig.~\ref{fig:distributions2} we consider two more kinematic
distributions. First, the opening angle of the charged lepton and the leading
bottom becomes $\Delta R_{\ell b_1} \approx \pi$ for boosted $2 \to 2$
processes, like the part of $WH$ phase space sensitive to $\ope{Hq}^{(3)}$.
Second, the reconstructed rapidity of the $b\bar{b}$ system also becomes more
central for boosted $2 \to 2$ processes. The opening angle is useful for
discriminating against the $t\bar{t}$ and $tb$ backgrounds, where we do not
expect the two $b$-jets to recoil against the lepton, but they do not otherwise
appear too promising to gain additional information into one of the three
dimension-6 operators, or into the relation between linearized and quadratic
contributions.

Altogether, we have learned that when looking at 1D kinematic distributions
those which scale like momentum, like $p_T$ or $\mtot$, are best-suited to
separate the effects of the different operators. Including QCD backgrounds makes
the high-momentum regime even more attractive. The only problem is that the
relevant region, for instance in $p_{T,W}$, depends not only on the operator,
but also on whether we linearize the effective field theory for the cross
section or keep the  ${\cal{O}}({1/\Lambda^4})$ squared term. Furthermore,
the balance between linear and quadratic contributions to a combined analysis
will change with the LHC luminosity and the kinematic focus of the analysis.

\subsection{Information in STXS}
\label{sec:phys_2d}

Testing one effective operator at a time violates the basic assumptions of
effective theories, namely that we expect to see effects from all operators
allowed by the underlying theory. We now examine how the kinematic variables
introduced in the previous section can be used to set constraints on the full
multi-dimensional parameter space.

\begin{figure}[t]
\centering
\includegraphics[width=0.9\linewidth]{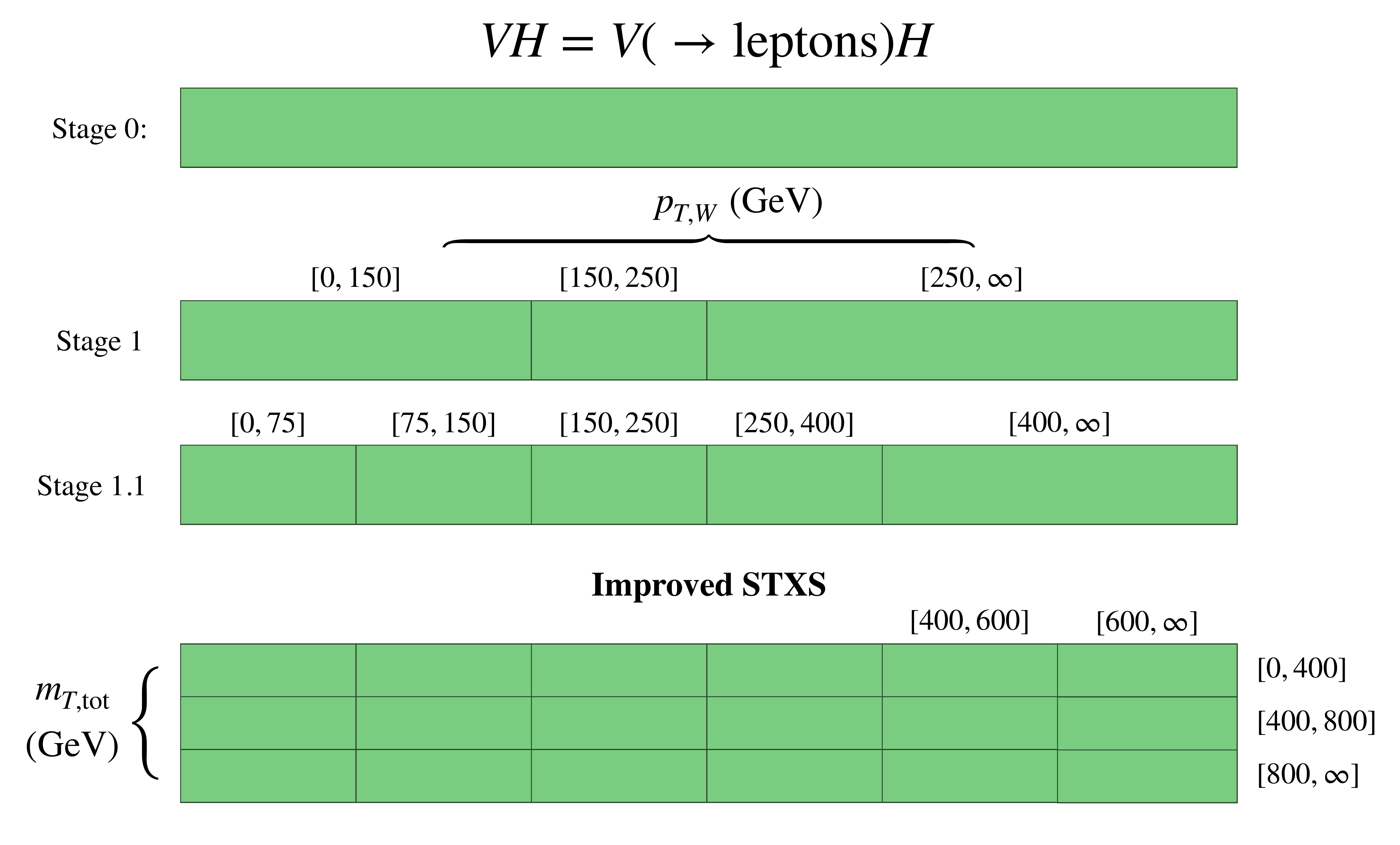}
\caption{Definition of the old STXS~\cite{Tackmann:2138079, Berger:2019wnu} and
our proposed improved version based on Eqs.~\eqref{eq:stxs+} and
\eqref{eq:stxs++}.}
\label{fig:stxs}
\end{figure}

Simplified template cross sections (STXS) have been proposed to provide
experimental results on kinematic distributions for global
analyses~\cite{Tackmann:2138079,Berger:2019wnu}. For the $WH$ production process
they include rate measurements binned in $p_{T,W}$. At stage~1, the three bins
are defined by $p_{T,W} = 0-150$~GeV, $p_{T,W} = 150-250$~GeV, and $p_{T,W} >
250$~GeV. For increased statistics, stage~1.1 proposes five bins, now split at
$p_{T,W} = 75, 150, 250, 400$~GeV, as shown in Fig.~\ref{fig:stxs}. Results
using this approach have recently been presented by ATLAS~\cite{Aaboud:2019nan}
using the 0-jet, single lepton data sample, to obtain limits on anomalous
couplings, assuming only one non-zero Wilson coefficient at a time.
Unfortunately, they do not consider the most interesting 4-point vertex from
$\ope{Hq}^{(3)}$, for which boosted $VH$ production provides the strongest
constraints~\cite{Biekotter:2018rhp}.

We evaluate these proposals by calculating the Fisher information in the STXS
bins.  From Fig.~\ref{fig:distributions1}, it is clear that  the stage~1.1 STXS
version will not capture the full information on $\ope{Hq}^{(3)}$, so we
consider a 6-bin setup with
\begin{align}
p_{T,W}^\text{bins} = ( \, 0-75, \, 75-150, \, 150-250, \, 250-400, \, 400-600, \, 600-\infty \, )~\gev \; .
\label{eq:stxs+}
\end{align}
We compare the results to the full information in the high-dimensional phase
space using the \sally method discussed in Sec.~\ref{sec:stats_score}. In
addition, we train a neural network with the \sally method on just $p_{T,W}$ as
input variable, which lets us calculate the information in the full $p_{T,W}$
distribution in the infinite-bin limit; see App.~\ref{sec:binning} for more
details.\bigskip

\begin{figure}%
\includegraphics[width=.315\linewidth]{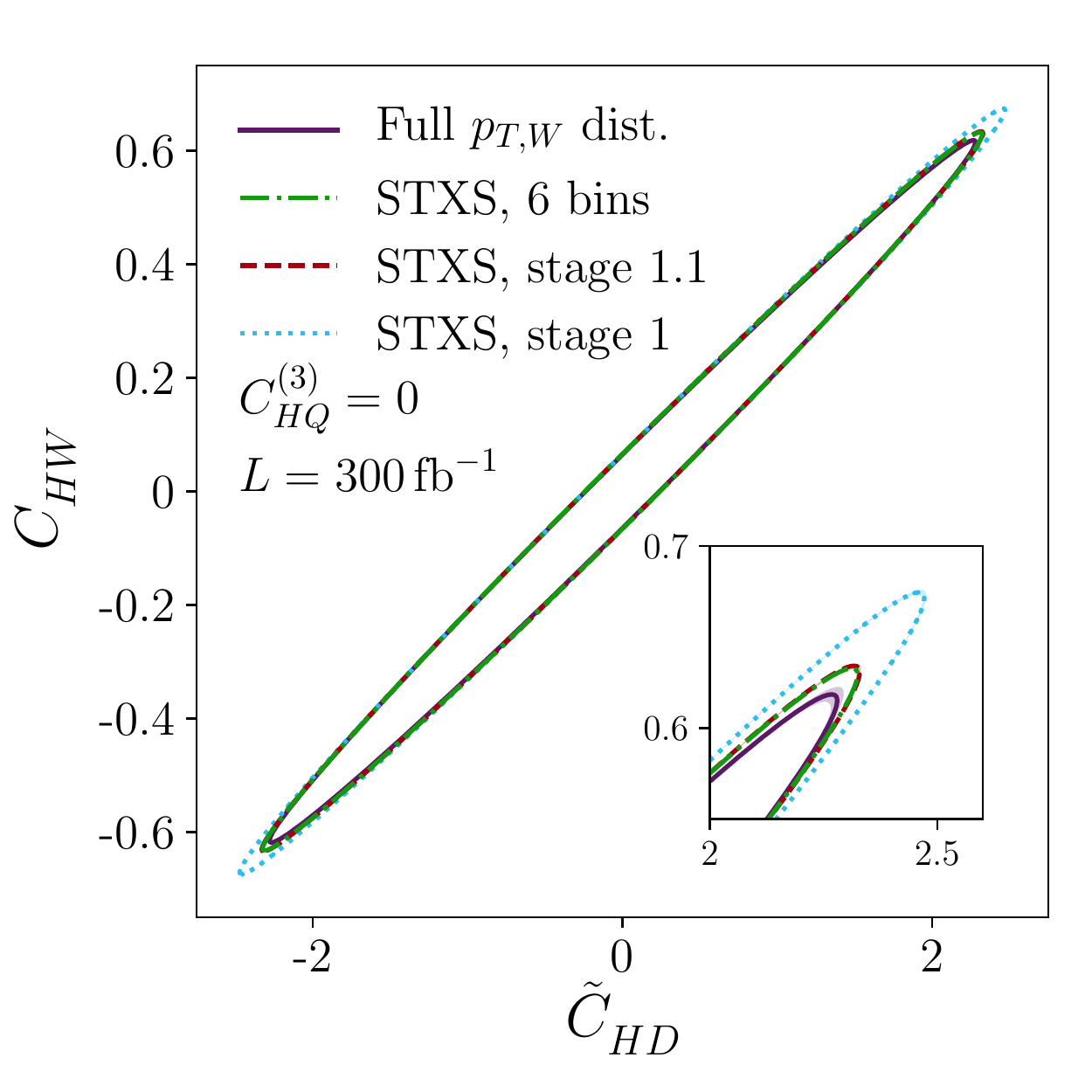}%
\includegraphics[width=.315\linewidth]{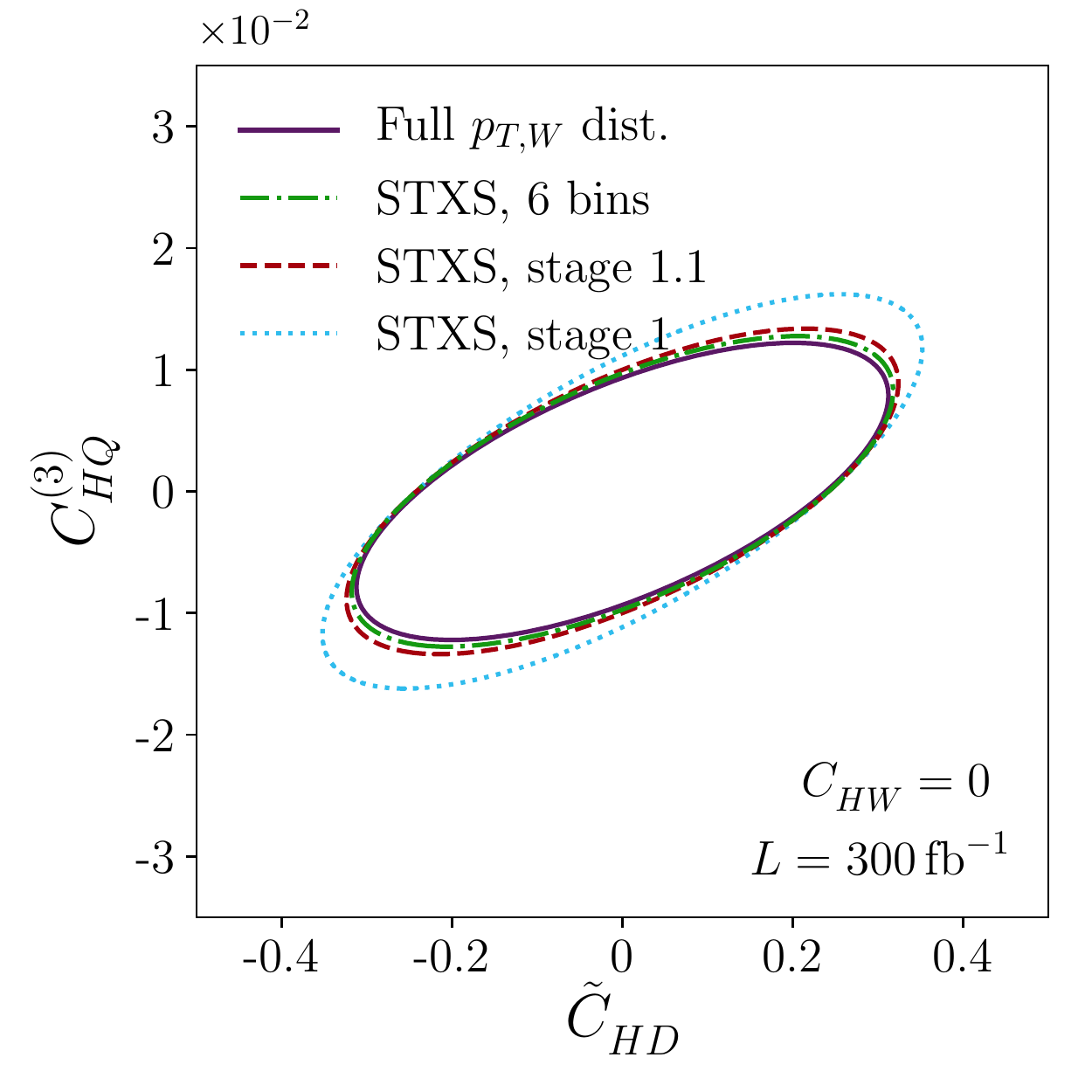}%
\includegraphics[width=.315\linewidth]{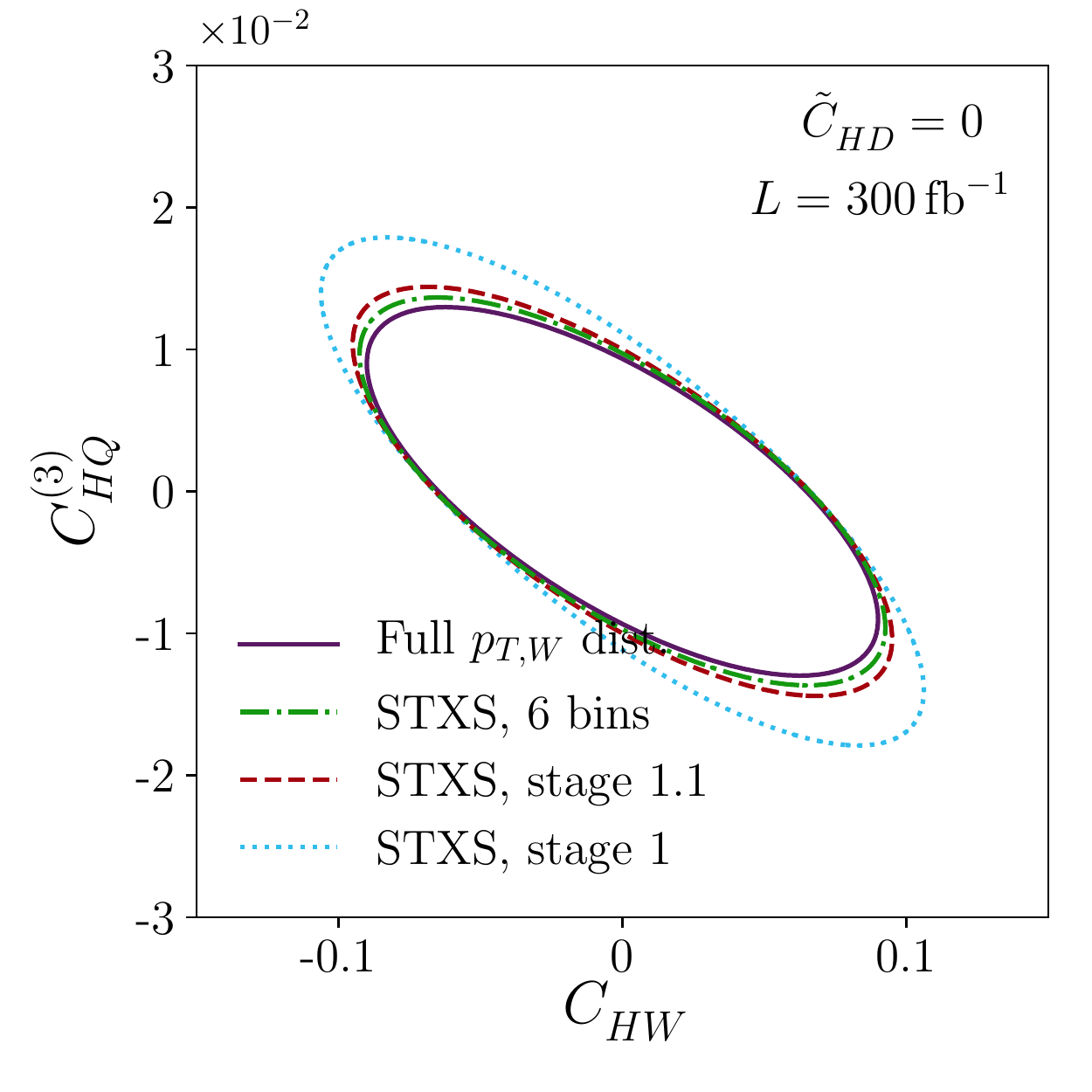}\\[-1mm]%
\includegraphics[width=.315\linewidth]{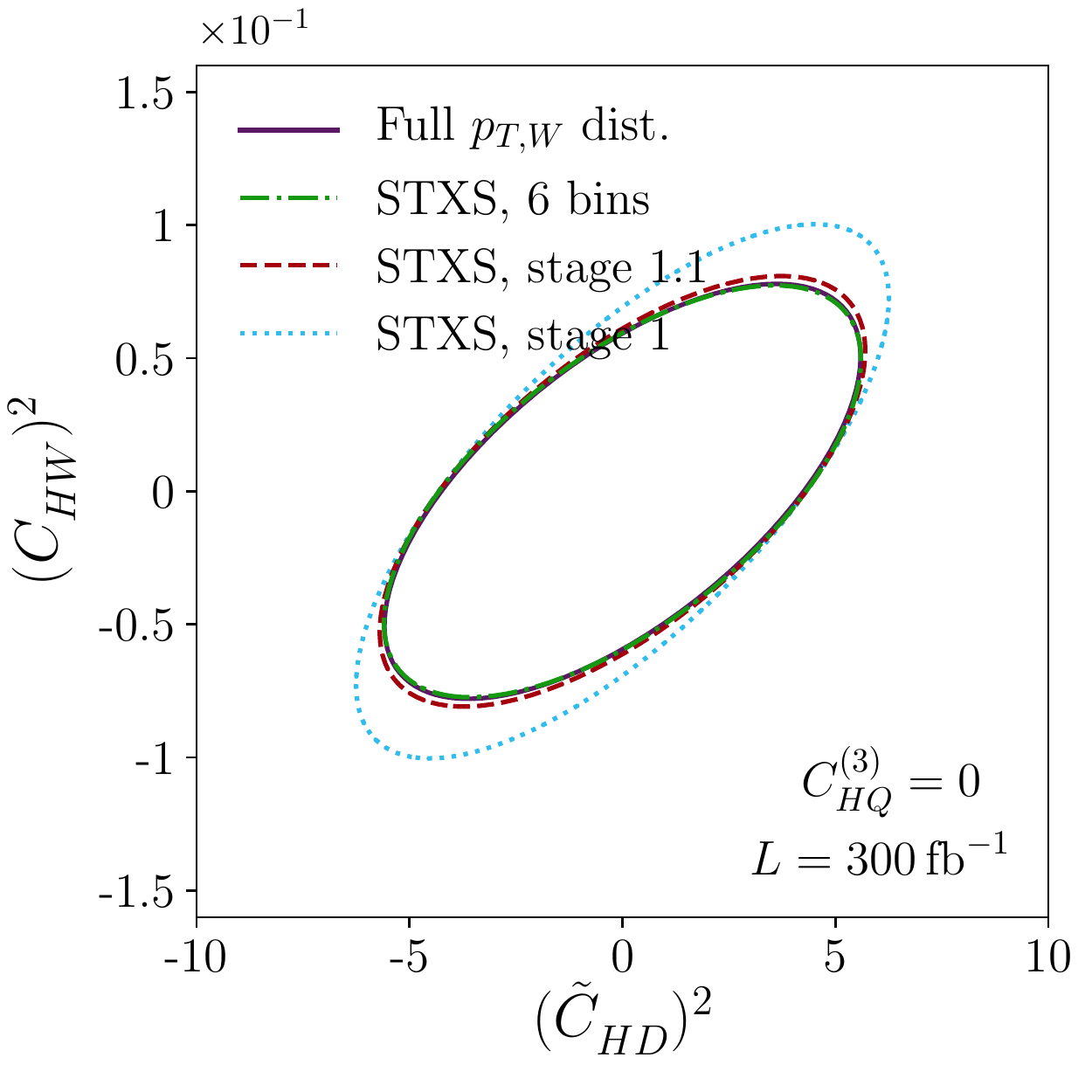}%
\includegraphics[width=.315\linewidth]{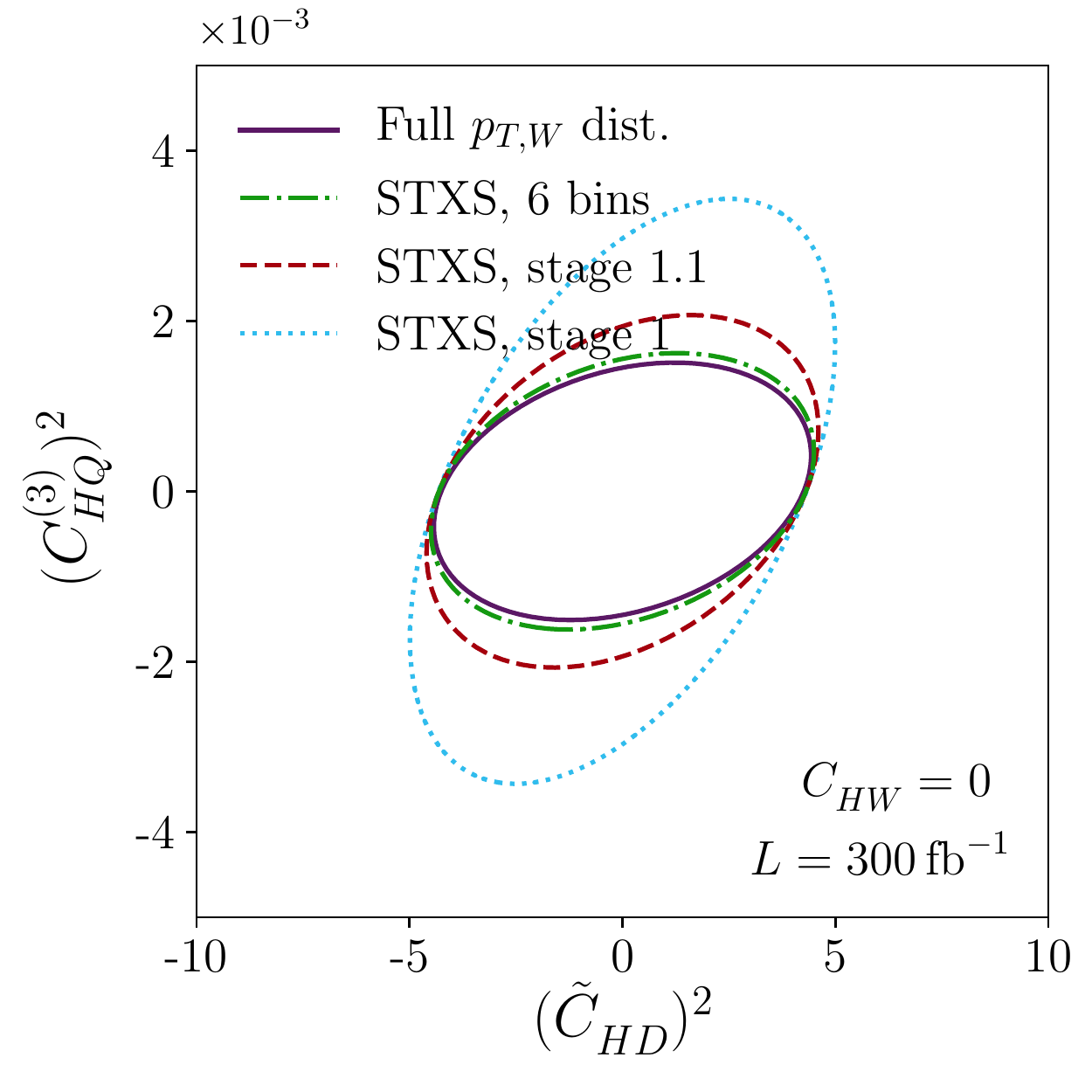}%
\includegraphics[width=.315\linewidth]{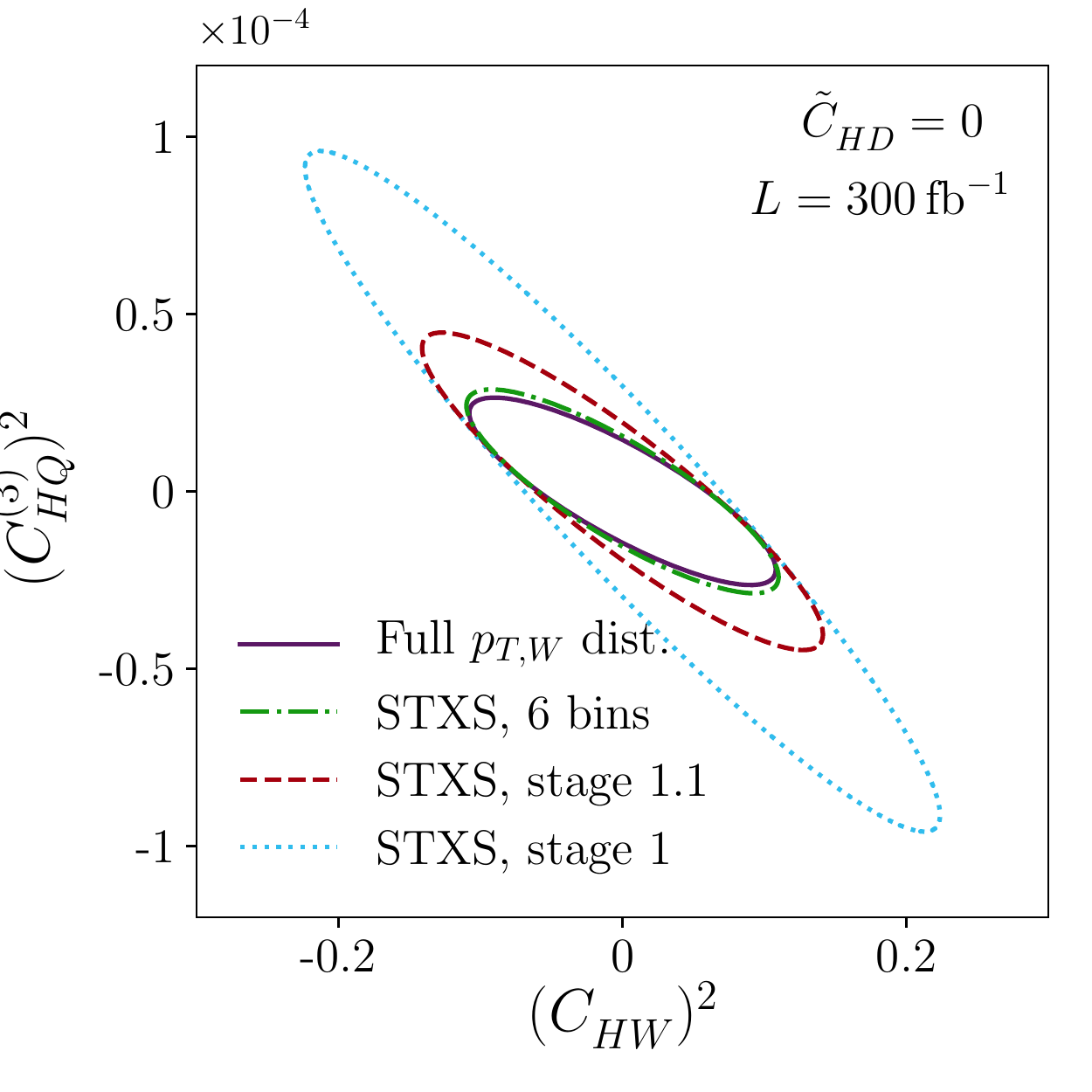}\\[-1mm]%
\includegraphics[width=.315\linewidth]{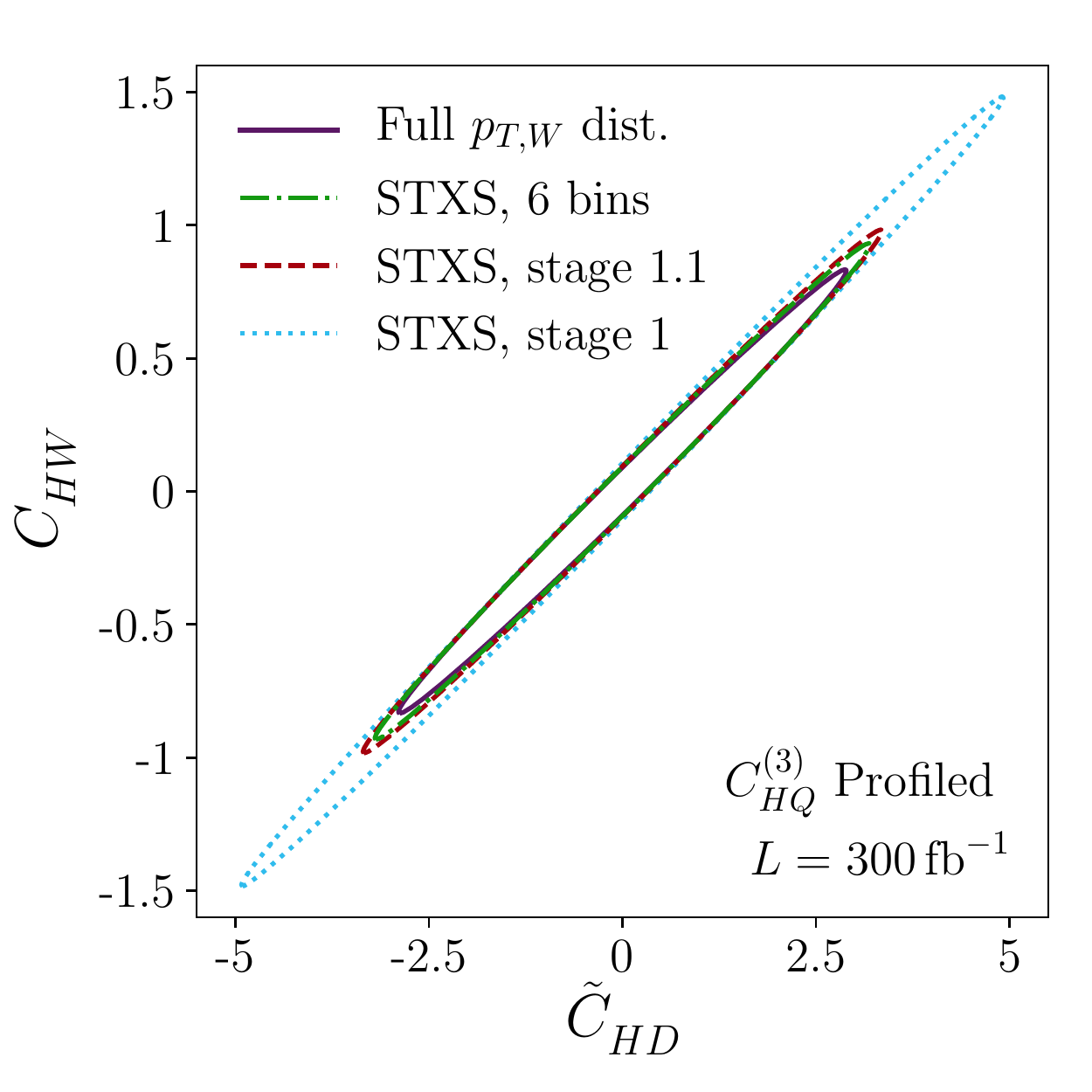}%
\includegraphics[width=.315\linewidth]{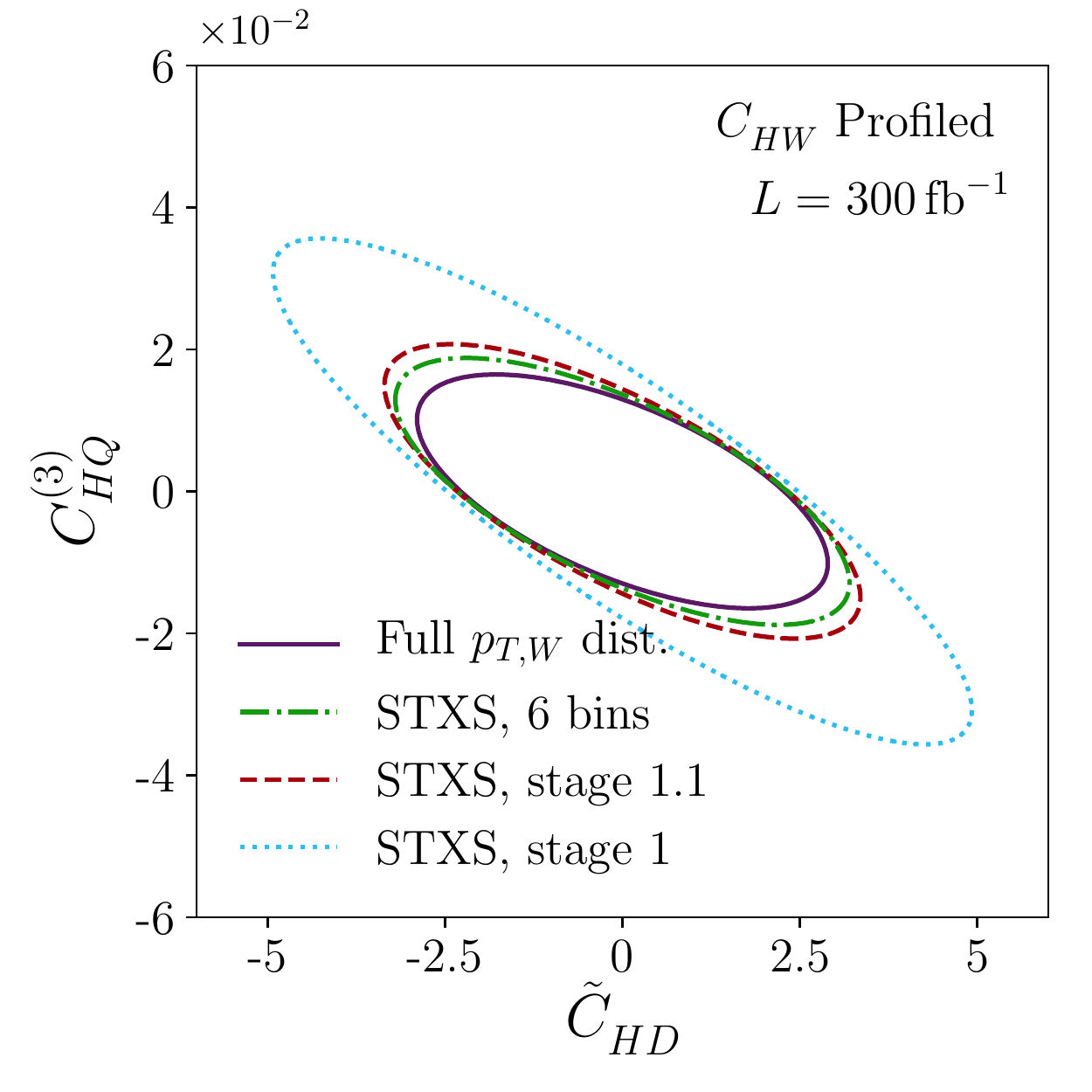}%
\includegraphics[width=.315\linewidth]{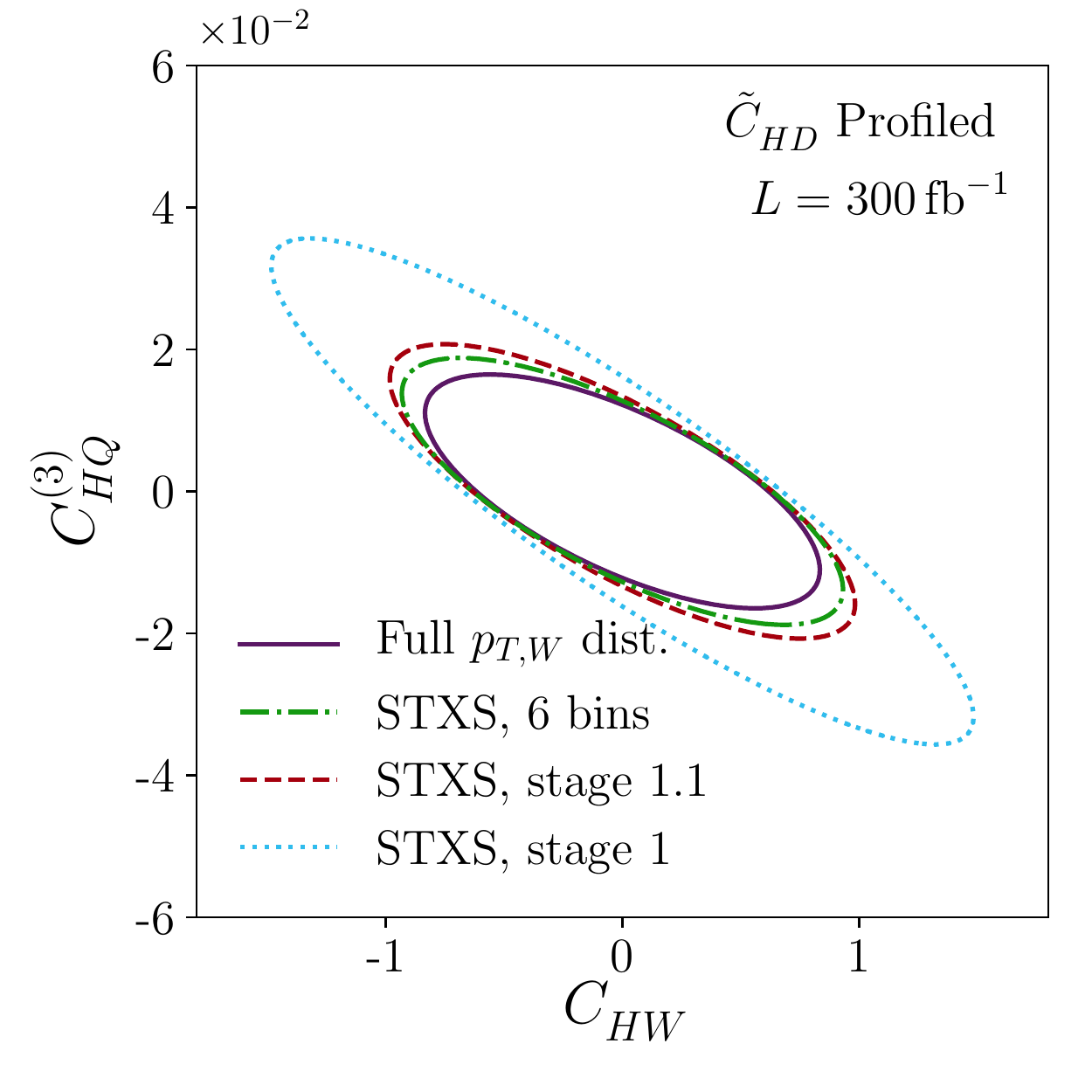}\\[-1mm]%
\includegraphics[width=.315\linewidth]{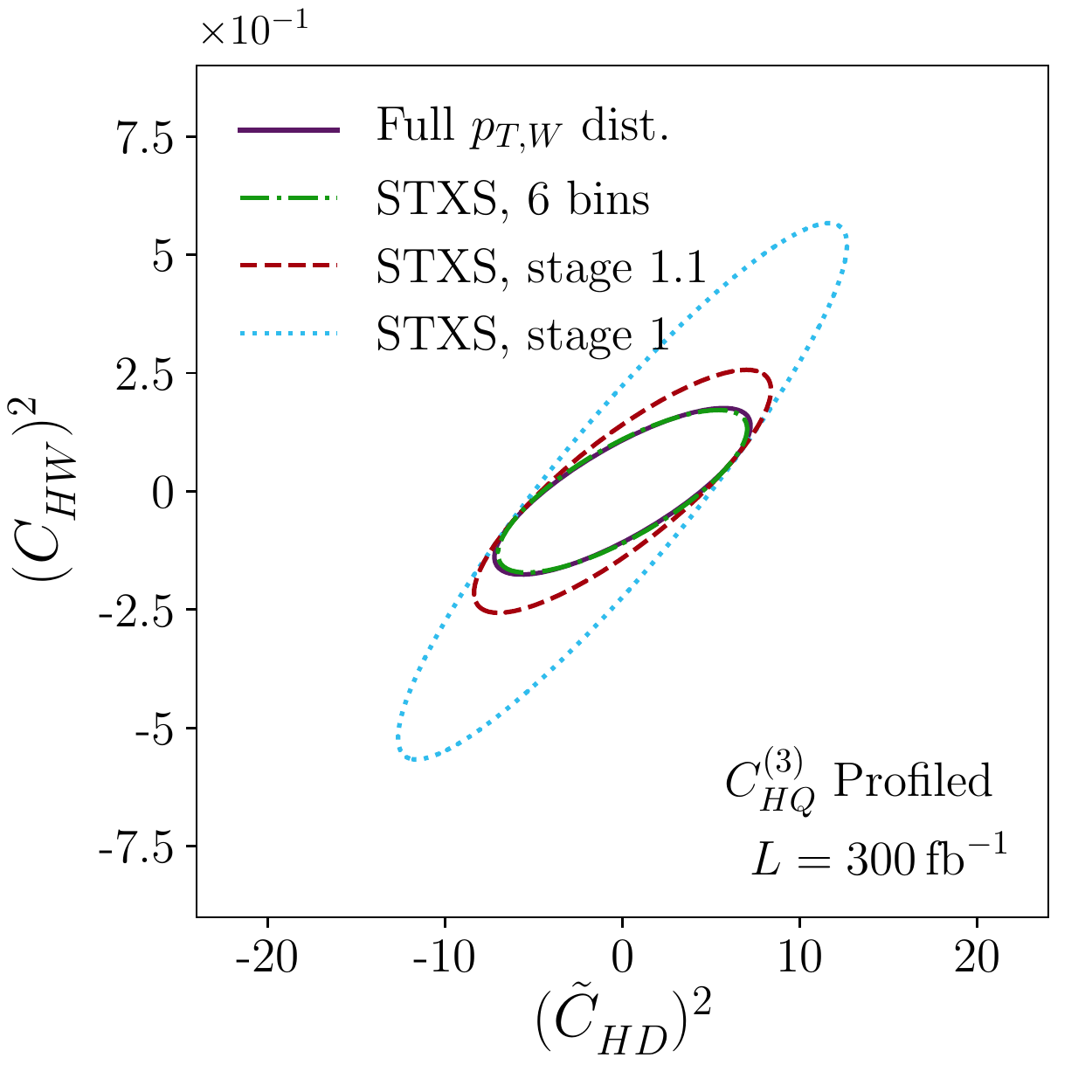}%
\includegraphics[width=.315\linewidth]{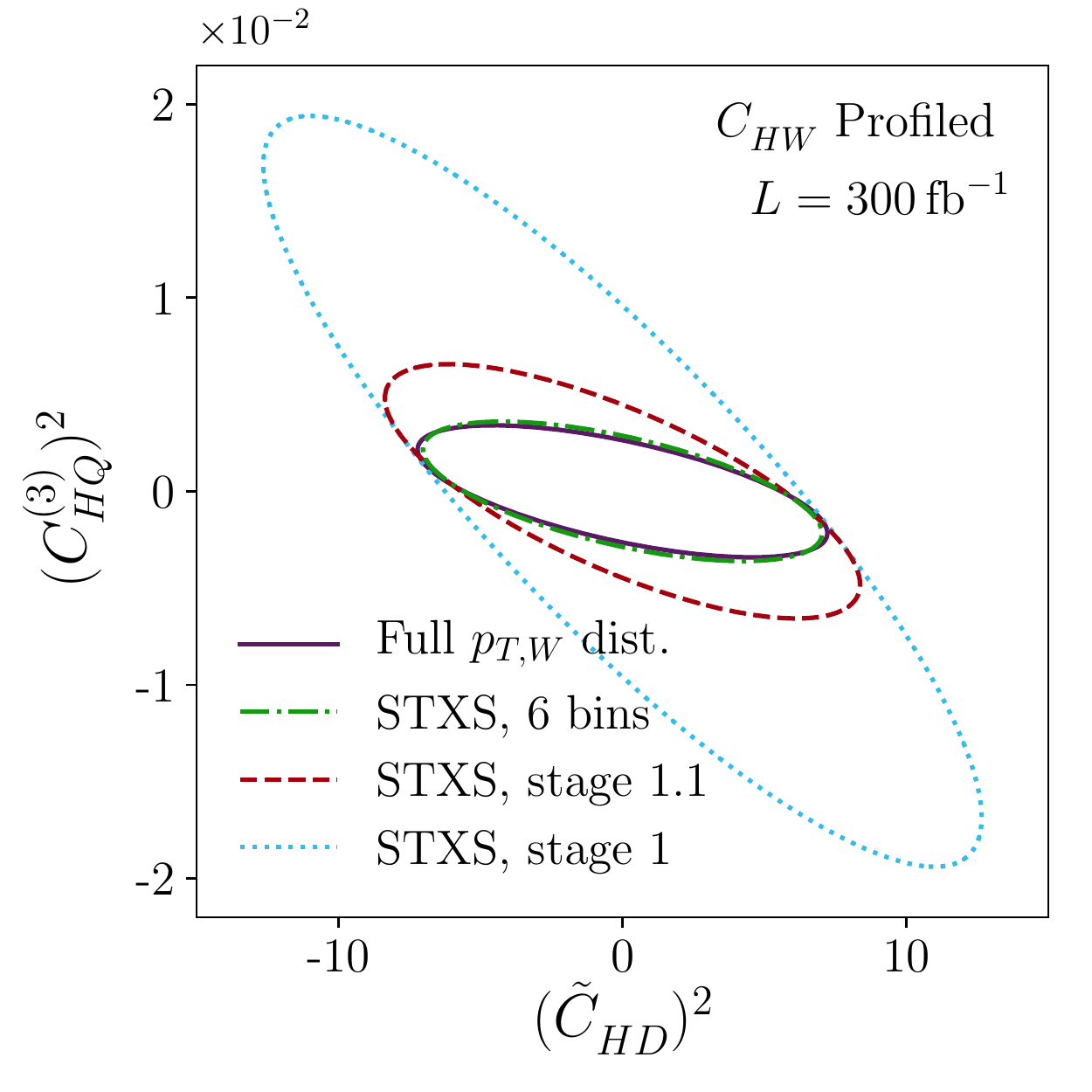}%
\includegraphics[width=.315\linewidth]{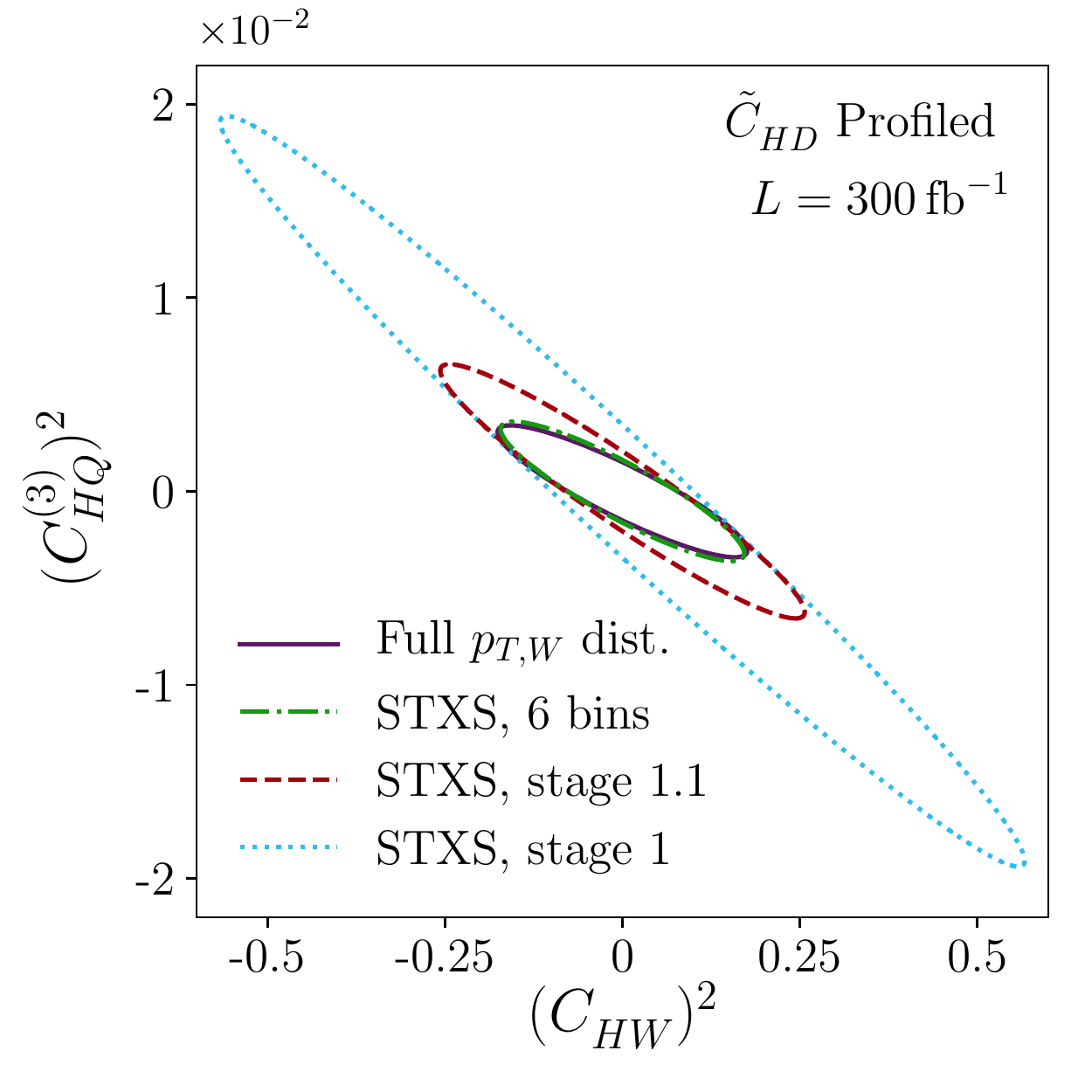}%
\caption{95\% CL constraints based on $p_{T,W}$ in the Fisher information
approximation. We show linearized and squared-only results setting the third
operator to zero (top six panels) and profiling over the third operator (bottom
six panels). The blue dotted line shows the STXS stage 1 (3 bins), red dashed
stage 1.1 (5 bins), and green dot-dashed and purple solid show the results when
adding a 6th bin or including the full $p_{T,W}$ distribution.}%
\label{fig:contours_ptw}%
\end{figure}

In Fig.~\ref{fig:contours_ptw}, we show the expected limits on pairs of Wilson
coefficients from the $p_{T,W}$ distribution. First, we see the almost flat
direction in the $\opet{HD}-\ope{HW}$ plane, reflecting the fact that both
operators are constrained  at low transverse momenta, as long as we only
consider linearized predictions. This flat direction will be broken by other
observables in a proper global analysis~\cite{Biekotter:2018rhp}, so it is of
less interest for our purposes. The situation changes once we consider the
correlation of either of these two operators with $\ope{Hq}^{(3)}$, because now
the two directions are tested by distinctly different $p_{T,W}$ regimes. The
only requirement in this case is that we include enough $p_{T,W}$ bins to
distinguish the two regimes, as is provided by the stage~1.1 setup. Indeed, this
framework collects essentially all information on our set of three operators
affecting the $p_{T,W}$ distribution.

In the second row of Fig.~\ref{fig:contours_ptw} we switch from the linear terms
in the Wilson coefficient to the squared terms alone, as defined 
in Section~\ref{sec:squared}. This allows us to analyze the relative sensitivity of 
the different analysis strategies and binnings on the squared terms which, as we
saw in Figs.~\ref{fig:distributions1} and \ref{fig:distributions2}, have very different
kinematic behaviors compared to the linear terms. Showing the information on the
squared terms as contours also allows us to study the correlations between different
squared parameters, which, as we will see, are often different than the correlations
between linear terms.
In the left panel we see
hardly any effect of the different binnings for $\opet{HD}$, but the flat
direction has vanished and the reach in $\ope{HW}$ is increased dramatically by
including the high $p_{T,W}$ bins. Because this sensitivity to $\ope{HW}$ comes
from high momentum transfer, we start to observe a strong anti-correlation with
$\ope{Hq}^{(3)}$. To break it, we need to resolve the high-$p_{T,W}$ range, and
for this purpose the additional bin in Eq.~\eqref{eq:stxs+} is crucial. Indeed,
we see that it improves the situation considerably.

In the lower six panels of Fig.~\ref{fig:contours_ptw} we show the same kind of
information, but now profiled over the respective third operator. The results
are generally much weaker, and the flat directions are more significant. As
mentioned above, this is not a big concern, since our $WH$ analysis has to be
embedded in a global analysis. An interesting aspect is that with three
operators actually contributing to the analysis the gap between the few-bins
STXS approach and the full information from the entire $p_{T,W}$ distribution
widens. Some details on the scaling with larger number of bins and on the number
of bins needed to saturate the information in a kinematic distribution are given
in App.~\ref{sec:binning}. Altogether, it becomes even more important that we
extract as much information for instance on $\ope{Hq}^{(3)}$ by sufficiently
covering the dedicated phase-space regions.\bigskip

\begin{figure}%
\includegraphics[width=.315\linewidth]{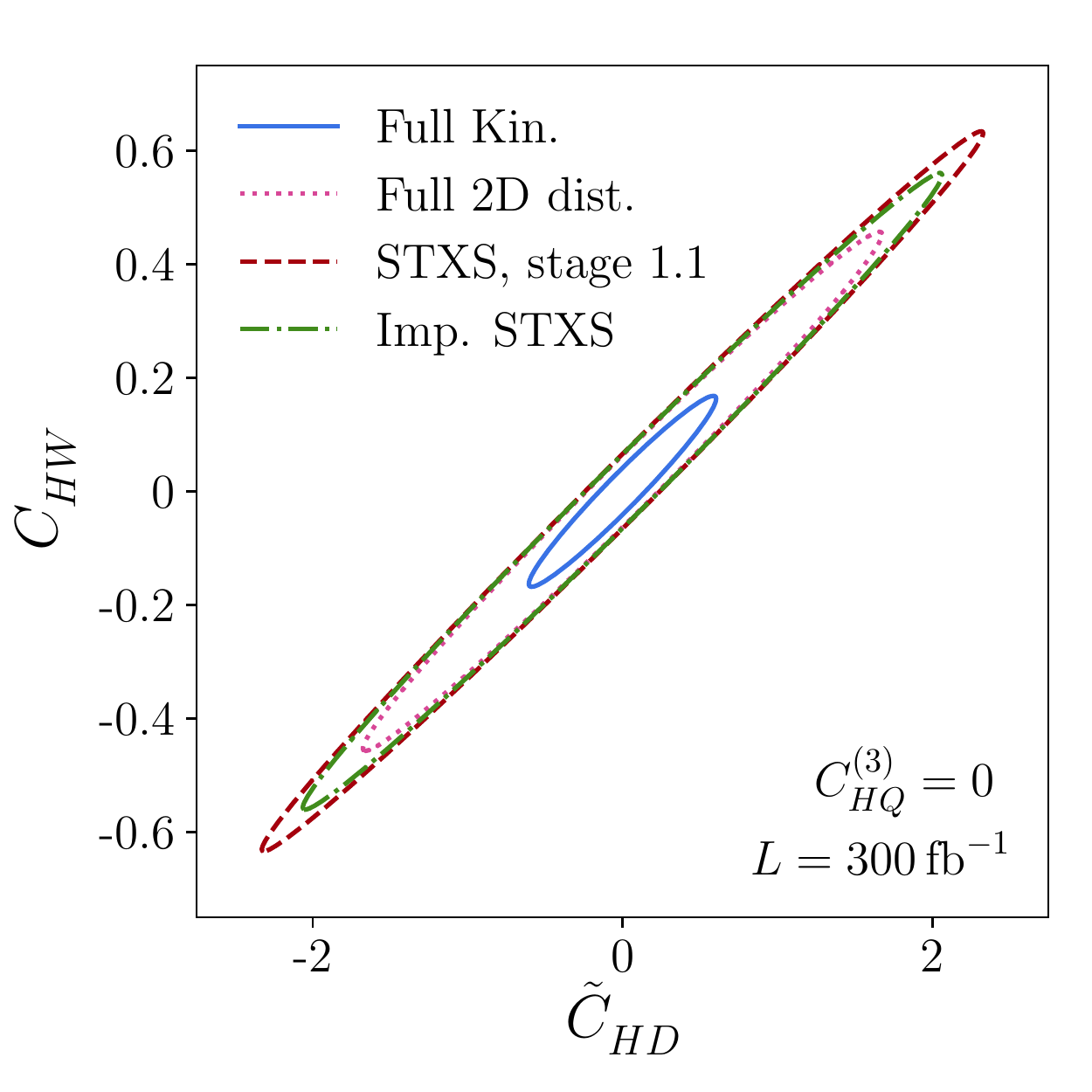}%
\includegraphics[width=.315\linewidth]{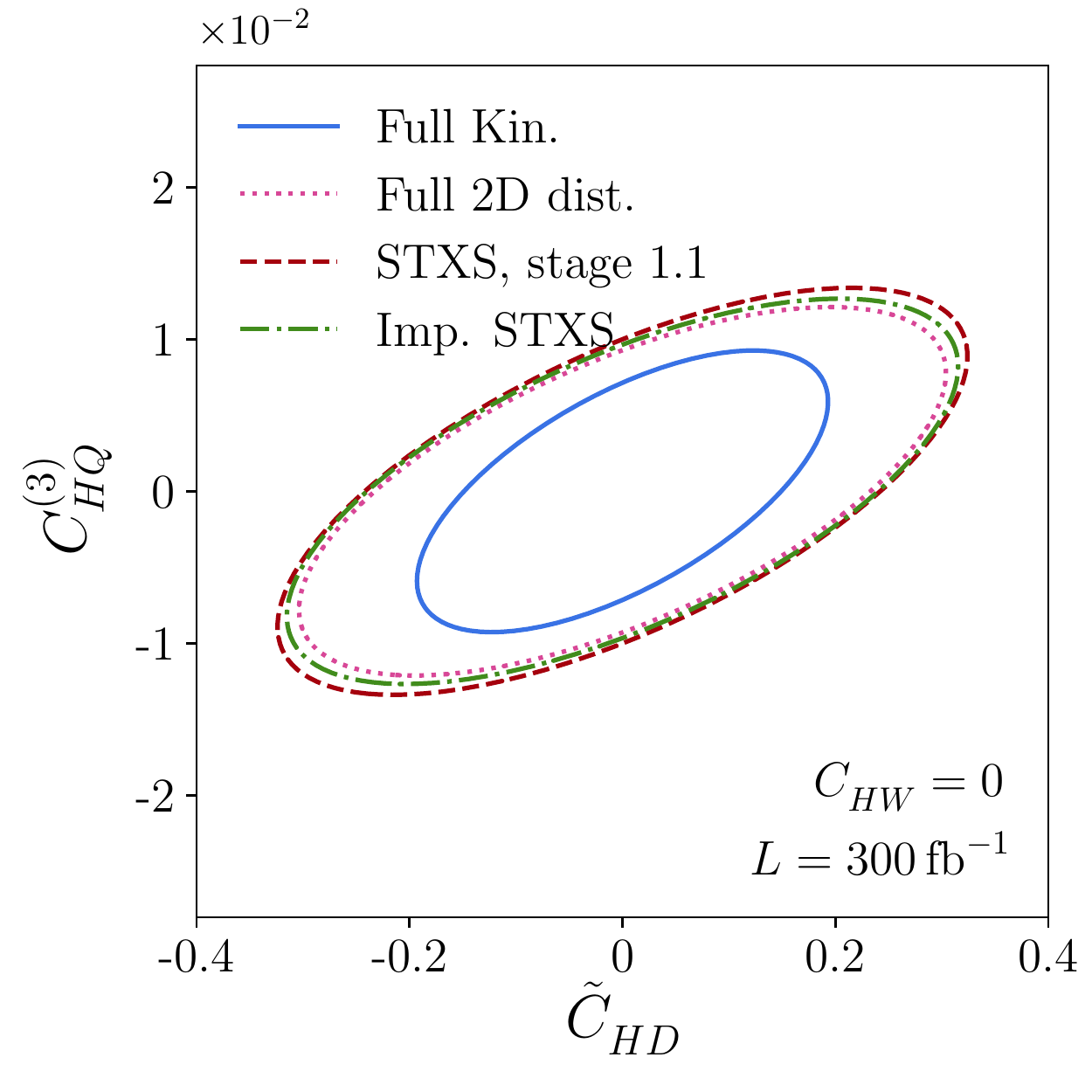}%
\includegraphics[width=.315\linewidth]{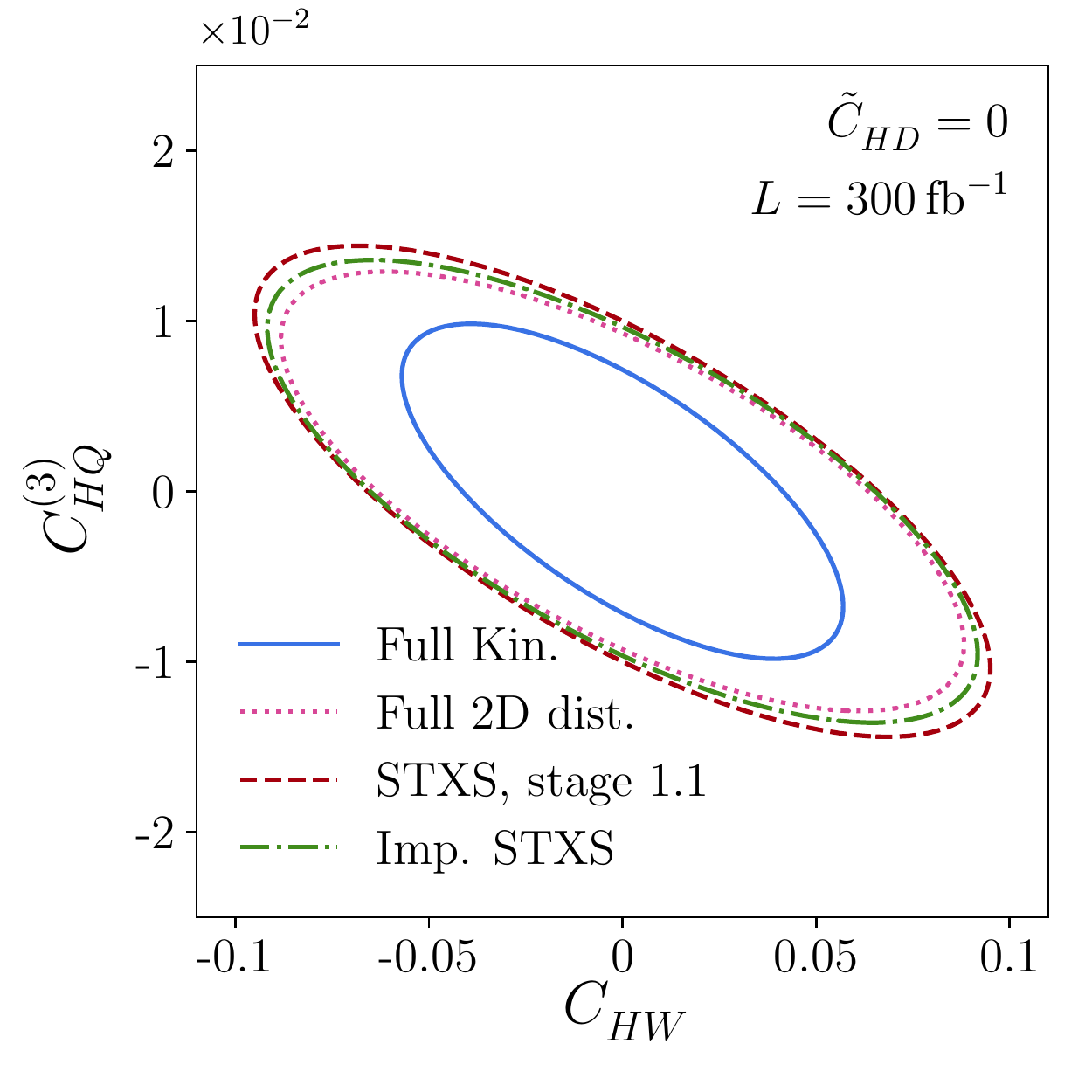}\\[-1mm]%
\includegraphics[width=.315\linewidth]{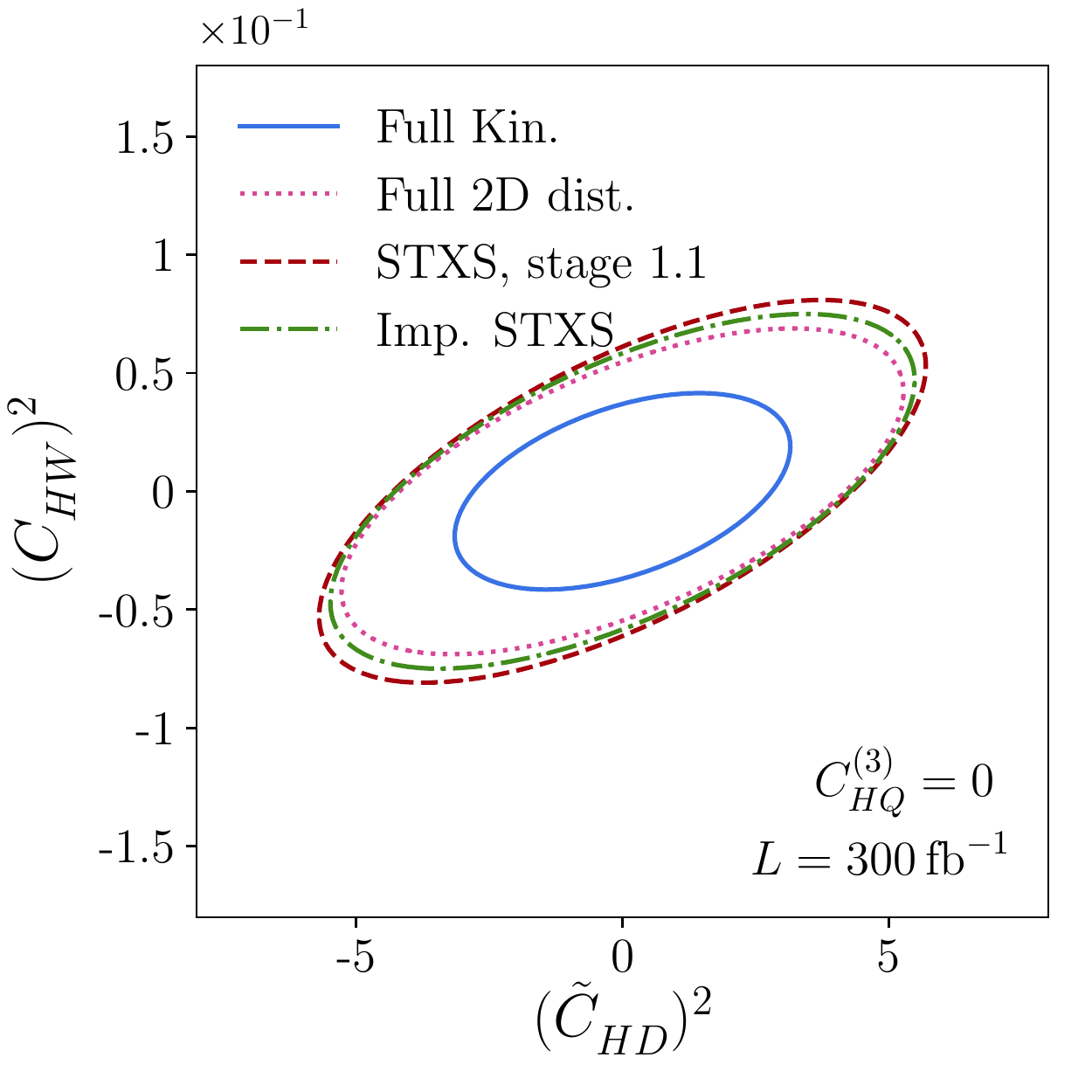}%
\includegraphics[width=.315\linewidth]{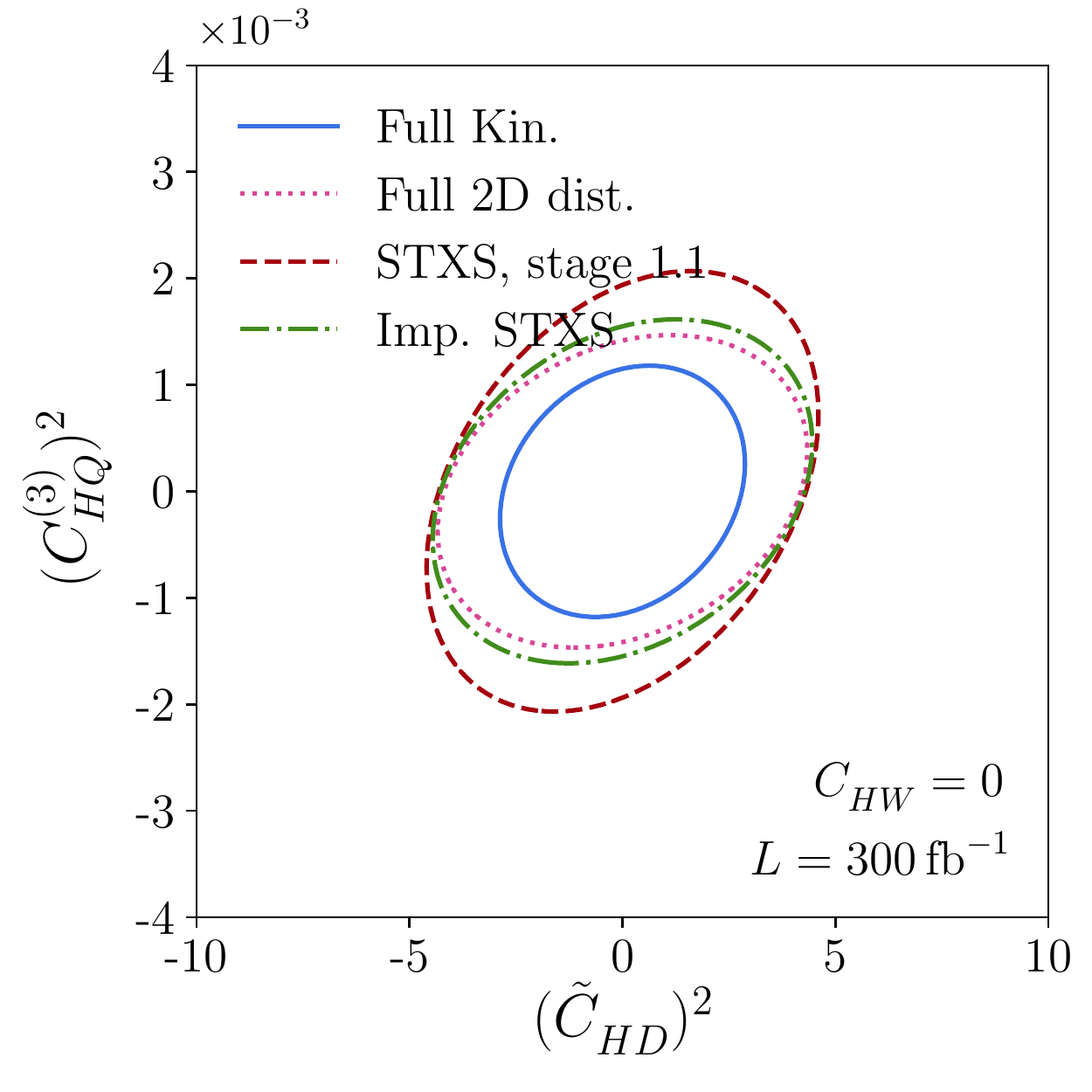}%
\includegraphics[width=.315\linewidth]{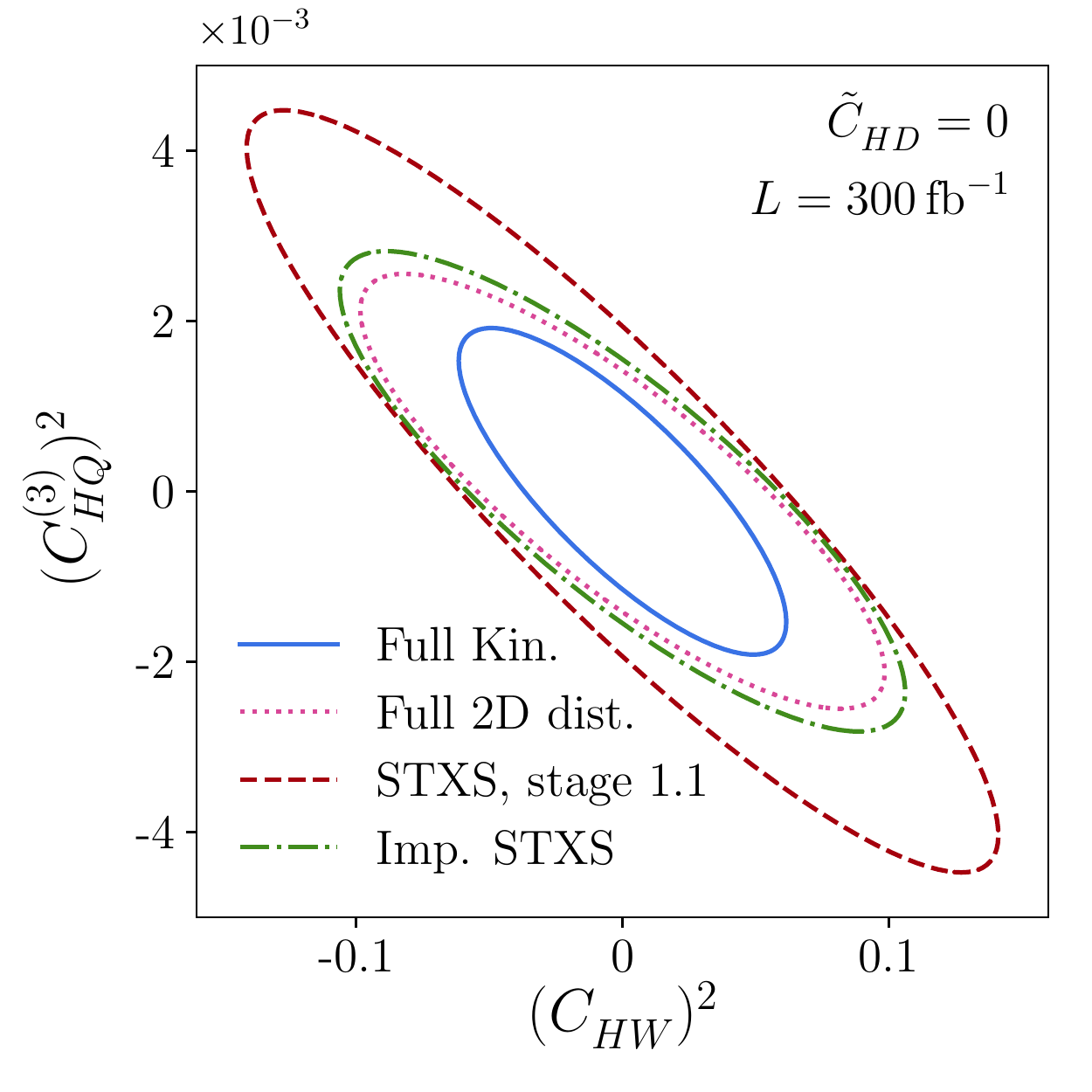}\\[-1mm]%
\includegraphics[width=.315\linewidth]{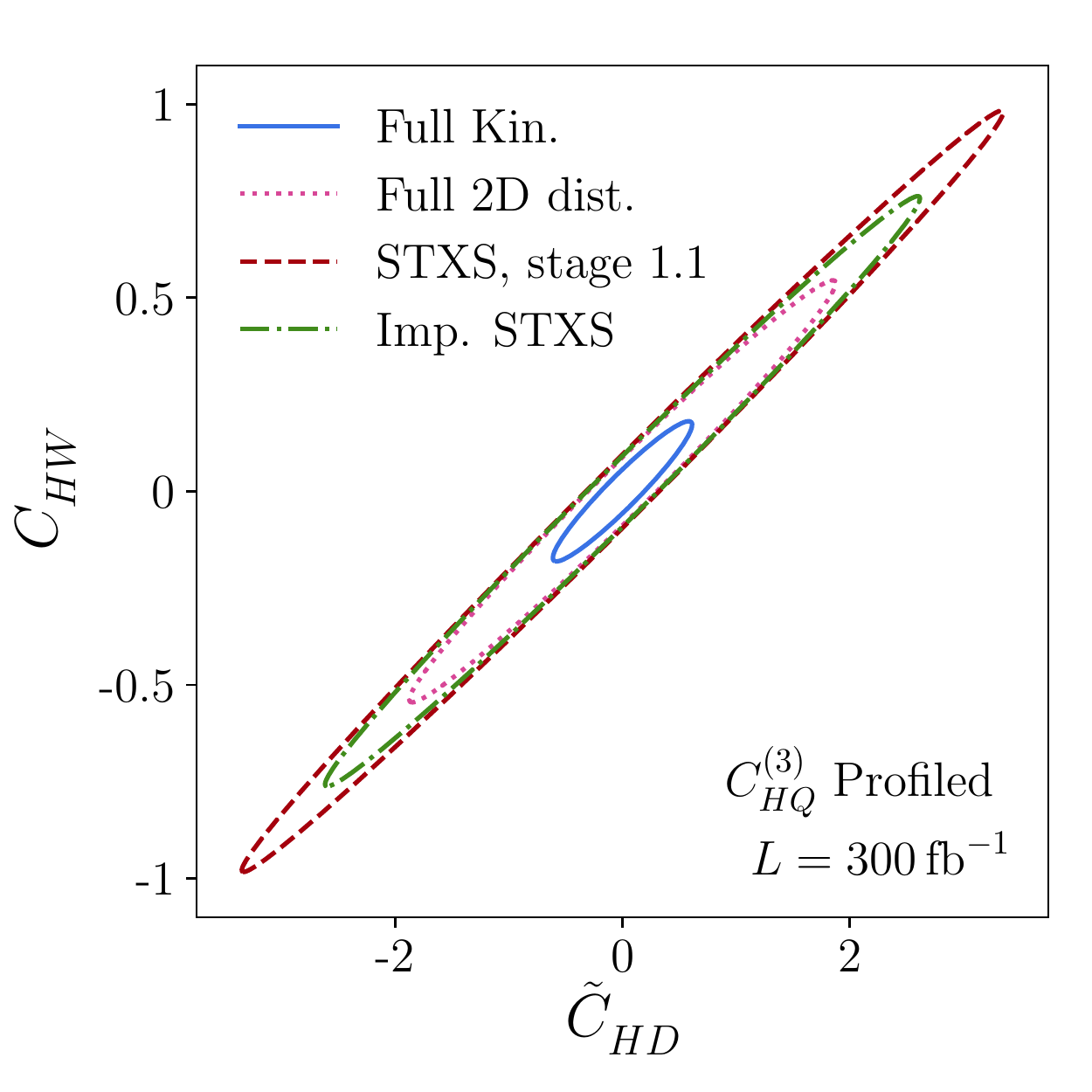}%
\includegraphics[width=.315\linewidth]{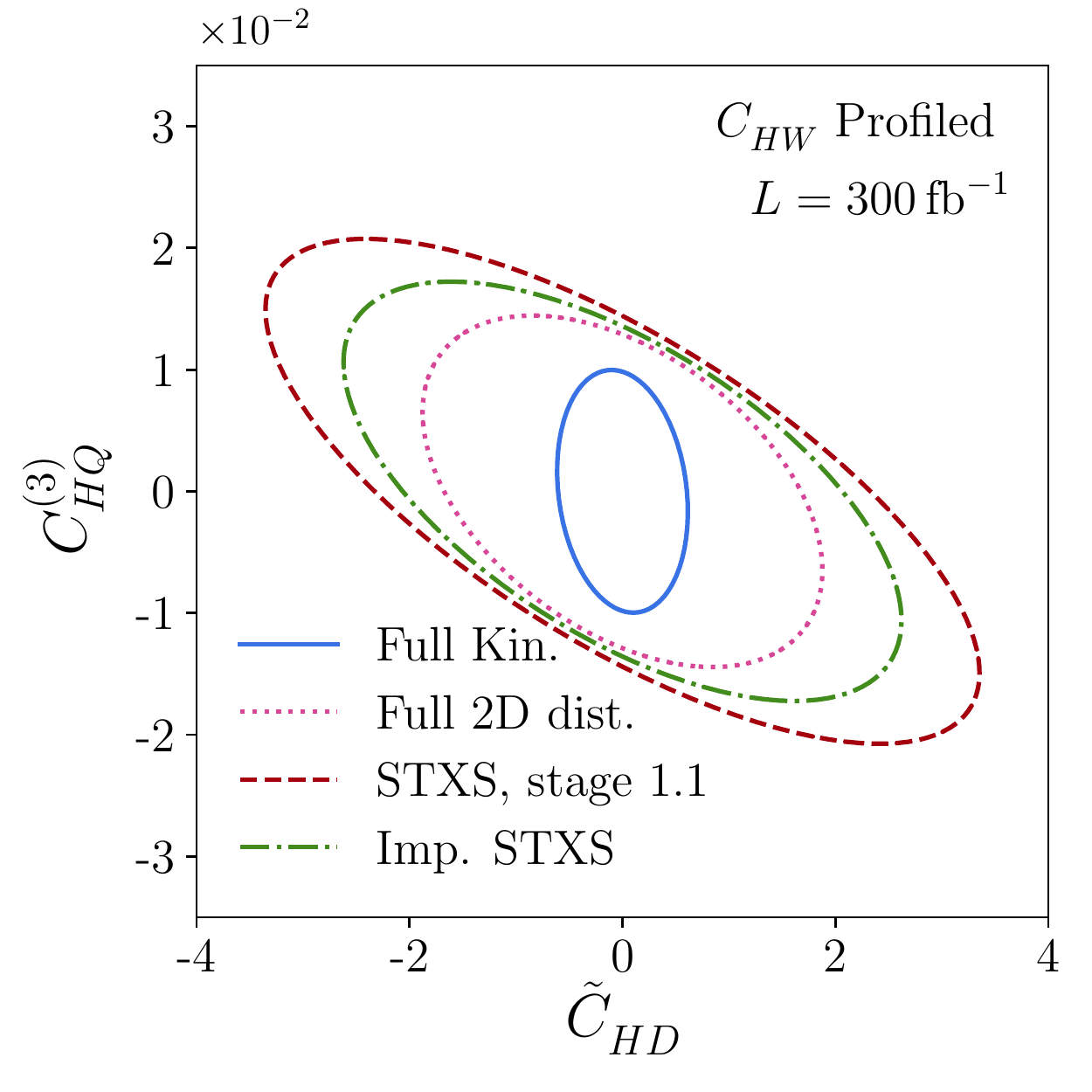}%
\includegraphics[width=.315\linewidth]{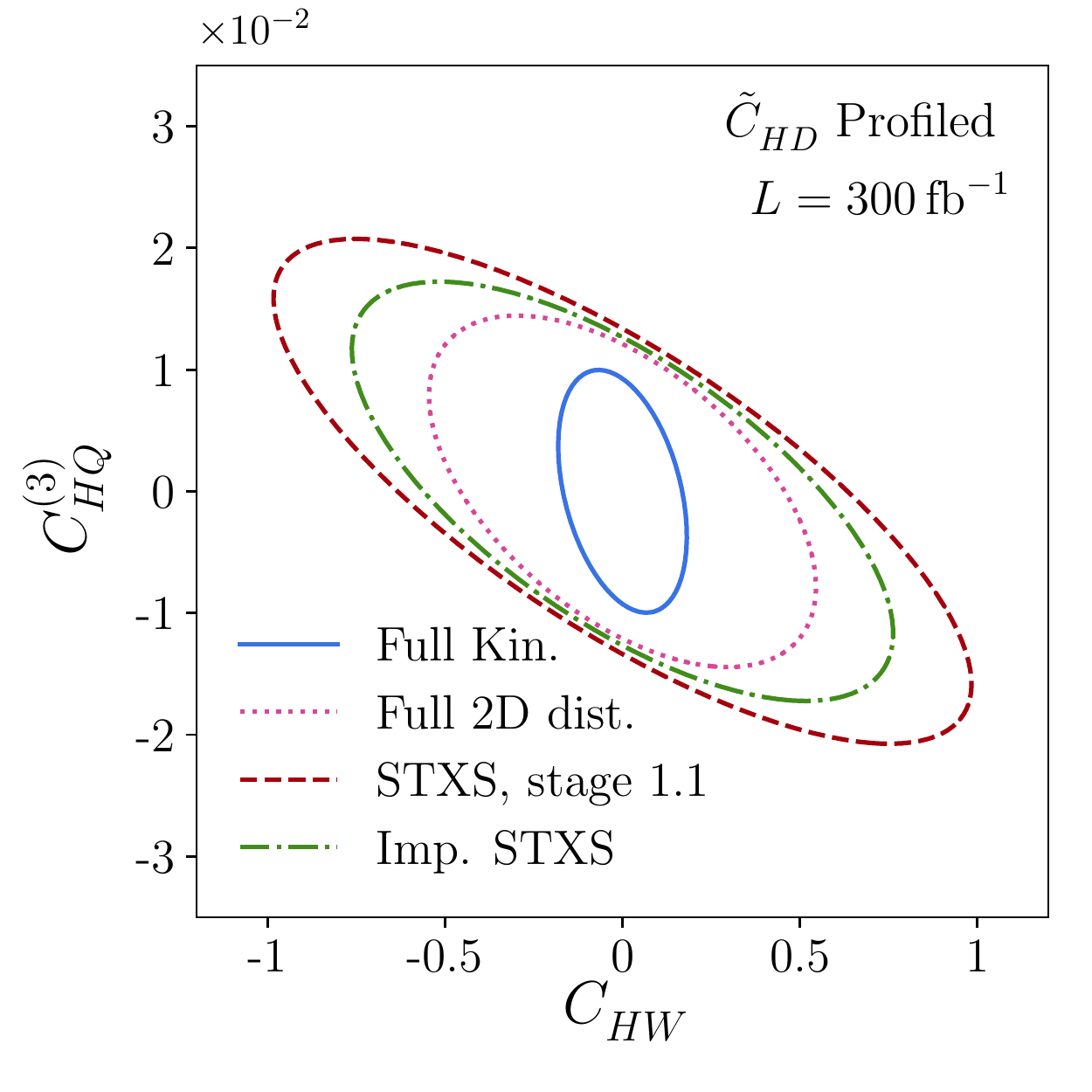}\\[-1mm]%
\includegraphics[width=.315\linewidth]{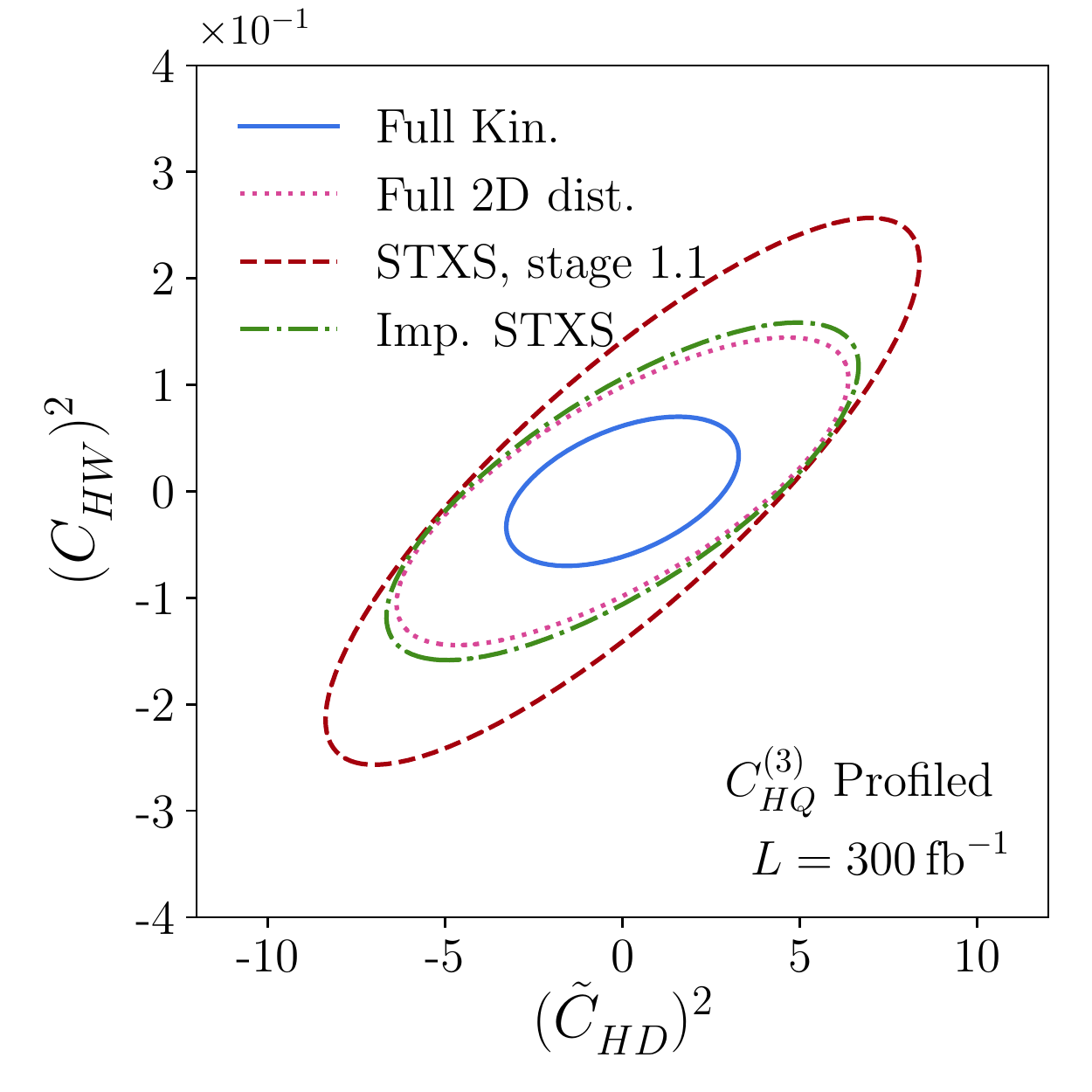}%
\includegraphics[width=.315\linewidth]{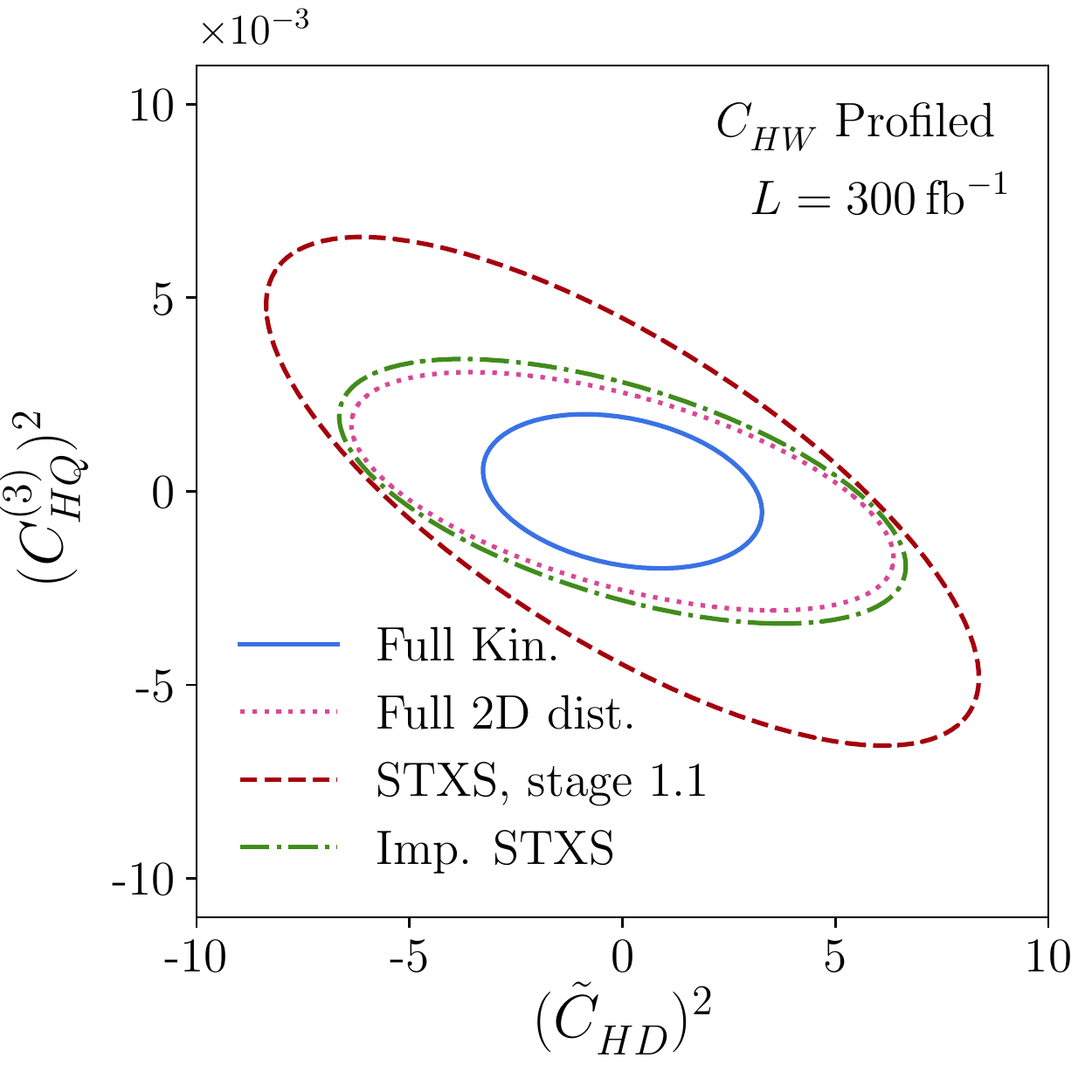}%
\includegraphics[width=.315\linewidth]{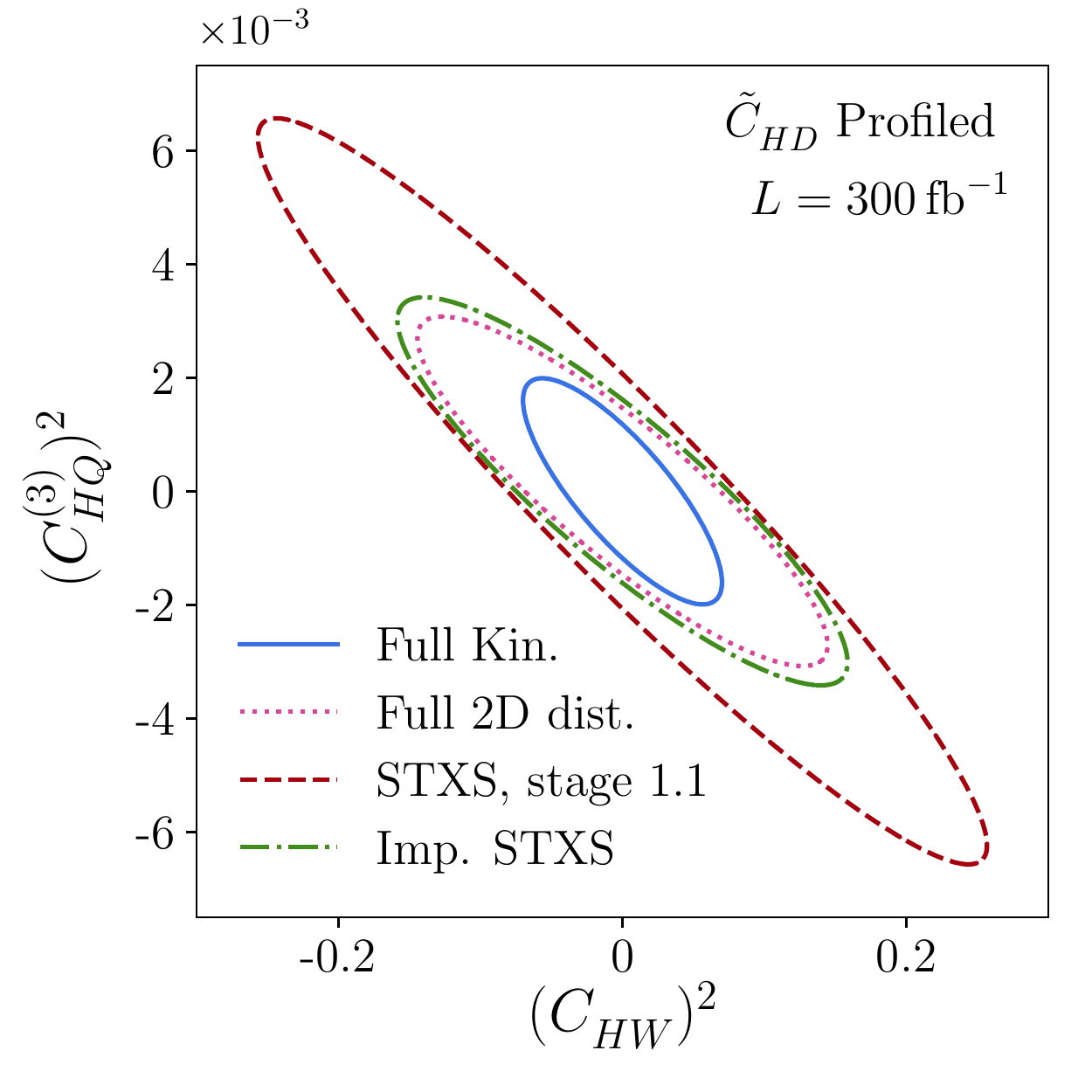}%
\caption{As Fig.~\ref{fig:contours_ptw}, now comparing various approaches: The
solid blue curve shows the limits from training on the full phase space. The
dotted pink line shows training only on the $p_{T,W}$ vs $\mtot$ histogram.  The
limits obtained from the STXS stage~1.1 are shown in dashed red, and the
dot-dashed green curve shows the limits obtained by six $p_{T,W}$ bins
subdivided into three $\mtot$ bins.}%
\label{fig:contours}%
\end{figure}

After having established that the currently used STXS for $WH$ production need
to be supplemented by another high-energy bin,  we can test how much of the
available information of the full phase space is captured in the binned
$p_{T,W}$ distribution. The reasoning behind picking an additional observable
scaling like momentum is that the signal process is dominantly $2 \to 2$
scattering, described by two kinematic observables, and the scattering angle or
rapidity are not found to be particularly useful, as shown in
Fig.~\ref{fig:distributions2}. On the other hand, if the signal from dimension-6
operators populates phase space far away from the SM we will become sensitive to
the background kinematics, and the simple picture of a $2 \to 2$ process can
completely change~\cite{Plehn:2013paa, Kling:2016lay}.

In the top row of Fig.~\ref{fig:contours} we show the limits from the \sally
approach trained on linearized effects in the Wilson coefficients (the same as in 
Fig.~\ref{fig:validation}) as a blue line. The dotted pink line comes from limiting the 
information of the \sally training to the observables $p_{T,W}$ and $\mtot$, which can be understood as
the constraints computed from the infinite-bin limit of a 2-dimensional
histogram. It is not clear a priori which second observable complements
$p_{T,W}$ best,  but we found that $\mtot$ as defined in Eq.~\eqref{eq:mttot}
works well. In the linearized approach, we confirm that compared to the full
phase space we still lose a significant amount of information on the three
operators. The next question is how much of the information from the
two-dimensional contribution we can capture in a reasonable number of STXS bins.
Motivated by Fig~\ref{fig:distributions1}, we propose to further split each of
the six bins in $p_{T,W}$ into three $\mtot$ bins in
\begin{align}
\mtot^\text{bins} = ( \, 0-400, \, 400-800, \, 800-\infty \, )~\gev \; ,
\label{eq:stxs++}
\end{align}
as illustrated in Fig.~\ref{fig:stxs}. Because $\mtot$ and $p_{T,W}$ are by no
means uncorrelated, this captures the effects of the $2 \to 3$ background phase
space and adds an implicit sub-division in $p_{T,W}$. In the top row of
Fig.~\ref{fig:contours},  we see that for the linearized case our new proposed
STXS definition performs extremely well, but  the STXS stage~1.1  also does
fine. In the second row, we show the same effects, but now for the squared new
physics terms only and not including the imposed  positivity condition. The blue
curve is now generated consistently based on the squared terms, using the \harry
method. As expected, the improved 2-dimensional STXS outperform the stage 1.1
results when we test $\ope{Hq}^{(3)}$, as a better understanding of the
information in high $p_{T,W}$ or $\mtot$ bins is effective. The lower panels of
Fig.~\ref{fig:stxs} show the same effect, but profiled over the full range of
the respective third Wilson coefficient. The large effect related to
$\ope{Hq}^{(3)}$ now propagates through the entire set of measurements.
Moreover, the limited number of bins in the STXS stage~1.1 become an issue.

\subsection{Limits with consistent squared terms}
\label{sec:phys_limits}

\begin{figure}[]
\centering
\includegraphics[width=.32\linewidth]{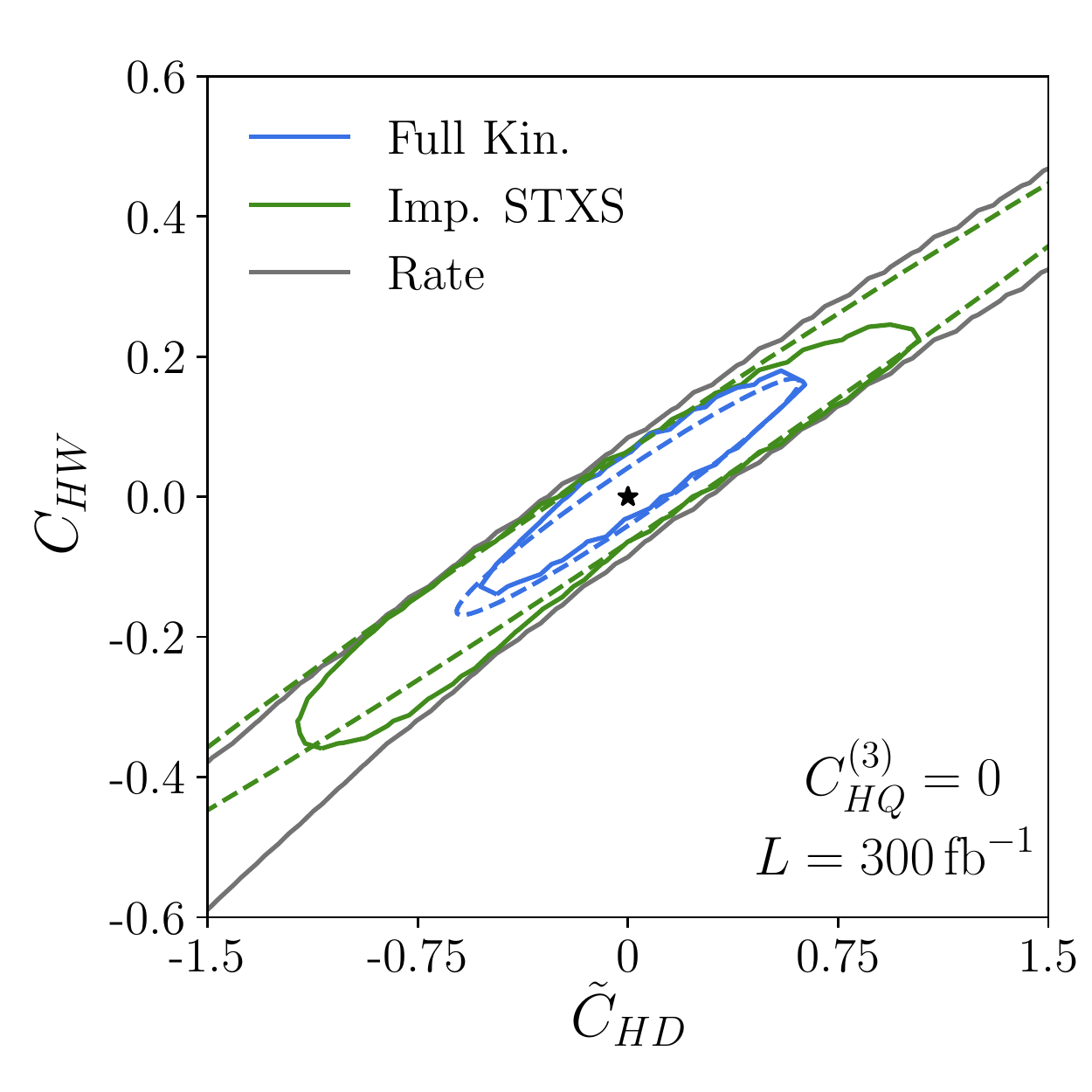}
\includegraphics[width=.32\textwidth]{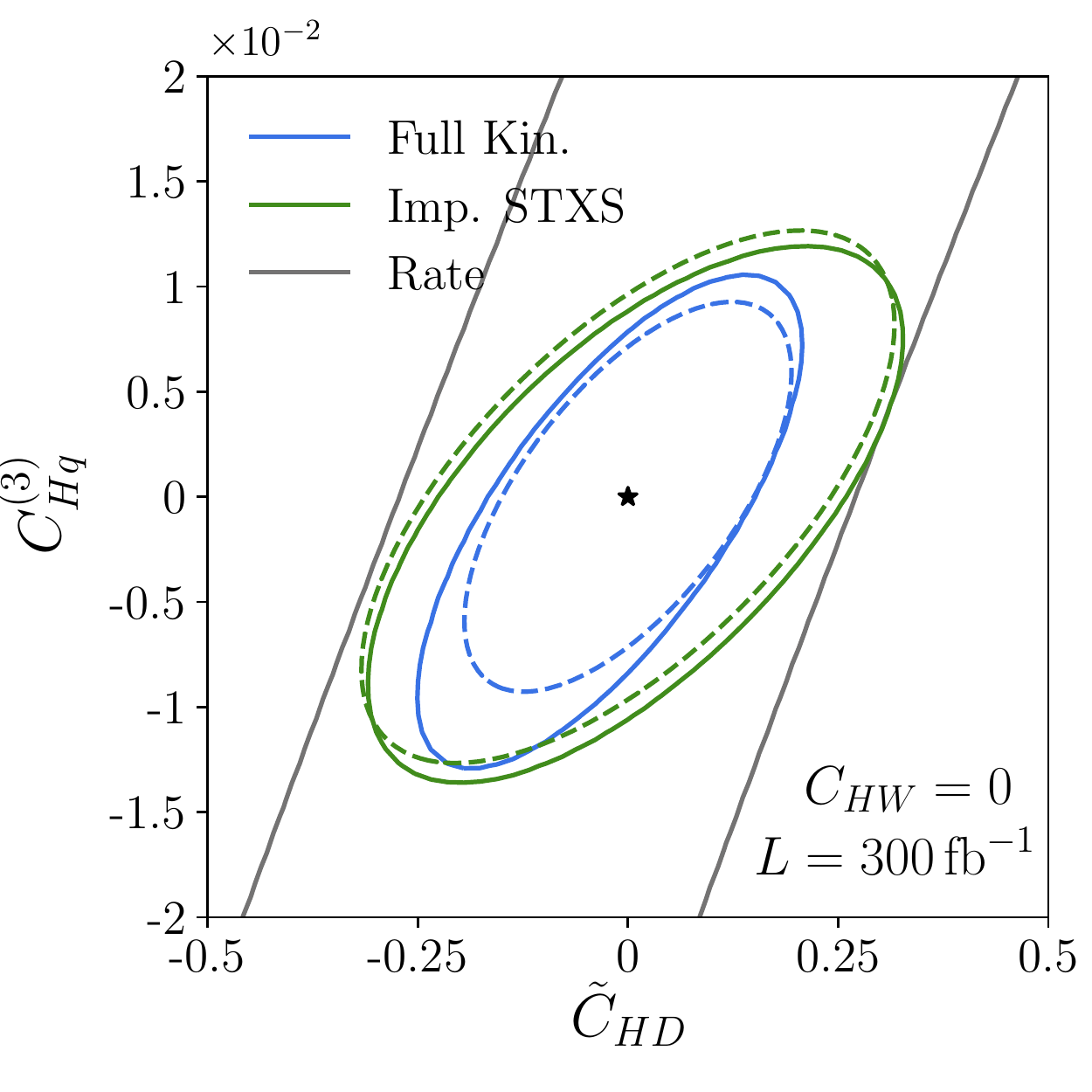}
\includegraphics[width=.32\linewidth]{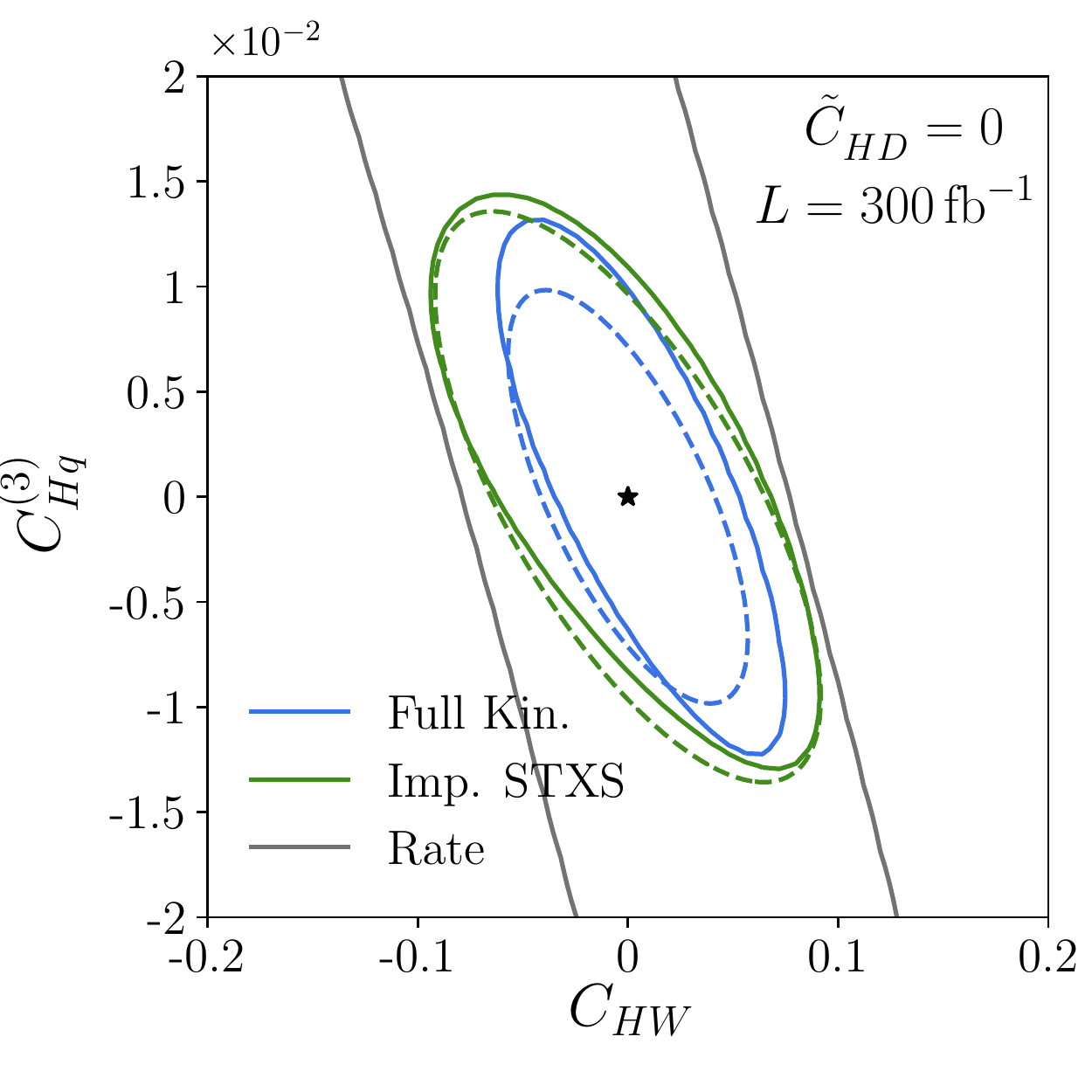}
\caption{Expected exclusion limits at 95\% CL based on a rate-only analysis
(grey), from the improved STXS (green), and from a multivariate analysis of the
full phase-space information with the \sally{} algorithm. We show (linearized)
limits based on the Fisher information (dashed) and full exclusion limits that
take into account the effects of linearized and squared new physics effects
(solid). In each panel, we set the operator not shown to zero.}
\label{fig:expected}
\end{figure}

Finally, we go beyond the linearized Fisher-information approach and calculate
expected exclusion limits on the dimension-6 Wilson coefficients with the
squared terms included in a physically consistent way. Generally, power counting
in $1/\Lambda^2$ suggests that the linear effects should be dominant. However,
in some directions of parameter space the interference contribution to the
analyzed distributions is small, the analysis is only sensitive further away
from the SM, and the squared terms are the dominant effects of new physics. We
thus expect the squared terms to break the flat directions of the linearized
results.

This is exactly what we find in Fig.~\ref{fig:expected}, where we show the
expected limits for a rate-only analysis~(gray), the improved STXS defined in
the previous section~(green), and a multivariate analysis of the full
phase-space information based on the \sally{} technique described in
Sec.~\ref{sec:stats_score}~(blue). In the plane spanned by $\tilde{C}_{HD}$ and
$C_{HW}$ (left panel) we can clearly see that the squared contributions remove
the approximate flat direction in the improved STXS limits. More generally, the
full exclusion limits based on the \sally method differ from those obtained from
the linearized  Fisher information, indicating the importance of squared new
physics effects in this region of parameter space. In these parameter-space
regions we expect that even stronger limits can be constructed with techniques
that estimate the full likelihood function to all orders in the Wilson
coefficients~\cite{Brehmer:2018hga, Brehmer:2018kdj, Brehmer:2018eca,
Brehmer:2019xox}. Closer to the SM, we find that the full limits are well
approximated by the Fisher information, confirming that the linear  operator
effects dominate there. Generally, these full limit calculations confirm the
main  result of the last section: the simplified template cross sections, in
particular our improved version, are substantially more informative than a
simple rate measurement, but fall short of capturing all of the kinematic
information in the high-dimensional final state.

There are numerous global studies that compare the sensitivity to various coefficients at LEP and at 
the LHC~\cite{Ellis:2018gqa,Berthier:2016tkq,Almeida:2018cld,Grojean:2018dqj,Biekotter:2018rhp}, as well as studies examining the reach of the $VH$ processes separately~\cite{Banerjee:2018bio,Franceschini:2017xkh,Freitas:2019hbk}.  The global Warsaw basis fit of Ref.~\cite{Ellis:2018gqa} finds single parameter fits which are roughly
consistent with  our improved STXS projections of Fig.~\ref{fig:expected}. Our results  should be interpreted 
as demonstrating the potential improvements in precision fits that could be obtained 
by improving the binning or by adding information from additional kinematic
variables.

\section{Conclusions}
\label{sec:conc}

We have, for the first time, performed a comprehensive benchmarking of
simplified template cross sections (STXS) in the $WH$ channel with a leptonic
$W$ decay. The Fisher information allowed us to quantify the reach of different
analysis strategies and to identify which phase-space regions are sensitive to
the three dimension-6 operators $\opet{HD} $, $\ope{HW}$, and $\ope{Hq}^{(3)}$.
We compared the STXS to a machine-learning-based analysis of the full,
high-dimensional final state, using the \sally technique of
Refs.~\cite{Brehmer:2018hga, Brehmer:2018kdj, Brehmer:2018eca} implemented in
\madminer{}~\cite{Brehmer:2019xox} to calculate the statistically optimal
observables and the maximal new physics reach of an analysis.

We  found that  $p_{T,W}$ is the most promising kinematic observable, but that
the STXS stage~1.1  criteria benefit from being supplemented by an additional
high-energy bin. The ability to distinguish different operator signatures is
further enhanced when we include a second observable in the definition of the
STXS, $\mtot$ being one workable option. We showed that for the $WH$ process
such a two-dimensional approach is promising, even though it cannot obtain all
of the kinematic information in the process that can be unearthed with
machine-learning-based techniques. Our results suggest, however,  that the current STXS
results can be significantly improved by the two-dimensional binning. More generally, this study presents a
blueprint for a systematic benchmarking of STXS or any other method of
publishing results needed for a global Higgs analysis.

\subsection*{Acknowledgments}

We would like to thank Anke Biek\"otter, Sebastian Bruggisser, and Lukas Heinrich
for useful discussions. We are grateful to Kyle Cranmer, whose ideas laid the
foundation for our statistical analysis. We acknowledge the use of
\toolfont{Delphes~3}~\cite{deFavereau:2013fsa}, \toolfont{Jupyter}
notebooks~\cite{Kluyver2016JupyterN},
\toolfont{MadGraph5\_aMC}~\cite{Alwall:2014hca},
\toolfont{MadMiner}~\cite{Brehmer:2019xox,madminer},
\toolfont{Matplotlib}~\cite{Hunter:2007}, \toolfont{NumPy}~\cite{numpy},
\toolfont{pylhe}~\cite{lukas_2018_1217032},
\toolfont{Pythia~8}~\cite{Sjostrand:2014zea},
\toolfont{Python}~\cite{van1995python},
\toolfont{PyTorch}~\cite{paszke2017automatic},
\toolfont{scikit-hep}~\cite{Rodrigues:2019nct},
\toolfont{scikit-learn}~\cite{scikit-learn}, and
\toolfont{uproot}~\cite{jim_pivarski_2019_3256257}.

The work of JB is supported by the U.\,S.\ National Science Foundation (NSF)
under the awards ACI-1450310, OAC-1836650, and OAC-1841471, and by the
Moore-Sloan data science environment at NYU. The work of SH is supported by the
NSF grant PHY-1620628 and the U.S. Department of Energy, Office of Science,
Office of Workforce Development for Teachers and Scientists, Office of Science
Graduate Student Research (SCGSR) program. The SCGSR program is administered by
the Oak Ridge Institute for Science and Education (ORISE) for the DOE. ORISE is
managed by ORAU under contract number DE-SC0014664. The work of FK is supported
by NSF grant PHY-1620638 and U. S. Department of Energy grant DE-AC02-76SF00515.
TP's research is supported by the German DFG Collaborative Research Centre P3H:
Particle Physics Phenomenology after the Higgs Discovery (CRC TRR 257). This
work was supported through the NYU IT High Performance Computing resources,
services, and staff expertise.  The work of SD is supported by the U.\,S.\
Department of Energy under Grant DE-SC0012704.

Our analysis code is available online at Ref.~\cite{repo}.

\appendix
\section{Detector effects}
\label{sec:detector}

\begin{figure}
\centering
\includegraphics[width=.49\linewidth]{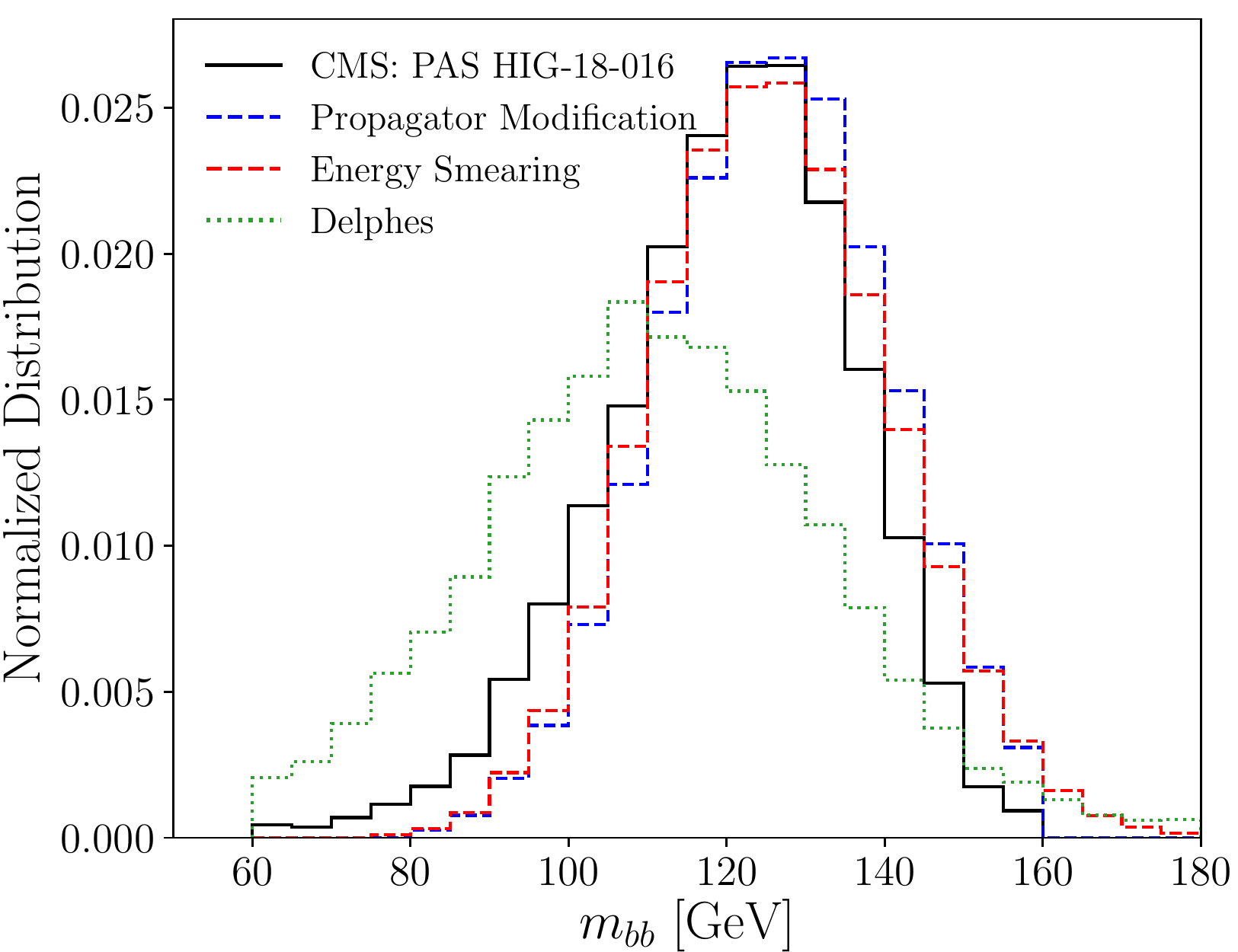}
\includegraphics[width=.49\linewidth]{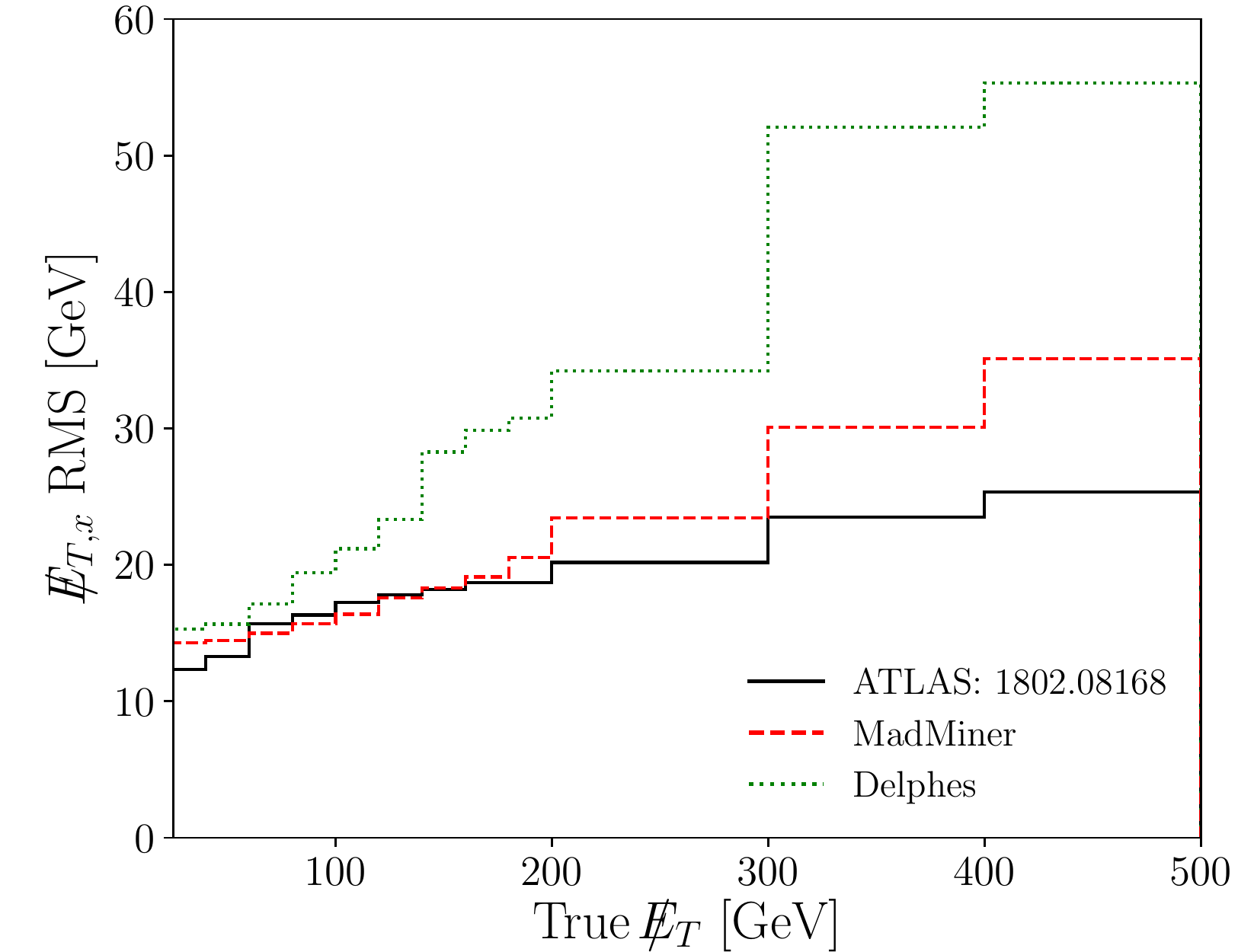}
\caption{\textbf{Left:} Invariant di-jet mass $m_{bb}$ from Higgs decay obtained
by CMS~\cite{CMS:2018abb} (solid black), the parton level energy smearing
(dashed red), the modified Higgs propagator smearing (dashed blue), and a
simulation of shower and detector response with \textsc{Pythia 8} and
\textsc{Delphes 3}  (green dotted). \textbf{Right:} RMS of the missing
transverse momentum smearing in the x-direction, $\slash \!\!\!\! E_{T,x}$, as a
function of the true missing energy $\met$ for the signal process obtained by
ATLAS~\cite{Aaboud:2018tkc} (solid black), the MET smearing in \madminer,
(dashed red) and by a simulation of shower and detector response with
\textsc{Pythia 8} and \textsc{Delphes 3} (dotted green). }
\label{fig:mass}
\end{figure}

The leading detector effect relevant for this analysis is the smearing of the
di-jet invariant mass peak. The distribution has been carefully simulated by
CMS~\cite{CMS:2018abb} (see Fig.~\ref{fig:mass}) and we aim to reproduce it for
our analysis. In the \textsc{MadMiner} framework, this smearing can be simulated
in three different ways:
i) In the simplest approach, we explicitly smear the parton level $b$-quark
energies after event generation. This is parameterized by a gaussian transfer
function with width $\sigma_E/E = 0.1 $.
ii) Alternatively, we can include the smearing already in the event generation
process by modifying the Higgs
propagator~\cite{Brehmer:2017lrt,Brehmer:2016nyr,Kling:2016lay}. To reproduce
the $m_{bb}$ distribution obtained by CMS, the Breit-Wigner propagator is simply
replaced by the square-root of the Gaussian with mean $m_H$, and width
$\sigma=15$ GeV. As a result, the joint scores already include the smearing and
can therefore be used as estimator for the score, making this approach a useful
tool for validation (see Appendix~\ref{sec:background_reweighting}).
iii) Finally, it is also possible to simulate parton shower, hadronization and
detector response using \textsc{Pythia} and \textsc{Delphes}.

In the left panel of Fig.~\ref{fig:mass} we compare the three approaches to the
di-jet mass distribution obtained by CMS~\cite{CMS:2018abb} (solid black).  We
can see that both the parton level energy smearing (dashed red) and propagator
modifcation (dashed blue) approach can reproduce the CMS mass spectrum, while a
fast detector simulation using \textsc{Pythia 8}~\cite{Sjostrand:2014zea} and
\textsc{Delphes 3}~\cite{deFavereau:2013fsa} with default settings (dotted
green) systematically underestimates the di-jet mass.

Another important detector effect is the smearing of the missing transverse
energy $\met$. In this work, we aim to reproduce the most recent ATLAS
performance~\cite{Aaboud:2018tkc} and explicitly smear the transverse components
of the missing energy using a Gaussian transfer function with width
$\sigma_{\met} = 12.5~\gev$. An additional smearing at higher energies is
induced through the smearing of the $b$-quark energies. As shown in the right
panel of Fig.~\ref{fig:mass}, this procedure reproduces the experimental results
well.  In contrast, a fast detector simulation using \textsc{Pythia 8} and
\textsc{Delphes 3} once again fails to accurately reproduce the experimental
performance without tuning the simulation parameters, as indicated by the green dotted line.

\section{Fisher information in the presence of backgrounds}
\label{sec:background_reweighting}

A key feature of the machine learning approach in \madminer is that one can
reliably estimate the score in the presence of backgrounds. As explained in
Ref.~\cite{Brehmer:2019xox}, this also works when signal and background samples
are generated separately from each other, implicitly inducing  an additional
discrete latent variable which labels each training event as either signal or
background. This latent variable is then integrated out in the machine learning
step alongside with all other unobservable latent variables, such as the
longitudinal neutrino momenta, initial state quark flavours or additional
unobserved particles in reducible backgrounds (such as the jets in top pair
production). Reference~\cite{Brehmer:2019xox} contains examples to validate the
performance of \madminer's machine learning approach in the case of either a
single observable or a single process. In the following, we will consider
another simplified scenario in which we can validate \madminer's reach estimate
in the presence of both a high-dimensional observable space and irreducible
backgrounds.

As mentioned in Sec.~\ref{sec:stats_fisher}, for a fixed initial and final state
at parton level, we can calculate the Fisher information directly from the event
weights and hence cross check the machine learning based results. In particular,
we consider the process $u \bar{d} \to \mu^+\nu  \, b\bar{b}$, with
contributions coming from both the $WH$ signal and the irreducible $Wb\bar{b}$
background, and restrict ourselves to only two effective operators $\ope{HD}$
and $\ope{Hq}^{(3)}$. We first generate a set of signal events $\{ x_k \}$,
described by event weights $\Delta \sigma_{S,k}(\theta)$. In contrast to the
results in Section~\ref{sec:analysis}, where the smearing of the $m_{bb}$ peak
was done at analysis level, here we generate the signal events using the
propagator modification method described in Appendix~\ref{sec:detector}. We then
use the \textsc{MadGraph} reweighting tool~\cite{Mattelaer:2016gcx} to obtain
the background weights $\Delta \sigma_{B,k}$ for the same set of events. As
shown in Ref.~\cite{Brehmer:2016nyr}, the Fisher information at parton level for
a set of events can then be computed as
\begin{equation}
I_{ij}= \sum_{\text{events, }k} \frac{\mathcal{L}}{\Delta \sigma_k(0)} \frac{\partial \Delta \sigma_k(\theta)} {\partial_{\theta_i}} \Bigg|_{\theta=0} \frac{\partial \Delta \sigma_k(\theta)}{\partial_{\theta_j}} \Bigg|_{\theta=0} \; .
\label{eq:fi_truth_discrete}
\end{equation}

\begin{figure}
\centering
\includegraphics[width=.5\linewidth]{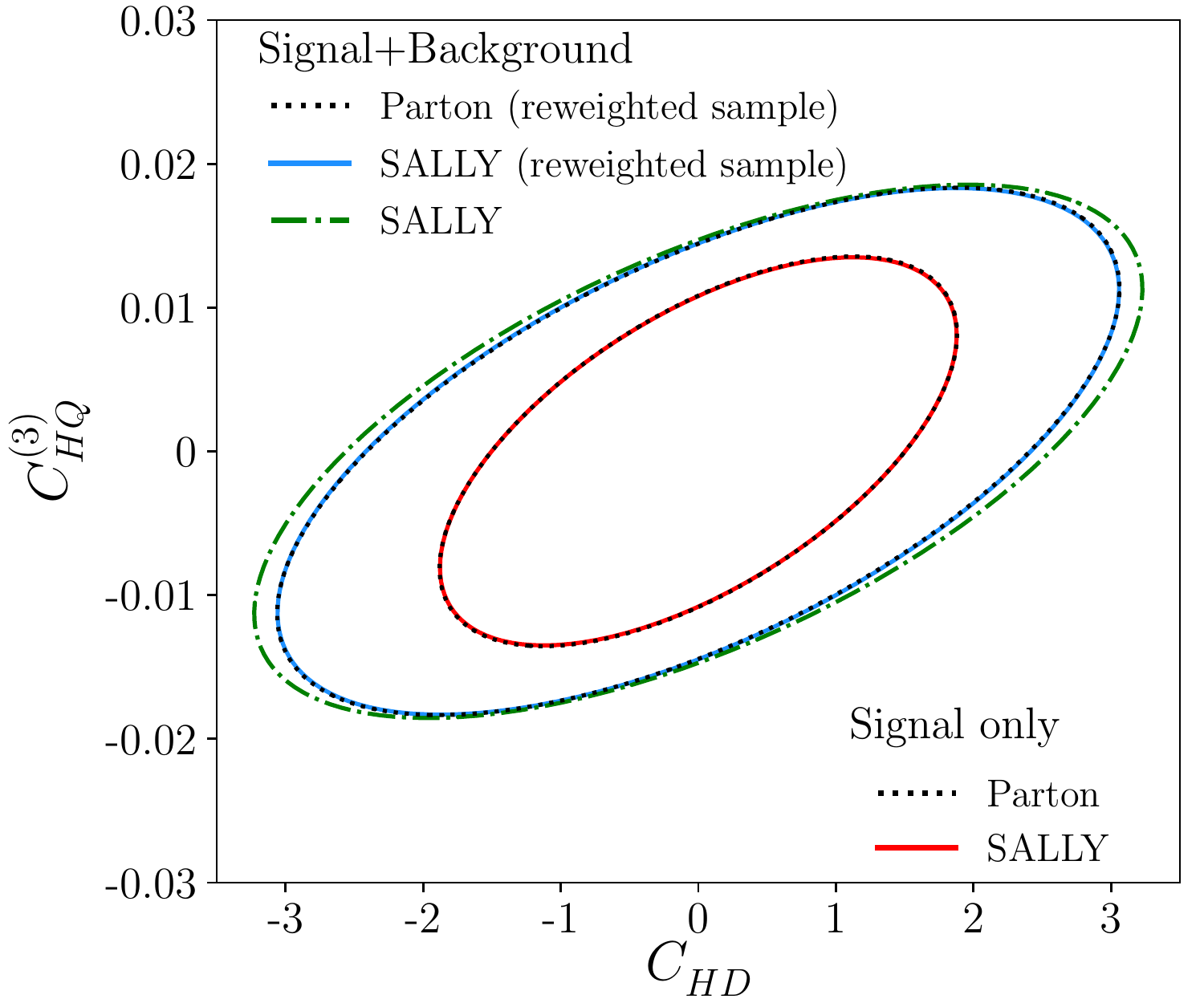}
\caption{
A comparison of the 95\% CL limits based on the Fisher information in the
two-parameter space for a simplified version of our process, comparing the
Fisher information obtained from a set of parton level event weights (dotted
lines), using a \sally estimator trained on the same sample events (solid
lines), and using a \sally estimator trained on separate signal and background
event samples (dot-dashed line). The inner ellipses show the results when
evaluated only on the signal sample, while the outer ellipses show the results
taking into account both signal and background.}
\label{fig:truth_validation}
\end{figure}

A comparison of the Fisher information obtained using machine learning results
with the Fisher information directly obtained from the parton level event
weights is shown in Fig.~\ref{fig:truth_validation}. The inner ellipses show the
reach in the absence of backgrounds ($\Delta \sigma_k = \Delta \sigma_{S,k}$),
while the outer ellipses take into account both signal and background ($\Delta
\sigma_k = \Delta \sigma_{S,k}+\Delta \sigma_{B,k}$). In both cases, the Fisher
information obtained at parton level (dotted curves) and using a \textsc{Sally}
estimator trained on the same sample (solid curves) are indistinguishable,
indicating excellent machine learning performance. This is then compared to the
standard way of estimating the Fisher information, using a \textsc{Sally}
estimator trained with separately generated signal and background samples, which
is indicated by the green dot-dashed ellipse. This result agrees remarkably well
with the other results, confirming that the treatment of backgrounds in the
\madminer algorithm works correctly.

\section{Systematics}
\label{sec:systematics}

An important question that must be addressed when comparing the expected limits
from different sets of kinematic data from experiments is the effect of
systematic uncertainties on the expected limits. While a treatment of all
experimental systematics is beyond the scope of this study, an important step in
this direction is to understand the effects of theory uncertainties, namely the
dependence on the choice of renormalization and factorization scales as well as
the PDFs. These uncertainties are fully integrated in the \madminer
framework~\cite{Brehmer:2019xox} and parameterized through nuisance parameters,
which are then marginalized over to set limits.

\begin{figure}[t]
\centering
\includegraphics[width=.513\linewidth]{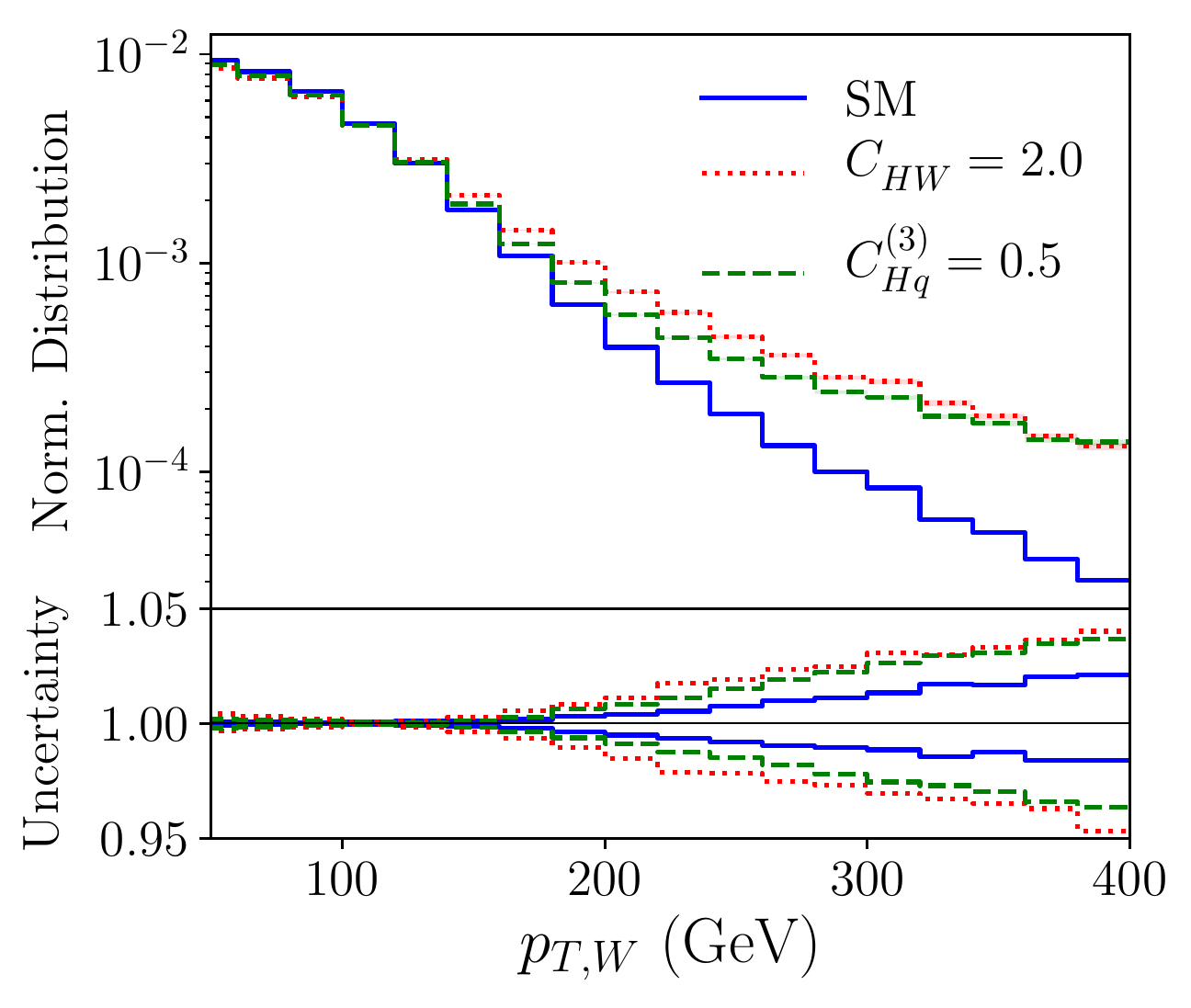}
\includegraphics[width=.45\linewidth]{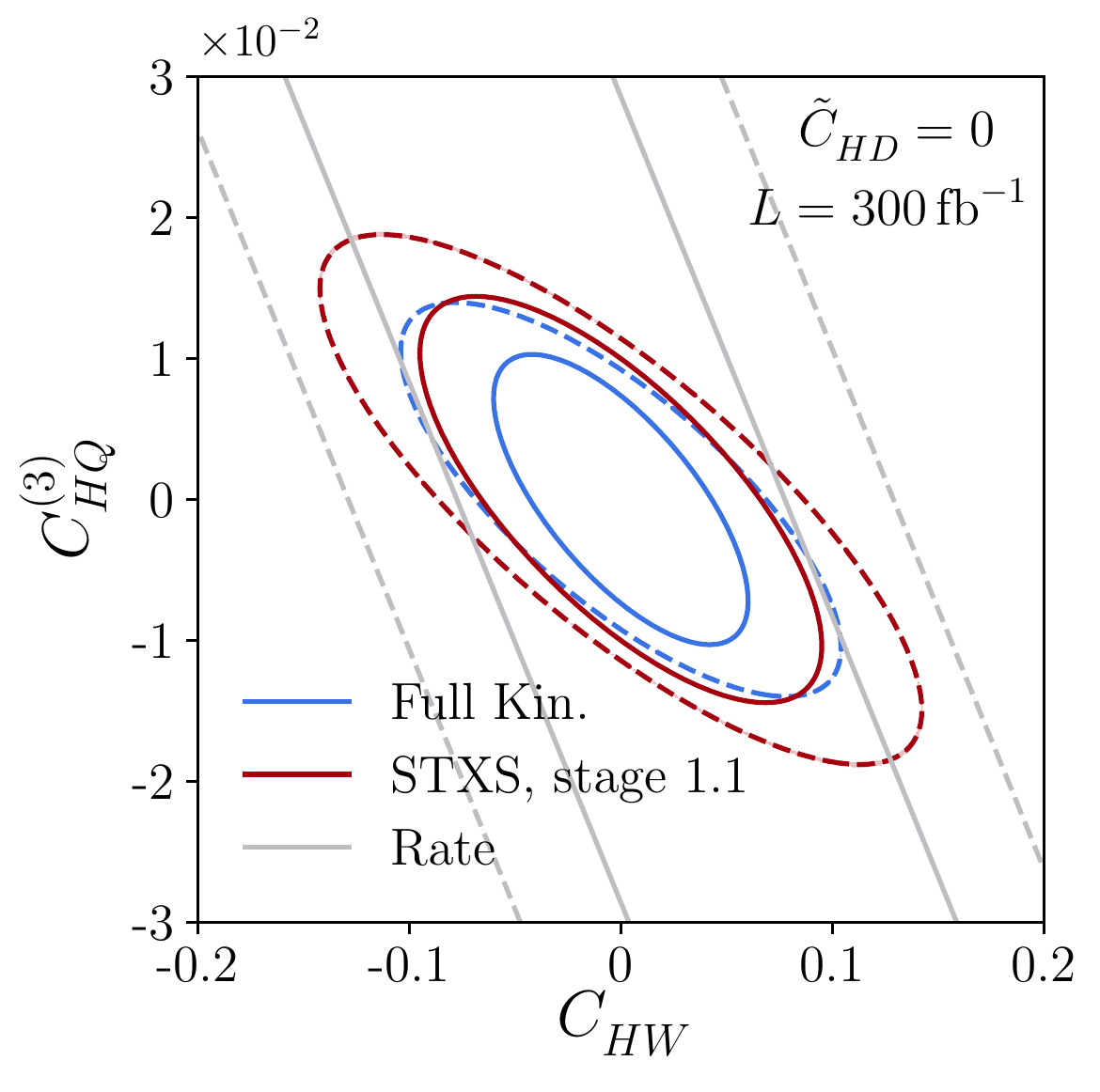}
\caption{
\textbf{Left:} The normalized distributions of $p_{T,W}$ for signal and
background weights with various choices of the theory parameters and (barely
visible) shaded bands showing the effects of the PDF and scale uncertainties.
\textbf{Right:} The 95\% CL bounds on $C_{HW}$ and $C_{Hq}^{(3)}$ in the Fisher
information approximation obtained with the \sally estimator evaluated using the
full set of physical variables (blue), the Stage 1.1 simplified template cross
sections (red), and measurements of the rate (grey), with $\tilde{C}_{HD}$ fixed
to zero. The solid lines show the projected limits obtained in the absence of
systematic uncertainties, while the dashed lines show the reach contours after
marginalizing over the additional nuisance parameters.
}
\label{fig:systematics}
\end{figure}

In the following we only consider the theory uncertainties on the $WH$ signal.
Since uncertainties on the backgrounds are typically mitigated through the use
of sideband analyses, rather than through direct simulations, including the
scale and PDF uncertainties on the backgrounds would dramatically overestimate
the actual systematic uncertainty in a realistic measurement.

In the left panel of Fig.~\ref{fig:systematics} we show the normalized
distribution of the $W$ boson's transverse momentum including scale and PDF
uncertainties for several benchmark parameter points. In particular we vary the
renormalization and factorization scale by a factor of two and use the
\textsc{PDF4LHC15\_nlo\_30} set to estimate PDF uncertainties. We see that the
relative uncertainty is small in the background-dominated low-$p_{T,W}$ regime,
and increases to $\mathcal{O}(5\%)$ at large transverse momentum. In $WH$
production, the PDF shape uncertainties dominate over the scale uncertainties,
given that the scale dependence of the quark PDF is small.

In the right panel of Fig.~\ref{fig:systematics} we show how including
systematics changes the expected limits. While the solid contours show the reach
neglecting systematic uncertainties, the dashed lines account for the systematic
uncertainties by profiling over the nuisance parameters. We can see that the
presence of systematic uncertainties mainly reduce the sensitivity in the
rate-sensitive direction, rotating the orientation of the ellipses. The overall
effect is sizable, indicating the importance of understanding these systematic
uncertainties. In  Fig.~\ref{fig:systematics}, we see that at   $p_{T,W}\sim 300-400~GeV$, the energy scaling of 
$C_{HW}$ and $C_{HQ}^{(3)}$  is roughly the same, due to the fact that  we are not yet in the high $p_{T,W}$
regime where the longitudinal modes dominate.  In this energy range, the 
 transverse modes, that 
have the same scaling with $\sqrt{s}/M_W$, are clearly important.

\section{Binning distributions}
\label{sec:binning}

\begin{figure}[t]
\includegraphics[width=.32\linewidth]{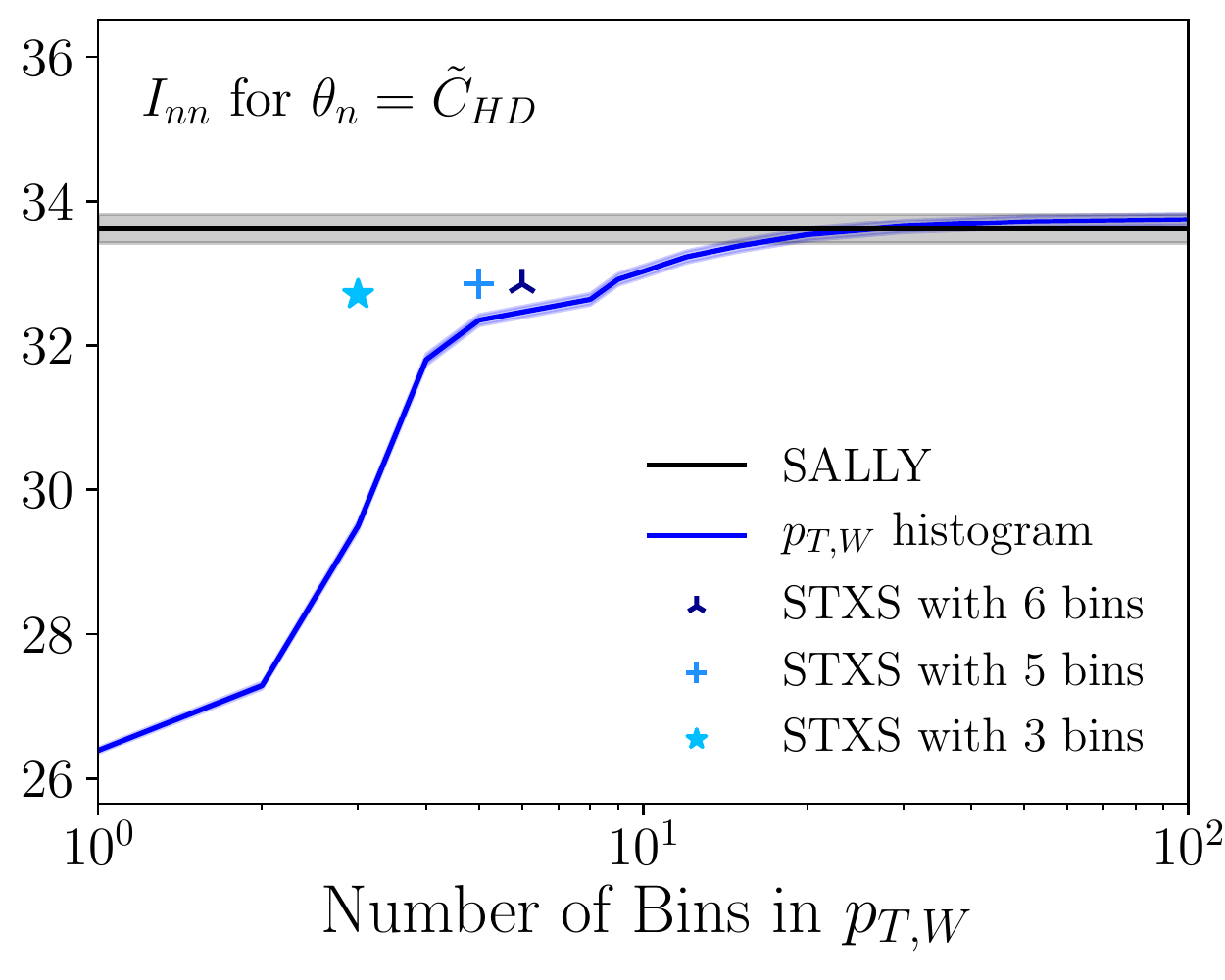}
\includegraphics[width=.32\linewidth]{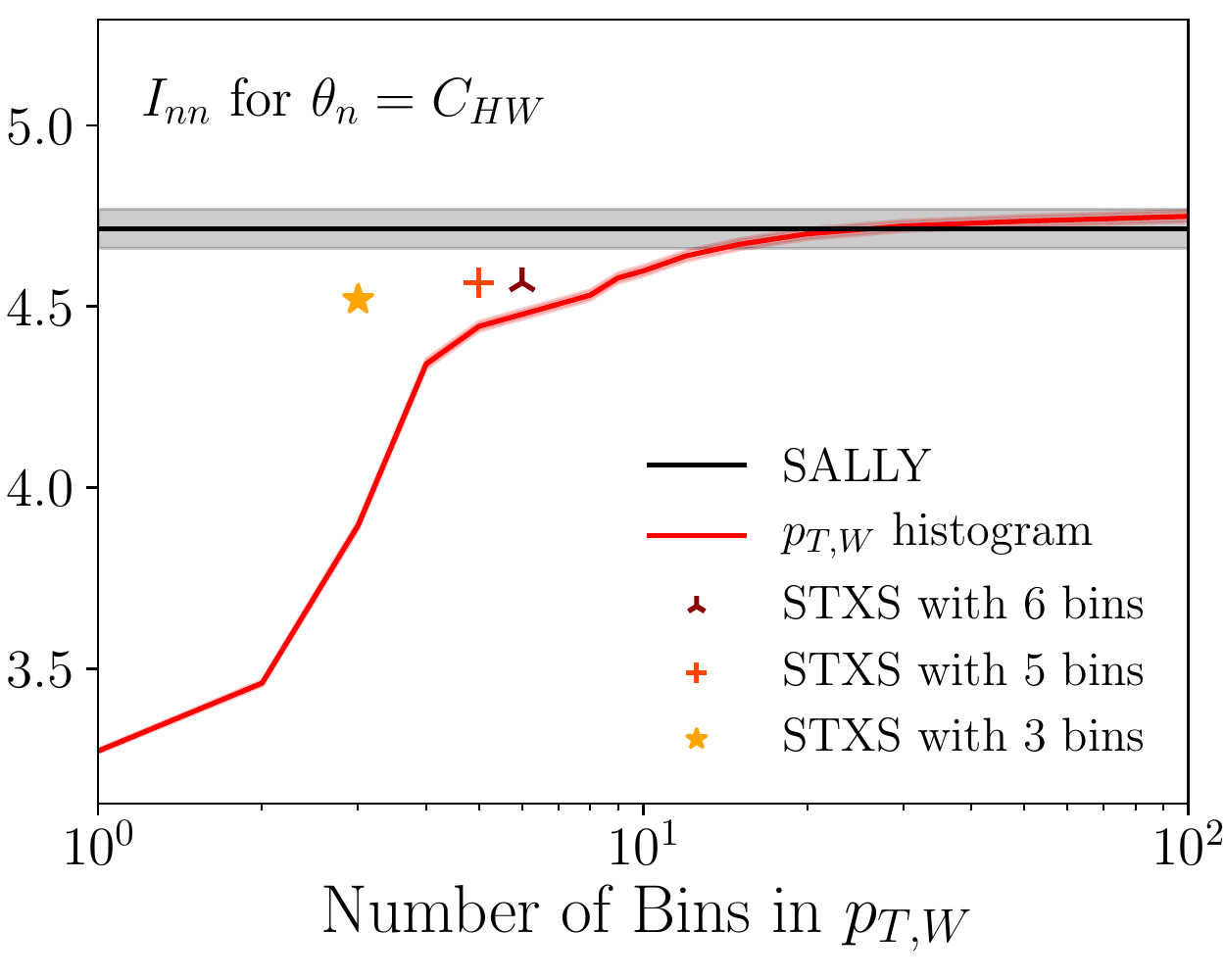}
\includegraphics[width=.32\linewidth]{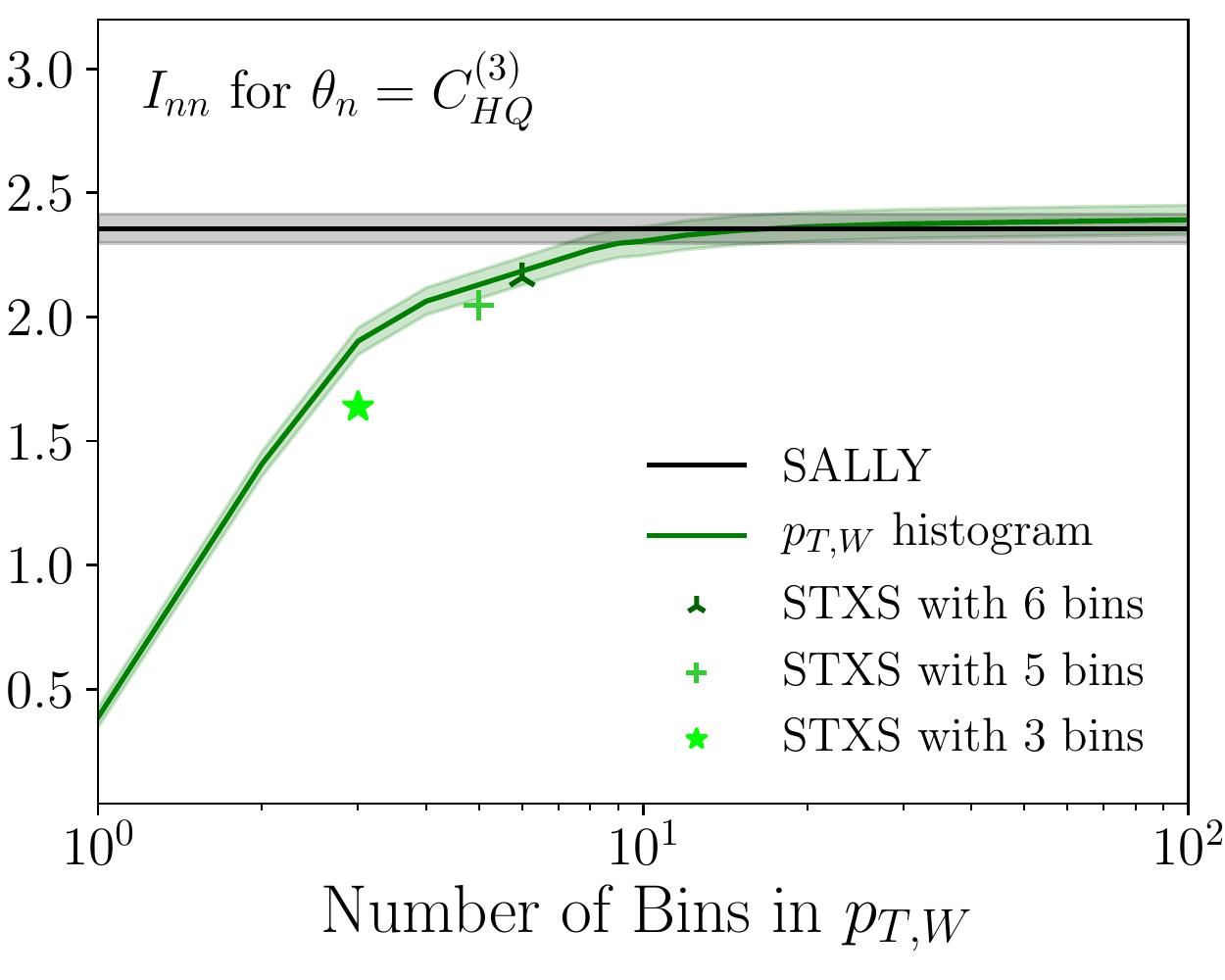}
\caption{Diagonal elements of Fisher information obtained from a $p_{T,W}$
histogram with varying number of bins. The horizontal black line shows the Fisher
information obtained using the \textsc{Sally} method trained only with $p_{T,W}$
as input observable. The markers show the information in the STXS with 3, 5 and
6 bins, respectively.}
\label{fig:info_vs_nbins}
\end{figure}

In view of an appropriate definition of simplified template cross sections, it
is interesting to ask how finely we need to bin a kinematic distribution to
extract as much of the available information as possible. This is illustrated in
Fig.~\ref{fig:info_vs_nbins}, where we show the Fisher information corresponding
to a $p_{T,W}$ histogram for varying number of bins with equal size in the range
$[0,800~\text{GeV}]$. The three panels correspond to the diagonal element of the
Fisher information for $\tilde{C}_{HD}$ (left), $C_{HW}$ (center) and
$C_{QH}^{(3)}$ (right).

This is compared to the information obtained using a \textsc{Sally} score
estimator trained with only $p_{T,W}$ as input observable (solid black) and the
information in the STXS with 3 bins (stars), 5 bins (crosses) and 3 bins
(triangles). The shaded regions indicated the uncertainties of the Fisher
information, where the uncertainty for the \textsc{Sally} method was estimated
using an ensemble of estimators with varying architecture. We can see that the
information in histograms approaches the information in the distribution in the
limit of large number of bins, showing that the \textsc{Sally} approach gives
the correct large-$n_{\text{bin}}$ limit. For all three operators, a histogram
with an $\mathcal{O}(10)$ number of bins will essentially extract the full
information of the $p_{T,W}$ distribution. The STXS with 3 bins already collects
almost all information on $\tilde{C}_{HD}$ and $C_{HW}$, while for
$C_{QH}^{(3)}$ a significant improvement is obtained when increasing the number
of bins to 6.

\bibliographystyle{JHEP}
\bibliography{references}

\providecommand{\href}[2]{#2}\begingroup\raggedright\begin{thebibliography}{10}

\bibitem{Dawson:2018dcd}
S.~Dawson, C.~Englert and T.~Plehn, \emph{{Higgs Physics: It ain't over till
  it's over}}, \href{https://doi.org/10.1016/j.physrep.2019.05.001}{\emph{Phys.
  Rept.} {\bfseries 816} (2019) 1}
  [\href{https://arxiv.org/abs/1808.01324}{{\ttfamily 1808.01324}}].

\bibitem{Brivio:2017vri}
I.~Brivio and M.~Trott, \emph{{The Standard Model as an Effective Field
  Theory}}, \href{https://doi.org/10.1016/j.physrep.2018.11.002}{\emph{Phys.
  Rept.} {\bfseries 793} (2019) 1}
  [\href{https://arxiv.org/abs/1706.08945}{{\ttfamily 1706.08945}}].

\bibitem{Butter:2016cvz}
A.~Butter, O.~J.~P. Eboli, J.~Gonzalez-Fraile, M.~C. Gonzalez-Garcia, T.~Plehn
  and M.~Rauch, \emph{{The Gauge-Higgs Legacy of the LHC Run I}},
  \href{https://doi.org/10.1007/JHEP07(2016)152}{\emph{JHEP} {\bfseries 07}
  (2016) 152} [\href{https://arxiv.org/abs/1604.03105}{{\ttfamily
  1604.03105}}].

\bibitem{Brivio:2016fzo}
I.~Brivio, J.~Gonzalez-Fraile, M.~C. Gonzalez-Garcia and L.~Merlo, \emph{{The
  complete HEFT Lagrangian after the LHC Run I}},
  \href{https://doi.org/10.1140/epjc/s10052-016-4211-9}{\emph{Eur. Phys. J.}
  {\bfseries C76} (2016) 416}
  [\href{https://arxiv.org/abs/1604.06801}{{\ttfamily 1604.06801}}].

\bibitem{Banerjee:2013apa}
S.~Banerjee, S.~Mukhopadhyay and B.~Mukhopadhyaya, \emph{{Higher dimensional
  operators and the LHC Higgs data: The role of modified kinematics}},
  \href{https://doi.org/10.1103/PhysRevD.89.053010}{\emph{Phys. Rev.}
  {\bfseries D89} (2014) 053010}
  [\href{https://arxiv.org/abs/1308.4860}{{\ttfamily 1308.4860}}].

\bibitem{DiVita:2017vrr}
S.~Di~Vita, G.~Durieux, C.~Grojean, J.~Gu, Z.~Liu, G.~Panico et~al., \emph{{A
  global view on the Higgs self-coupling at lepton colliders}},
  \href{https://doi.org/10.1007/JHEP02(2018)178}{\emph{JHEP} {\bfseries 02}
  (2018) 178} [\href{https://arxiv.org/abs/1711.03978}{{\ttfamily
  1711.03978}}].

\bibitem{Ellis:2018gqa}
J.~Ellis, C.~W. Murphy, V.~Sanz and T.~You, \emph{{Updated Global SMEFT Fit to
  Higgs, Diboson and Electroweak Data}},
  \href{https://doi.org/10.1007/JHEP06(2018)146}{\emph{JHEP} {\bfseries 06}
  (2018) 146} [\href{https://arxiv.org/abs/1803.03252}{{\ttfamily
  1803.03252}}].

\bibitem{Almeida:2018cld}
E.~da~Silva~Almeida, A.~Alves, N.~Rosa~Agostinho, O.~J.~P. {\'E}boli and M.~C.
  Gonzalez-Garcia, \emph{{Electroweak Sector Under Scrutiny: A Combined
  Analysis of LHC and Electroweak Precision Data}},
  \href{https://doi.org/10.1103/PhysRevD.99.033001}{\emph{Phys. Rev.}
  {\bfseries D99} (2019) 033001}
  [\href{https://arxiv.org/abs/1812.01009}{{\ttfamily 1812.01009}}].

\bibitem{Biekotter:2018rhp}
A.~Biek{\"o}tter, T.~Corbett and T.~Plehn, \emph{{The Gauge-Higgs Legacy of the
  LHC Run II}},
  \href{https://doi.org/10.21468/SciPostPhys.6.6.064}{\emph{SciPost Phys.}
  {\bfseries 6} (2019) 064} [\href{https://arxiv.org/abs/1812.07587}{{\ttfamily
  1812.07587}}].

\bibitem{ATL-PHYS-PUB-2019-029}
{\scshape ATLAS} collaboration, \emph{{Reproducing searches for new physics
  with the ATLAS experiment through publication of full statistical
  likelihoods}},  Tech. Rep. ATL-PHYS-PUB-2019-029, CERN, Geneva, Aug, 2019.

\bibitem{Collaboration:2242860}
{\scshape CMS} collaboration, \emph{{Simplified likelihood for the
  re-interpretation of public CMS results}},  Tech. Rep. CMS-NOTE-2017-001.
  CERN-CMS-NOTE-2017-001, CERN, Geneva, Jan, 2017.

\bibitem{Cranmer:2013hia}
K.~Cranmer, S.~Kreiss, D.~Lopez-Val and T.~Plehn, \emph{{Decoupling Theoretical
  Uncertainties from Measurements of the Higgs Boson}},
  \href{https://doi.org/10.1103/PhysRevD.91.054032}{\emph{Phys. Rev.}
  {\bfseries D91} (2015) 054032}
  [\href{https://arxiv.org/abs/1401.0080}{{\ttfamily 1401.0080}}].

\bibitem{Maguire:2017ypu}
E.~Maguire, L.~Heinrich and G.~Watt, \emph{{HEPData: a repository for high
  energy physics data}},
  \href{https://doi.org/10.1088/1742-6596/898/10/102006}{\emph{J. Phys. Conf.
  Ser.} {\bfseries 898} (2017) 102006}
  [\href{https://arxiv.org/abs/1704.05473}{{\ttfamily 1704.05473}}].

\bibitem{Leung:1984ni}
C.~N. Leung, S.~T. Love and S.~Rao, \emph{{Low-Energy Manifestations of a New
  Interaction Scale: Operator Analysis}},
  \href{https://doi.org/10.1007/BF01588041}{\emph{Z. Phys.} {\bfseries C31}
  (1986) 433}.

\bibitem{Buchmuller:1985jz}
W.~Buchmuller and D.~Wyler, \emph{{Effective Lagrangian Analysis of New
  Interactions and Flavor Conservation}},
  \href{https://doi.org/10.1016/0550-3213(86)90262-2}{\emph{Nucl. Phys.}
  {\bfseries B268} (1986) 621}.

\bibitem{GonzalezGarcia:1999fq}
M.~C. Gonzalez-Garcia, \emph{{Anomalous Higgs couplings}},
  \href{https://doi.org/10.1142/S0217751X99001494}{\emph{Int. J. Mod. Phys.}
  {\bfseries A14} (1999) 3121}
  [\href{https://arxiv.org/abs/hep-ph/9902321}{{\ttfamily hep-ph/9902321}}].

\bibitem{Grojean:2018dqj}
C.~Grojean, M.~Montull and M.~Riembau, \emph{{Diboson at the LHC vs LEP}},
  \href{https://doi.org/10.1007/JHEP03(2019)020}{\emph{JHEP} {\bfseries 03}
  (2019) 020} [\href{https://arxiv.org/abs/1810.05149}{{\ttfamily
  1810.05149}}].

\bibitem{deBlas:2016nqo}
J.~de~Blas, M.~Ciuchini, E.~Franco, S.~Mishima, M.~Pierini, L.~Reina et~al.,
  \emph{{Electroweak precision constraints at present and future colliders}},
  \href{https://doi.org/10.22323/1.282.0690}{\emph{PoS} {\bfseries ICHEP2016}
  (2017) 690} [\href{https://arxiv.org/abs/1611.05354}{{\ttfamily
  1611.05354}}].

\bibitem{Tackmann:2138079}
F.~Tackmann, K.~Tackmann, K.~Tackmann, M.~Duehrssen-Debling and P.~Francavilla,
  \emph{{Simplified template cross sections}},
  \texttt{\href{https://cds.cern.ch/record/2138079}{cds.cern.ch/record/2138079}}.

\bibitem{Berger:2019wnu}
N.~Berger et~al., \emph{{Simplified Template Cross Sections - Stage 1.1}},
  \href{https://arxiv.org/abs/1906.02754}{{\ttfamily 1906.02754}}.

\bibitem{Brehmer:2016nyr}
J.~Brehmer, K.~Cranmer, F.~Kling and T.~Plehn, \emph{{Better Higgs boson
  measurements through information geometry}},
  \href{https://doi.org/10.1103/PhysRevD.95.073002}{\emph{Phys. Rev.}
  {\bfseries D95} (2017) 073002}
  [\href{https://arxiv.org/abs/1612.05261}{{\ttfamily 1612.05261}}].

\bibitem{Brehmer:2017lrt}
J.~Brehmer, F.~Kling, T.~Plehn and T.~M.~P. Tait, \emph{{Better Higgs-CP Tests
  Through Information Geometry}},
  \href{https://doi.org/10.1103/PhysRevD.97.095017}{\emph{Phys. Rev.}
  {\bfseries D97} (2018) 095017}
  [\href{https://arxiv.org/abs/1712.02350}{{\ttfamily 1712.02350}}].

\bibitem{Brehmer:2018eca}
J.~Brehmer, K.~Cranmer, G.~Louppe and J.~Pavez, \emph{{A Guide to Constraining
  Effective Field Theories with Machine Learning}},
  \href{https://doi.org/10.1103/PhysRevD.98.052004}{\emph{Phys. Rev.}
  {\bfseries D98} (2018) 052004}
  [\href{https://arxiv.org/abs/1805.00020}{{\ttfamily 1805.00020}}].

\bibitem{Brehmer:2018kdj}
J.~Brehmer, K.~Cranmer, G.~Louppe and J.~Pavez, \emph{{Constraining Effective
  Field Theories with Machine Learning}},
  \href{https://doi.org/10.1103/PhysRevLett.121.111801}{\emph{Phys. Rev. Lett.}
  {\bfseries 121} (2018) 111801}
  [\href{https://arxiv.org/abs/1805.00013}{{\ttfamily 1805.00013}}].

\bibitem{Brehmer:2018hga}
J.~Brehmer, G.~Louppe, J.~Pavez and K.~Cranmer, \emph{{Mining gold from
  implicit models to improve likelihood-free inference}},
  \href{https://arxiv.org/abs/1805.12244}{{\ttfamily 1805.12244}}.

\bibitem{Brehmer:2019xox}
J.~Brehmer, F.~Kling, I.~Espejo and K.~Cranmer, \emph{{MadMiner: Machine
  learning-based inference for particle physics}},
  \href{https://arxiv.org/abs/1907.10621}{{\ttfamily 1907.10621}}.

\bibitem{Freitas:2019hbk}
F.~F. Freitas, C.~K. Khosa and V.~Sanz, \emph{{Exploring SMEFT in VH with
  Machine Learning}},  \href{https://arxiv.org/abs/1902.05803}{{\ttfamily
  1902.05803}}.

\bibitem{Corbett:2012ja}
T.~Corbett, O.~J.~P. Eboli, J.~Gonzalez-Fraile and M.~C. Gonzalez-Garcia,
  \emph{{Robust Determination of the Higgs Couplings: Power to the Data}},
  \href{https://doi.org/10.1103/PhysRevD.87.015022}{\emph{Phys. Rev.}
  {\bfseries D87} (2013) 015022}
  [\href{https://arxiv.org/abs/1211.4580}{{\ttfamily 1211.4580}}].

\bibitem{Grzadkowski:2010es}
B.~Grzadkowski, M.~Iskrzynski, M.~Misiak and J.~Rosiek, \emph{{Dimension-Six
  Terms in the Standard Model Lagrangian}},
  \href{https://doi.org/10.1007/JHEP10(2010)085}{\emph{JHEP} {\bfseries 10}
  (2010) 085} [\href{https://arxiv.org/abs/1008.4884}{{\ttfamily 1008.4884}}].

\bibitem{Brivio:2017bnu}
I.~Brivio and M.~Trott, \emph{{Scheming in the SMEFT... and a
  reparameterization invariance!}},
  \href{https://doi.org/10.1007/JHEP05(2018)136,
  10.1007/JHEP07(2017)148}{\emph{JHEP} {\bfseries 07} (2017) 148}
  [\href{https://arxiv.org/abs/1701.06424}{{\ttfamily 1701.06424}}], [Addendum:
  JHEP05,136(2018)].

\bibitem{Berthier:2016tkq}
L.~Berthier, M.~Bjorn and M.~Trott, \emph{{Incorporating doubly resonant
  $W^\pm$ data in a global fit of SMEFT parameters to lift flat directions}},
  \href{https://doi.org/10.1007/JHEP09(2016)157}{\emph{JHEP} {\bfseries 09}
  (2016) 157} [\href{https://arxiv.org/abs/1606.06693}{{\ttfamily
  1606.06693}}].

\bibitem{Haller:2018nnx}
J.~Haller, A.~Hoecker, R.~Kogler, K.~Monig, T.~Peiffer and J.~Stelzer,
  \emph{{Update of the global electroweak fit and constraints on
  two-Higgs-doublet models}},
  \href{https://doi.org/10.1140/epjc/s10052-018-6131-3}{\emph{Eur. Phys. J.}
  {\bfseries C78} (2018) 675}
  [\href{https://arxiv.org/abs/1803.01853}{{\ttfamily 1803.01853}}].

\bibitem{Dedes:2017zog}
A.~Dedes, W.~Materkowska, M.~Paraskevas, J.~Rosiek and K.~Suxho, \emph{{Feynman
  rules for the Standard Model Effective Field Theory in R$_{\xi}$ -gauges}},
  \href{https://doi.org/10.1007/JHEP06(2017)143}{\emph{JHEP} {\bfseries 06}
  (2017) 143} [\href{https://arxiv.org/abs/1704.03888}{{\ttfamily
  1704.03888}}].

\bibitem{Nakamura:2017ihk}
J.~Nakamura, \emph{{Polarisations of the $Z$ and $W$ bosons in the processes
  $pp \to ZH$ and $pp \to W^{\pm}_{}H$}},
  \href{https://doi.org/10.1007/JHEP08(2017)008}{\emph{JHEP} {\bfseries 08}
  (2017) 008} [\href{https://arxiv.org/abs/1706.01816}{{\ttfamily
  1706.01816}}].

\bibitem{Alwall:2014hca}
J.~Alwall, R.~Frederix, S.~Frixione, V.~Hirschi, F.~Maltoni, O.~Mattelaer
  et~al., \emph{{The automated computation of tree-level and next-to-leading
  order differential cross sections, and their matching to parton shower
  simulations}}, \href{https://doi.org/10.1007/JHEP07(2014)079}{\emph{JHEP}
  {\bfseries 07} (2014) 079} [\href{https://arxiv.org/abs/1405.0301}{{\ttfamily
  1405.0301}}].

\bibitem{Butterworth:2015oua}
J.~Butterworth et~al., \emph{{PDF4LHC recommendations for LHC Run II}},
  \href{https://doi.org/10.1088/0954-3899/43/2/023001}{\emph{J. Phys.}
  {\bfseries G43} (2016) 023001}
  [\href{https://arxiv.org/abs/1510.03865}{{\ttfamily 1510.03865}}].

\bibitem{Hirschi:2015iia}
V.~Hirschi and O.~Mattelaer, \emph{{Automated event generation for loop-induced
  processes}}, \href{https://doi.org/10.1007/JHEP10(2015)146}{\emph{JHEP}
  {\bfseries 10} (2015) 146}
  [\href{https://arxiv.org/abs/1507.00020}{{\ttfamily 1507.00020}}].

\bibitem{Brivio:2017btx}
I.~Brivio, Y.~Jiang and M.~Trott, \emph{{The SMEFTsim package, theory and
  tools}}, \href{https://doi.org/10.1007/JHEP12(2017)070}{\emph{JHEP}
  {\bfseries 12} (2017) 070}
  [\href{https://arxiv.org/abs/1709.06492}{{\ttfamily 1709.06492}}].

\bibitem{Aaboud:2018zhk}
{\scshape ATLAS} collaboration, \emph{{Observation of $H \rightarrow b\bar{b}$
  decays and $VH$ production with the ATLAS detector}},
  \href{https://doi.org/10.1016/j.physletb.2018.09.013}{\emph{Phys. Lett.}
  {\bfseries B786} (2018) 59}
  [\href{https://arxiv.org/abs/1808.08238}{{\ttfamily 1808.08238}}].

\bibitem{Sirunyan:2018kst}
{\scshape CMS} collaboration, \emph{{Observation of Higgs boson decay to bottom
  quarks}}, \href{https://doi.org/10.1103/PhysRevLett.121.121801}{\emph{Phys.
  Rev. Lett.} {\bfseries 121} (2018) 121801}
  [\href{https://arxiv.org/abs/1808.08242}{{\ttfamily 1808.08242}}].

\bibitem{Aaboud:2018tkc}
{\scshape ATLAS} collaboration, \emph{{Performance of missing transverse
  momentum reconstruction with the ATLAS detector using proton-proton
  collisions at $\sqrt{s}$ = 13 TeV}},
  \href{https://doi.org/10.1140/epjc/s10052-018-6288-9}{\emph{Eur. Phys. J.}
  {\bfseries C78} (2018) 903}
  [\href{https://arxiv.org/abs/1802.08168}{{\ttfamily 1802.08168}}].

\bibitem{Cranmer:2006zs}
K.~Cranmer and T.~Plehn, \emph{{Maximum significance at the LHC and Higgs
  decays to muons}},
  \href{https://doi.org/10.1140/epjc/s10052-007-0309-4}{\emph{Eur. Phys. J.}
  {\bfseries C51} (2007) 415}
  [\href{https://arxiv.org/abs/hep-ph/0605268}{{\ttfamily hep-ph/0605268}}].

\bibitem{Atwood:1991ka}
D.~Atwood and A.~Soni, \emph{{Analysis for magnetic moment and electric dipole
  moment form-factors of the top quark via $e^+ e^- \to t \bar{t}$}},
  \href{https://doi.org/10.1103/PhysRevD.45.2405}{\emph{Phys. Rev.} {\bfseries
  D45} (1992) 2405}.

\bibitem{Davier:1992nw}
M.~Davier, L.~Duflot, F.~Le~Diberder and A.~Rouge, \emph{{The Optimal method
  for the measurement of tau polarization}},
  \href{https://doi.org/10.1016/0370-2693(93)90101-M}{\emph{Phys. Lett.}
  {\bfseries B306} (1993) 411}.

\bibitem{Diehl:1993br}
M.~Diehl and O.~Nachtmann, \emph{{Optimal observables for the measurement of
  three gauge boson couplings in $e^+ e^- \to W^+ W^-$}},
  \href{https://doi.org/10.1007/BF01555899}{\emph{Z. Phys.} {\bfseries C62}
  (1994) 397}.

\bibitem{Kondo:1988yd}
K.~Kondo, \emph{{Dynamical Likelihood Method for Reconstruction of Events With
  Missing Momentum. 1: Method and Toy Models}},
  \href{https://doi.org/10.1143/JPSJ.57.4126}{\emph{J. Phys. Soc. Jap.}
  {\bfseries 57} (1988) 4126}.

\bibitem{fisher1935detection}
R.~A. Fisher, \emph{The detection of linkage with ``dominant'' abnormalities},
  {\emph{Annals of Eugenics} {\bfseries 6} (1935) 187}.

\bibitem{Brehmer:2019bvj}
J.~Brehmer, K.~Cranmer, I.~Espejo, F.~Kling, G.~Louppe and J.~Pavez,
  \emph{{Effective LHC measurements with matrix elements and machine
  learning}},  in \emph{{19th International Workshop on Advanced Computing and
  Analysis Techniques in Physics Research: Empowering the revolution: Bringing
  Machine Learning to High Performance Computing (ACAT 2019) Saas-Fee,
  Switzerland, March 11-15, 2019}}, 2019,
  \href{https://arxiv.org/abs/1906.01578}{{\ttfamily 1906.01578}}.

\bibitem{2014arXiv1412.6980K}
D.~P. {Kingma} and J.~{Ba}, \emph{{Adam: A Method for Stochastic
  Optimization}}, {\emph{arXiv e-prints} (2014) arXiv:1412.6980}
  [\href{https://arxiv.org/abs/1412.6980}{{\ttfamily 1412.6980}}].

\bibitem{Wilks:1938dza}
S.~S. Wilks, \emph{{The Large-Sample Distribution of the Likelihood Ratio for
  Testing Composite Hypotheses}},
  \href{https://doi.org/10.1214/aoms/1177732360}{\emph{Annals Math. Statist.}
  {\bfseries 9} (1938) 60}.

\bibitem{Wald}
A.~Wald, \emph{Tests of statistical hypotheses concerning several parameters
  when the number of observations is large}, {\emph{Transactions of the
  American Mathematical Society} {\bfseries 54} (1943) 426}.

\bibitem{Panico:2017frx}
G.~Panico, F.~Riva and A.~Wulzer, \emph{{Diboson Interference Resurrection}},
  \href{https://doi.org/10.1016/j.physletb.2017.11.068}{\emph{Phys. Lett.}
  {\bfseries B776} (2018) 473}
  [\href{https://arxiv.org/abs/1708.07823}{{\ttfamily 1708.07823}}].

\bibitem{Stoye:2018ovl}
M.~Stoye, J.~Brehmer, G.~Louppe, J.~Pavez and K.~Cranmer,
  \emph{{Likelihood-free inference with an improved cross-entropy estimator}},
  \href{https://arxiv.org/abs/1808.00973}{{\ttfamily 1808.00973}}.

\bibitem{SFitter_top}
I.~Brivio, S.~Bruggisser, F.~Maltoni, R.~Moutafis, T.~Plehn, E.~Vryonidou
  et~al., \emph{{Another Fit of the Dimension-6 Top Sector}}, {\emph{in
  preparation} (2019) }.

\bibitem{Biekotter:2016ecg}
A.~Biekotter, J.~Brehmer and T.~Plehn, \emph{{Extending the limits of Higgs
  effective theory}},
  \href{https://doi.org/10.1103/PhysRevD.94.055032}{\emph{Phys. Rev.}
  {\bfseries D94} (2016) 055032}
  [\href{https://arxiv.org/abs/1602.05202}{{\ttfamily 1602.05202}}].

\bibitem{Plehn:2013paa}
T.~Plehn, P.~Schichtel and D.~Wiegand, \emph{{Where boosted significances come
  from}}, \href{https://doi.org/10.1103/PhysRevD.89.054002}{\emph{Phys. Rev.}
  {\bfseries D89} (2014) 054002}
  [\href{https://arxiv.org/abs/1311.2591}{{\ttfamily 1311.2591}}].

\bibitem{Aaboud:2019nan}
{\scshape ATLAS} collaboration, \emph{{Measurement of VH, $ \mathrm{H}\to
  \mathrm{b}\overline{\mathrm{b}} $ production as a function of the
  vector-boson transverse momentum in 13 TeV pp collisions with the ATLAS
  detector}}, \href{https://doi.org/10.1007/JHEP05(2019)141}{\emph{JHEP}
  {\bfseries 05} (2019) 141}
  [\href{https://arxiv.org/abs/1903.04618}{{\ttfamily 1903.04618}}].

\bibitem{Kling:2016lay}
F.~Kling, T.~Plehn and P.~Schichtel, \emph{{Maximizing the significance in
  Higgs boson pair analyses}},
  \href{https://doi.org/10.1103/PhysRevD.95.035026}{\emph{Phys. Rev.}
  {\bfseries D95} (2017) 035026}
  [\href{https://arxiv.org/abs/1607.07441}{{\ttfamily 1607.07441}}].

\bibitem{Banerjee:2018bio}


\bibitem{Franceschini:2017xkh}
R.~Franceschini, G.~Panico, A.~Pomarol, F.~Riva and A.~Wulzer,
  \emph{{Electroweak Precision Tests in High-Energy Diboson Processes}},
  \href{https://doi.org/10.1007/JHEP02(2018)111}{\emph{JHEP} {\bfseries 02}
  (2018) 111} [\href{https://arxiv.org/abs/1712.01310}{{\ttfamily
  1712.01310}}].

\bibitem{deFavereau:2013fsa}
{\scshape DELPHES 3} collaboration, \emph{{DELPHES 3, A modular framework for
  fast simulation of a generic collider experiment}},
  \href{https://doi.org/10.1007/JHEP02(2014)057}{\emph{JHEP} {\bfseries 02}
  (2014) 057} [\href{https://arxiv.org/abs/1307.6346}{{\ttfamily 1307.6346}}].

\bibitem{Kluyver2016JupyterN}
T.~Kluyver, B.~Ragan-Kelley, F.~P{\'e}rez, B.~E. Granger, M.~Bussonnier,
  J.~Frederic et~al., \emph{Jupyter notebooks - a publishing format for
  reproducible computational workflows},  in \emph{ELPUB}, 2016.

\bibitem{madminer}
J.~Brehmer, F.~Kling, I.~Espejo and K.~Cranmer, ``{MadMiner: an inference
  toolkit for particle physics.}.''
\newblock 10.5281/zenodo.1489147.

\bibitem{Hunter:2007}
J.~D. Hunter, \emph{Matplotlib: A 2d graphics environment},
  \href{https://doi.org/10.1109/MCSE.2007.55}{\emph{Computing in Science \&
  Engineering} {\bfseries 9} (2007) 90}.

\bibitem{numpy}
T.~Oliphant, ``{NumPy}: A guide to {NumPy}.'' USA: Trelgol Publishing, 2006--.

\bibitem{lukas_2018_1217032}
Lukas, \emph{lukasheinrich/pylhe v0.0.4},  Apr., 2018.
\newblock 10.5281/zenodo.1217032.

\bibitem{Sjostrand:2014zea}
T.~Sjstrand, S.~Ask, J.~R. Christiansen, R.~Corke, N.~Desai, P.~Ilten et~al.,
  \emph{{An Introduction to PYTHIA 8.2}},
  \href{https://doi.org/10.1016/j.cpc.2015.01.024}{\emph{Comput. Phys. Commun.}
  {\bfseries 191} (2015) 159}
  [\href{https://arxiv.org/abs/1410.3012}{{\ttfamily 1410.3012}}].

\bibitem{van1995python}
G.~Van~Rossum and F.~L. Drake~Jr, \emph{Python tutorial}. Centrum voor Wiskunde
  en Informatica Amsterdam, The Netherlands, 1995.

\bibitem{paszke2017automatic}
A.~Paszke, S.~Gross, S.~Chintala, G.~Chanan, E.~Yang, Z.~DeVito et~al.,
  \emph{Automatic differentiation in pytorch},  in \emph{NIPS-W}, 2017.

\bibitem{Rodrigues:2019nct}
E.~Rodrigues, \emph{{The Scikit-HEP Project}},  in \emph{{23rd International
  Conference on Computing in High Energy and Nuclear Physics (CHEP 2018) Sofia,
  Bulgaria, July 9-13, 2018}}, 2019,
  \href{https://arxiv.org/abs/1905.00002}{{\ttfamily 1905.00002}}.

\bibitem{scikit-learn}
F.~Pedregosa, G.~Varoquaux, A.~Gramfort, V.~Michel, B.~Thirion, O.~Grisel
  et~al., \emph{Scikit-learn: Machine learning in {P}ython}, {\emph{Journal of
  Machine Learning Research} {\bfseries 12} (2011) 2825}.

\bibitem{jim_pivarski_2019_3256257}
J.~Pivarski, P.~Das, D.~Smirnov, C.~Burr, M.~Feickert, N.~Biederbeck et~al.,
  \emph{scikit-hep/uproot: 3.7.2},  June, 2019.
\newblock 10.5281/zenodo.3256257.

\bibitem{repo}
J.~Brehmer, S.~Dawson, S.~Homiller, F.~Kling and T.~Plehn, ``{Code repository
  for this paper}.''
\newblock
  \texttt{\href{https://github.com/shomiller/wh-madminer}{github.com/shomiller/wh-madminer}}.

\bibitem{CMS:2018abb}
{\scshape CMS} collaboration, \emph{{Observation of Higgs boson decay to bottom
  quarks}}, .

\bibitem{Mattelaer:2016gcx}
O.~Mattelaer, \emph{{On the maximal use of Monte Carlo samples: re-weighting
  events at NLO accuracy}},
  \href{https://doi.org/10.1140/epjc/s10052-016-4533-7}{\emph{Eur. Phys. J.}
  {\bfseries C76} (2016) 674}
  [\href{https://arxiv.org/abs/1607.00763}{{\ttfamily 1607.00763}}].

\end{thebibliography}\endgroup

\end{document}